\documentclass[acmsmall]{acmart}

\usepackage{paralist}
\usepackage{cleveref}
\usepackage{colortbl}
\usepackage{multirow}
\usepackage{subcaption}
\usepackage{tcolorbox}
\usepackage[flushleft]{threeparttable}
\usepackage{xspace}

\graphicspath{{figures/}}
\tcbset{arc=2pt, boxrule=\heavyrulewidth, colback=gray!1, colframe=black}

\newcommand{\rqi}{How much do the studied projects use Stale bot to deal with their PR backlog?\xspace}
\newcommand{\rqii}{What is the impact of Stale bot on pull-based development in the studied projects?\xspace}
\newcommand{\rqiii}{What kind of PRs are usually intervened by Stale bot in the studied projects?\xspace}

\begin{document}

\acmJournal{TOSEM}
\title{Understanding the Helpfulness of Stale Bot for Pull-based Development}
\subtitle{An Empirical Study of 20 Large Open-Source Projects}

\author{SayedHassan Khatoonabadi}
\orcid{0000-0003-0615-9242}
\affiliation{%
    \department[0]{Data-driven Analysis of Software (DAS) Lab}
    \department[1]{Department of Computer Science \& Software Engineering}
    \institution{Concordia University}
    \city{Montreal}
    \state{QC}
    \country{Canada}
}
\email{sayedhassan.khatoonabadi@mail.concordia.ca}

\author{Diego Elias Costa}
\orcid{0000-0001-7084-2594}
\affiliation{%
    \department[0]{LATECE Lab}
    \department[1]{Department of Computer Science}
    \institution{Université du Québec à Montréal (UQAM)}
    \city{Montreal}
    \state{QC}
    \country{Canada}
}
\email{costa.diego@uqam.ca}

\author{Suhaib Mujahid}
\orcid{0000-0003-2738-1387}
\affiliation{%
    \institution{Mozilla Corporation}
    \city{Montreal}
    \state{QC}
    \country{Canada}
}
\email{smujahid@mozilla.com}

\author{Emad Shihab}
\orcid{0000-0003-1285-9878}
\affiliation{%
    \department[0]{Data-driven Analysis of Software (DAS) Lab}
    \department[1]{Department of Computer Science \& Software Engineering}
    \institution{Concordia University}
    \city{Montreal}
    \state{QC}
    \country{Canada}
} 
\email{emad.shihab@concordia.ca}

\renewcommand{\shortauthors}{Khatoonabadi et al.}

\begin{abstract}
    Pull Requests (PRs) that are neither progressed nor resolved clutter the list of PRs, making it difficult for the maintainers to manage and prioritize unresolved PRs. To automatically track, follow up, and close such inactive PRs, Stale bot was introduced by GitHub. Despite its increasing adoption, there are ongoing debates on whether using Stale bot alleviates or exacerbates the problem of inactive PRs. To better understand if and how Stale bot helps projects in their pull-based development workflow, we perform an empirical study of 20 large and popular open-source projects. We find that Stale bot can help deal with a backlog of unresolved PRs as the projects closed more PRs within the first few months of adoption. Moreover, Stale bot can help improve the efficiency of the PR review process as the projects reviewed PRs that ended up merged and resolved PRs that ended up closed faster after the adoption. However, Stale bot can also negatively affect the contributors as the projects experienced a considerable decrease in their number of active contributors after the adoption. Therefore, relying solely on Stale bot to deal with inactive PRs may lead to decreased community engagement and an increased probability of contributor abandonment.
\end{abstract}

\ccsdesc[500]{Human-centered computing~Empirical studies in collaborative and social computing}
\ccsdesc[500]{Software and its engineering~Collaboration in software development}
\ccsdesc[500]{Human-centered computing~Open source software}
\ccsdesc[500]{Software and its engineering~Open source model}

\keywords{Software development bots, pull request abandonment, pull-based development, modern code review, social coding platforms, open-source software}

\maketitle

\clearpage
\section{Introduction}
Open-source projects widely adopt pull-based development as a more efficient and effective alternative to traditional methods for contributing and reviewing code changes \citep{gousios_exploratory_2014, zhu_effectiveness_2016}. In this development model, contributors submit a Pull Request (PR) to suggest changes for integration into the project. The PR is then reviewed by the project maintainers and updated by the contributor until it is ready to be merged. However, the review process of some PRs is left unfinished due to either the contributor not addressing the maintainers' comments or the maintainers not following up on the progress of the PR \citep{khatoonabadi_wasted_2023, li_are_2022, wang_why_2019}. If neither progressed nor resolved, such inactive PRs accumulate over time, clutter the list of PRs, and eventually make it difficult for the maintainers to manage and prioritize unresolved PRs \citep{li_are_2022, gousios_work_2015}. As a real-world example, a large backlog of unresolved PRs led the DefinitelyTyped project to declare ``bankruptcy'' in June 2016. Consequently, they closed all unresolved PRs submitted before May 2016 just to be able to start afresh \citep{sorensen_pull_2014}.

\begin{figure}
    \includegraphics[width=\textwidth]{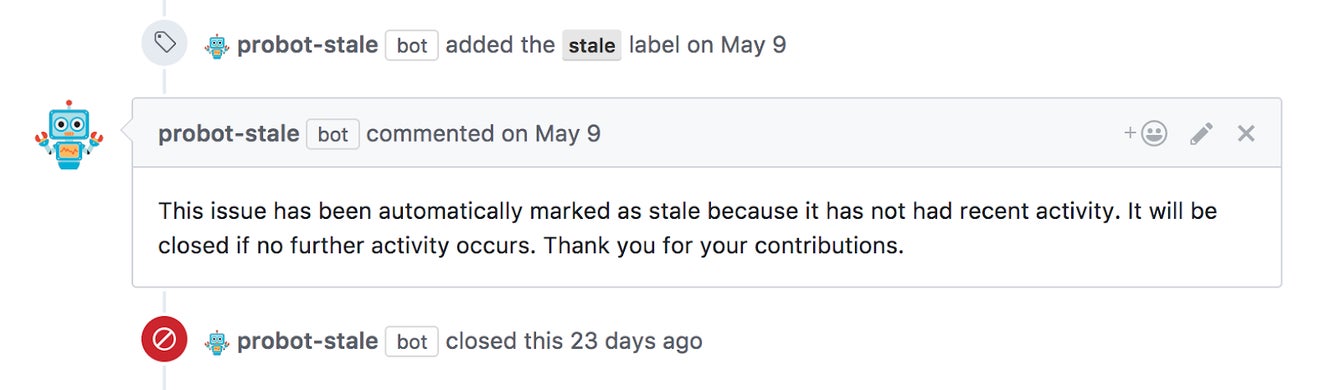}
    \caption{An example prompt by Stale bot}
    \label{fig:stale}
\end{figure}

Manually keeping track of inactive PRs, following up on their progress, and closing them if needed places an additional burden on the project maintainers who are already occupied with other development tasks \citep{khatoonabadi_wasted_2023, li_are_2022, gousios_work_2015}. To free the maintainers from manually triaging such PRs, Stale bot \citep{github_stale_2023} was released in 2017 and since then has been increasingly adopted by open-source projects on GitHub\footnote{We observed that on average 7\% more projects had adopted Stale bot each month till October 2021.}. As shown in \Cref{fig:stale}, Stale bot aims to make the status of open and not progressing PRs explicit by automatically labeling, commenting, and closing PRs after a pre-configured period of inactivity \citep{keepers_stale_2023}. Nevertheless, there are ongoing debates within the open-source community on whether using Stale bot alleviates or exacerbates the problem of inactive PRs. The creators of Stale bot claim that based on the experience of hundreds of projects and organizations, Stale bot is an effective method for focusing on the work that matters most \citep{keepers_stale_2023}. Conversely, part of the community regards Stale bot as ``harmful'' \citep{devault_github_2021} and a ``false economy'' \citep{winding_github_2021}. They argue that while Stale bot may initially seem helpful, it results in duplicated PRs, fragmented information, and eventually frustration in the community. Some studies have also incidentally mentioned that Stale bot can introduce noise and friction for both the contributors and the maintainers \citep{wessel_dont_2021, liu_understanding_2020, wessel_inconvenient_2020, farah_exploratory_2022, rahman_towards_2022}.

Despite all these positive and negative claims, the helpfulness of Stale bot has not yet been empirically validated. Therefore, we set out to better understand if and how adopting Stale bot helps open-source projects in their pull-based development workflow. This investigation is particularly important as Stale bot is commonly used to deal with inactive or abandoned PRs. Towards this goal, we perform an empirical study \citep{kitchenham_evidence-based_2015, creswell_research_2017} of 20 large and popular open-source projects on GitHub that have used Stale bot for at least one consecutive year. Specifically, we aim to answer the following research questions in this paper:

\smallskip
\begin{itemize}
    \item[\textbf{RQ\textsubscript{1}:}] \textbf{\rqi} We analyze the configuration and activity of Stale bot to understand the extent to which large open-source projects rely on Stale bot to automatically deal with their unresolved PRs. Our results show that the usage level of Stale bot widely varies among the studied projects. On average each month, Stale bot intervened in less than 25\% of open PRs in nine projects, between 25\% and 50\% of open PRs in five projects, and more than 50\% of open PRs in six projects. Unexpectedly, the projects with a larger backlog of unresolved PRs have typically not relied more aggressively on Stale bot to deal with their unresolved PRs.
    \smallskip
    \item[\textbf{RQ\textsubscript{2}:}] \textbf{\rqii} We apply interrupted time-series analysis \citep{wagner_segmented_2002} as a well-established quasi-experiment to understand if and how adopting Stale bot improves the efficiency and effectiveness of the pull-based development workflow in large open-source projects. Our results show that the studied projects closed more PRs within the first few months of adopting Stale bot, but overall closed and merged fewer PRs afterward. The adoption of Stale bot is also associated with faster first reviews in merged PRs, faster resolutions in closed PRs, slightly fewer updates in merged PRs, and considerably fewer active contributors in the projects.
    \smallskip
    \item[\textbf{RQ\textsubscript{3}:}] \textbf{\rqiii} We analyze the characteristics of PRs intervened by Stale bot, as well as their contributors and review processes, to understand the factors that are associated with a higher probability of getting intervened by Stale bot in large open-source projects. Our results show that Stale bot tends to intervene more in complex PRs, PRs from novice contributors, and PRs with lengthy review processes. Specifically, besides the resolution time of PRs, the largest differences are observed in the number of prior PRs by contributors, the mean response latency of PRs, the acceptance rate of contributors, and the contribution period of contributors.
\end{itemize}
\smallskip

Our findings imply that adopting Stale bot helped the studied projects deal with their accumulated backlog of unresolved PRs. Stale bot has also improved the efficiency of the review process of PRs by helping the maintainers focus on PRs that are more likely to get merged. Despite these advantages, the adoption of Stale bot also brings some disadvantages. For example, the projects experienced a decrease in their number of active contributors after the adoption. Besides, Stale bot also tends to intervene more in PRs submitted by novice contributors. However, such contributors are the ones who face the most barriers and thus need the most guidance from the maintainers \citep{steinmacher_social_2015, steinmacher_preliminary_2014, steinmacher_why_2013}. Prior studies have also highlighted the importance of attracting newcomers to ensure the sustainability of projects \citep{zhou_what_2012}. Therefore, relying solely on Stale bot to deal with inactive PRs may lead to decreased community engagement and an increased probability of contributor abandonment. In conclusion, our study provides a better understanding of the potential benefits and drawbacks of employing Stale bot within a pull-based development workflow.

\noindentparagraph{\emph{\textbf{Our Contributions.}}} In summary, we make the following contributions in this paper:

\begin{itemize}
    \item To the best of our knowledge, this is the first in-depth study investigating the helpfulness of Stale bot for the pull-based development workflow in large open-source projects.
    \item We provide empirical evidence on the reliance of projects on Stale bot to deal with their PR backlog, the impact of Stale bot on pull-based development, and the kind of PRs usually intervened by Stale bot.
    \item To promote the reproducibility of our study and facilitate future research on this topic, we publicly share our dataset online at \url{https://doi.org/10.5281/zenodo.7978381}.
\end{itemize}

\noindentparagraph{\emph{\textbf{Paper Organization.}}} The rest of the paper is organized as follows. \Cref{sec:related_work} reviews the related work and \Cref{sec:dataset} overviews the dataset used for our study. Then, \Cref{sec:rqi,sec:rqii,sec:rqiii} report our approach and findings to answer each research question, and \Cref{implications} discusses the implications of our study. Finally, \Cref{sec:limitations} describes the limitations of our study, and \Cref{sec:conclusion} concludes this paper with a summary of our study.

\section{Related Work}
\label{sec:related_work}
In the following, we first overview research regarding the usage and challenges of Stale bot before moving on to studies on the reasons and consequences of PR abandonment. Next, we review studies on the impact of tools on pull-based development and then summarize studies on the review latency and decision of PRs.

\noindentparagraph{\emph{\textbf{Usage and Challenges of Stale Bot.}}} Stale bot was released in 2017 to automatically triage abandoned issues and PRs and is adopted by many open-source projects on GitHub \citep{github_stale_2023, keepers_stale_2023, wessel_should_2019}. However, little is known about the usage and challenges of Stale bot in the literature. In a preliminary study, \citet{wessel_should_2019} investigated how projects adapt and maintain Stale bot over time by analyzing its configuration history in 765 open-source projects. They found that most projects use Stale bot to triage both their issues and PRs. Furthermore, they found that while most projects do not modify the configuration of Stale bot after its initial setup, the few that do rarely make more than three modifications subsequently. Therefore, they concluded that setting up and using Stale bot does not require much effort from the projects. Several studies \citep{wessel_dont_2021, liu_understanding_2020, wessel_inconvenient_2020, farah_exploratory_2022, rahman_towards_2022} have also incidentally mentioned that Stale bot introduces noise and friction for both the contributors and the maintainers.

Nevertheless, if and how Stale bot can actually be helpful to open-source projects has not yet been studied. To fill this knowledge gap, our paper empirically studies the helpfulness of Stale bot in the context of pull-based development. Specifically, we investigate the reliance of projects on Stale bot to deal with their PR backlog, the impact of Stale bot on pull-based development, and the kind of PRs usually intervened by Stale bot.

\noindentparagraph{\emph{\textbf{Reasons and Consequences of PR Abandonment.}}} PR abandonment as a challenge that wastes time and effort of both contributors and maintainers has only recently received attention from the literature. \citet{li_are_2022} attempted to understand the reasons for, the impacts of, and the coping strategies for abandoned PRs by qualitatively studying five popular open-source projects on GitHub. They observed that PRs are abandoned mostly due to the lack of maintainers' responsiveness and the lack of contributors' time and interest. They also reported that PR abandonment increases the efforts needed to manage and maintain projects because abandoned PRs clutter the list of PRs, waste review efforts, require additional attention for a careful closing, delay the landing of interdependent PRs, result in duplicated PRs, disorder project milestones, and finally leave a bad impression.

As a complementary study, \citet{khatoonabadi_wasted_2023} conducted a mixed-methods study using both quantitative and qualitative methods to better understand the underlying dynamics of PRs abandoned by their contributors in 10 popular and large open-source projects on GitHub. By investigating the influence of various factors related to PRs, contributors, review processes, and projects on the abandonment probability of PRs, they found that complex PRs, PRs from novice contributors, and PRs with long discussions are more likely to get abandoned. They also observed that the most frequent reasons for PR abandonment are related to the obstacles faced by the contributors, followed by the hurdles imposed by the maintainers during the review process of PRs. While these studies investigated the reasons and consequences of PR abandonment, our study focuses on the usage and impact of Stale bot as a common strategy to deal with abandoned PRs in open-source projects.

\noindentparagraph{\emph{\textbf{Impact of Tools on Pull-based Development.}}} Open-source projects are adopting various automation tools and bots to improve the efficiency and effectiveness of their pull-based development workflow \citep{wessel_bots_2022, wessel_power_2018, storey_disrupting_2016}. Several studies have evaluated the impact of adopting such tools in open-source projects by applying interrupted time-series analysis \citep{wagner_segmented_2002} (a.k.a. regression discontinuity design). \citet{zhao_impact_2017} was the first to apply this method for understanding the impact of adopting Travis CI. They found that the adoption slows down the increasing trend in the number of merge commits, accelerates the decreasing trend in the merge commit churn, and reverses the increasing trend in the number of closed PRs. \citet{cassee_silent_2020} further studied the impact of adopting Travis CI and found that it also slows down the increasing trend in the number of discussion comments in PRs.

\citet{wessel_quality_2022} studied the impact of adopting code review bots and found that it accelerates the increasing trend in the number of merged PRs, accelerates the decreasing trend in the number of closed PRs, decreases the number of comments in all PRs, increases the resolution time of merged PRs, and reverses the increasing trend in the resolution time of closed PRs. In another study, \citet{wessel_github_2022} investigated the impact of adopting GitHub Actions and found that it decreases the number of both merged PRs and closed PRs, increases the number of discussion comments in merged PRs, decreases the number of discussion comments in closed PRs, increases the resolution time of merged PRs, and decreases the resolution time of closed PRs. However, our study focuses on investigating the impact of adopting Stale bot to better understand its helpfulness for pull-based development.

\noindentparagraph{\emph{\textbf{Review Latency and Decision of PRs.}}} The influence of various technical, social, and personal factors on the review latency and decision of PRs has been extensively studied in the literature. \citet{gousios_exploratory_2014} was the first to investigate how technical factors can affect the merge decision and merge time of PRs. They found that the merge decision is mainly associated with whether the PR touches recently modified code, while the merge time is associated with the track record of the developer, as well as the size of the project, its test coverage, and its openness to external contributions. \citet{gousios_work_2015} reports that the decision of integrators to accept a contribution is based on its quality, including conformance to the project style and architecture, source code quality, and test coverage. \citet{tsay_influence_2014} showed that in addition to technical factors, social factors could influence the acceptance of PRs. They found that while PRs with lots of comments are associated with a lower probability of getting accepted, the prior interactions of submitters in the project moderate this effect. Additionally, they found that well-established projects are more conservative while evaluating contributions.

Since then, many studies have continued investigating the role of different technical and social factors \citep{soares_acceptance_2015, yu_determinants_2016, kononenko_studying_2018, pinto_who_2018, zou_how_2019, lenarduzzi_does_2021}, as well as various personal and demographic factors \citep{rastogi_biases_2016, terrell_gender_2017, rastogi_relationship_2018, furtado_how_2021, iyer_effects_2021, nadri_insights_2021, nadri_relationship_2022} on the review latency or decision of PRs. Recently, \citet{zhang_pull_2022} and \citet{zhang_pull_2023} conducted a large-scale empirical study of how a large set of factors identified through a systematic literature review can explain the review latency and decision of PRs, respectively. While these studies investigated what factors can make a PR get reviewed faster and accepted, our study investigates the characteristics of inactive PRs usually intervened by Stale bot to better understand its impacts.

\section{Dataset}
\label{sec:dataset}
For our study, we need large and popular open-source projects as their higher workload makes them more likely to benefit significantly from the adoption of Stale bot. The projects should also have a rich history of using Stale bot as part of their pull-based development workflow to ensure that we have enough data for our analyses. To identify such projects, we rely on GH Archive \citep{grigorik_gh_2023}, which archives all the public events happening on GitHub \citep{github_events_2023}. First, we query the GH Archive's public dataset on Google BigQuery \citep{google_bigquery_2023} and look for all the events performed by Stale bot on PRs (i.e., the actor name is \texttt{stale[bot]} and the payload string contains \texttt{"pull\_request"}). Then, we use the retrieved events to identify the projects that have ever used Stale bot in their pull-based development workflow. Next, we collect the timeline of activities for the identified projects using the \texttt{PyGithub} package \citep{jacques_pygithub_2023} on November 17th, 2021. The timeline of activities of PRs is provided by the GitHub API \citep{github_rest_2023} and includes the details (e.g., type, actor, and time) of all the events (e.g., commits, comments, labelings, and resolutions) that happened during their lifecycle \citep{github_timeline_2023, github_pulls_2023, github_issues_2023}.

We then determine the adoption time of each project by looking for the first event performed by Stale bot in the PRs of the project. Following the recommendation of \citet{wagner_segmented_2002}, we consider 12 months before and 12 months after the adoption (a total of two years) as our observation period. This period allows us to ensure that we have a sufficient number of observations for our analyses and to take into account the expected seasonal variations in the projects \citep{wagner_segmented_2002}. Therefore, we focus on the projects that have at least 12 months of pull-based development history before the adoption and that have also used Stale bot for at least 12 consecutive months in their pull-based development workflow after the adoption. Among these projects, we finally select the top 20 with the most PRs to focus on the largest and most popular projects.

\Cref{tab:projects} provides an overview of the projects that we selected for our study. In summary, the selected projects have thousands of PRs (median of 18,975), thousands of stars (median of 6,888), hundreds of contributors (median of 1,137), tens of maintainers (median of 41), years of pull-based development history (median of 81 months), and years of using Stale bot (median of 28 months). Additionally, these projects span multiple application domains and programming languages, providing a more diverse selection of projects for our study.

\begin{table}
    \caption{Overview of the projects selected to study the helpfulness of Stale bot for pull-based development.}
    \label{tab:projects}
    \resizebox{\textwidth}{!}{%
        \begin{tabular}{@{}l|cccccc|ll@{}}
            \toprule
            \multirow{2}{*}{\textbf{Project}} & \multirow{2}{*}{\textbf{PRs}} &\multirow{2}{*}{\textbf{Stars}} & \multirow{2}{*}{\textbf{Contributors}} & \multirow{2}{*}{\textbf{Maintainers}} & \textbf{Age}       & \textbf{Months}         & \multirow{2}{*}{\textbf{Domain}} & \multirow{2}{*}{\textbf{Language}} \\
                                              &                               &                                &                                        &                                       & \textbf{in Months} & \textbf{Since Adoption} &                                  &                                    \\
            \midrule
            nixos/nixpkgs                     & 121,998                       & 8,013                          & 4,629                                  & 300                                   & 111                & 17                      & Package Manager                  & Nix                                \\
            homebrew/homebrew-core            & 84,686                        & 10,203                         & 6,747                                  & 50                                    & 67                 & 54                      & Package Manager                  & Ruby                               \\
            ceph/ceph                         & 43,885                        & 9,826                          & 1,573                                  & 73                                    & 120                & 35                      & Storage Platform                 & C++                                \\
            automattic/wp-calypso             & 38,301                        & 11,922                         & 707                                    & 151                                   & 70                 & 29                      & WordPress Frontend               & JavaScript                         \\
            home-assistant/core               & 34,922                        & 47,437                         & 3,875                                  & 51                                    & 96                 & 19                      & Home Automation                  & Python                             \\
            cleverraven/cataclysm-dda         & 33,239                        & 5,762                          & 1,686                                  & 36                                    & 107                & 26                      & Video Game                       & C++                                \\
            homebrew/linuxbrew-core           & 23,259                        & 1,171                          & 285                                    & 13                                    & 64                 & 47                      & Package Manager                  & Ruby                               \\
            istio/istio                       & 21,236                        & 28,423                         & 1,053                                  & 78                                    & 58                 & 15                      & Microservice Mesh                & Go                                 \\
            qgis/qgis                         & 19,587                        & 5,092                          & 569                                    & 39                                    & 124                & 39                      & GIS System                       & C++                                \\
            devexpress/devextreme             & 19,412                        & 1,521                          & 152                                    & 21                                    & 53                 & 18                      & Web Framework                    & JavaScript                         \\
            grafana/grafana                   & 18,538                        & 44,786                         & 2,259                                  & 55                                    & 93                 & 21                      & Visualization Platform           & TypeScript                         \\
            wikia/app                         & 18,293                        & 191                            & 217                                    & 64                                    & 105                & 34                      & Wiki Engine                      & PHP                                \\
            helm/charts                       & 18,196                        & 15,305                         & 4,958                                  & 16                                    & 69                 & 34                      & Kubernetes Apps                  & Go                                 \\
            grpc/grpc                         & 17,939                        & 32,361                         & 1,221                                  & 49                                    & 81                 & 25                      & RPC Framework                    & C++                                \\
            solana-labs/solana                & 17,805                        & 5,351                          & 275                                    & 30                                    & 44                 & 25                      & Decentralized Blockchain         & Rust                               \\
            home-assistant/home-assistant.io  & 17,780                        & 2,253                          & 4,410                                  & 28                                    & 81                 & 46                      & Home Automation                  & HTML                               \\
            conda-forge/staged-recipes        & 16,119                        & 487                            & 2,443                                  & 32                                    & 72                 & 19                      & Package Manager                  & Python                             \\
            apache/beam                       & 15,924                        & 5,068                          & 961                                    & 33                                    & 68                 & 38                      & Data Pipelines                   & Java                               \\
            frappe/erpnext                    & 15,840                        & 9,970                          & 580                                    & 42                                    & 122                & 39                      & ERP System                       & Python                             \\
            riot-os/riot                      & 14,335                        & 3,985                          & 447                                    & 33                                    & 104                & 26                      & Operating System                 & C                                  \\
            \bottomrule
        \end{tabular}
    }
\end{table}

\section{RQ\texorpdfstring{\textsubscript{1}}{1}: \rqi}
\label{sec:rqi}
Open-source projects on GitHub are adopting Stale bot to automatically follow up on the status of their inactive PRs, warn the contributors and maintainers about the lack of activity, and eventually close such PRs if there is no further progress on them \citep{github_stale_2023, keepers_stale_2023, wessel_should_2019}. As our first research question, we aim to understand the extent to which large open-source projects rely on Stale bot to automatically deal with their unresolved PRs. In the following, we first explain our approach and then discuss our findings to answer this research question.

\subsection{Approach}
To investigate the usage of Stale bot, we analyze both its configuration and activity during its first year of adoption (i.e., our observation period) in the studied projects. For each project, we extract the configured number of days of inactivity before a PR is marked as stale (i.e., \texttt{daysUntilStale}) and the configured number of days of inactivity before a PR already marked as stale gets closed (i.e., \texttt{daysUntilClose}) from its configuration file of Stale bot (i.e., \texttt{.github/stale.yml}). We also measure the monthly activity of Stale bot in the PRs of each project as a proxy for the extent to which the project actually relies on Stale bot to deal with its unresolved PRs. Since Stale bot can either warn about the lack of activity in a PR or close a PR due to inactivity, we further distinguish between these two types of activities. For this purpose, we use the following definitions to classify Stale bot activities in a PR:

\begin{itemize}
    \item \textbf{Intervention:} Any commenting, labeling, unlabeling, or closing event performed by Stale bot.
    \item \textbf{Warning:} Any commenting or labeling event performed by Stale bot that is not immediately followed by a closing event from Stale bot within a minute (to account for the processing delay between the closing comment and the actual closure of a PR on GitHub).
    \item \textbf{Closure:} A closing event performed by Stale bot.
\end{itemize}

\subsection{Findings}
\Cref{tab:usage} provides an overview of the usage of Stale bot during its first year of adoption across the studied projects, indicating \begin{inparaenum} \item the number of open PRs immediately upon the adoption (Backlog), \item the average number of open PRs at the beginning of each month (\#~Open~PRs), \item the average ratio of open PRs intervened by Stale bot in each month (\%~Intervened~PRs), \item the average ratio of open PRs warned by Stale bot in each month (\%~Warned~PRs), \item the average ratio of open PRs closed by Stale bot in each month (\%~Closed~PRs), \item the average configured number of days to stale a PR in each month (Days~to~Stale), and \item the average configured number of days to close a stale PR in each month (Days~to~Close). \end{inparaenum} Note that the ratio of intervened PRs might not equal the sum of the ratios of warned PRs and closed PRs in a project. This discrepancy arises because intervention includes other activities (i.e., unlabeling events) aside from warnings and closures, and also because PRs might get intervened multiple times within a month (e.g., closure after a warning).

\begin{table}
    \caption{Usage of Stale bot during its first year of adoption across the studied projects. {\colorbox{gray}{Dark gray}} highlights activities in more than 50\% of monthly open PRs and {\colorbox{lightgray}{light gray}} highlights activities in between 25\% and 50\% of monthly open PRs.}
    \label{tab:usage}
    \resizebox{\textwidth}{!}{%
        \begin{tabular}{@{}cl|c|cccc|cc@{}}
            \toprule
                                               &                                  &                  & \multicolumn{6}{c}{\textbf{Monthly Average}}                                                                                                                     \\
            \textbf{Usage Level}               & \textbf{Project}                 & \textbf{Backlog} & \textbf{\# Open PRs} & \textbf{\% Intervened PRs}  & \textbf{\% Warned PRs}      & \textbf{\% Closed PRs}      & \textbf{Days to Stale} & \textbf{Days to Close} \\
            \midrule
            \multirow{6}{*}{\textbf{High}}     & solana-labs/solana               & 27               & 28                   & \cellcolor{gray}70.8\%      & \cellcolor{gray}65.5\%      & \cellcolor{lightgray}30.5\% & 15                     & 7                      \\
                                               & grafana/grafana                  & 139              & 109                  & \cellcolor{gray}61.3\%      & \cellcolor{lightgray}48.4\% & 11.0\%                      & 14                     & 30                     \\
                                               & istio/istio                      & 151              & 160                  & \cellcolor{gray}60.7\%      & \cellcolor{lightgray}47.3\% & 12.1\%                      & 14                     & 27                     \\
                                               & homebrew/linuxbrew-core          & 174              & 58                   & \cellcolor{gray}58.4\%      & \cellcolor{gray}51.8\%      & \cellcolor{lightgray}37.3\% & 21                     & 7                      \\
                                               & helm/charts                      & 423              & 366                  & \cellcolor{gray}54.8\%      & \cellcolor{lightgray}43.8\% & 23.0\%                      & 30                     & 14                     \\
                                               & frappe/erpnext                   & 14               & 44                   & \cellcolor{gray}51.0\%      & \cellcolor{lightgray}46.9\% & 13.7\%                      & 22                     & 7                      \\
            \midrule
            \multirow{5}{*}{\textbf{Moderate}} & qgis/qgis                        & 40               & 31                   & \cellcolor{lightgray}48.0\% & \cellcolor{lightgray}41.9\% & 18.4\%                      & 14                     & 7                      \\
                                               & wikia/app                        & 23               & 24                   & \cellcolor{lightgray}46.1\% & \cellcolor{lightgray}41.6\% & 19.8\%                      & 30                     & 7                      \\
                                               & homebrew/homebrew-core           & 72               & 82                   & \cellcolor{lightgray}32.5\% & \cellcolor{lightgray}28.7\% & 9.0\%                       & 21                     & 7                      \\
                                               & conda-forge/staged-recipes       & 543              & 226                  & \cellcolor{lightgray}29.7\% & 16.7\%                      & 13.0\%                      & 150                    & 30                     \\
                                               & home-assistant/core              & 228              & 270                  & \cellcolor{lightgray}29.0\% & \cellcolor{lightgray}26.8\% & 7.9\%                       & 65                     & 7                      \\
            \midrule
            \multirow{9}{*}{\textbf{Low}}      & grpc/grpc                        & 332              & 207                  & 22.2\%                      & 17.3\%                      & 11.5\%                      & 150                    & 6                      \\
                                               & ceph/ceph                        & 742              & 637                  & 20.2\%                      & 15.0\%                      & 3.7\%                       & 60                     & 90                     \\
                                               & apache/beam                      & 127              & 95                   & 15.4\%                      & 13.2\%                      & 7.7\%                       & 60                     & 7                      \\
                                               & cleverraven/cataclysm-dda        & 65               & 95                   & 13.2\%                      & 10.3\%                      & 1.6\%                       & 30                     & 30                     \\
                                               & riot-os/riot                     & 555              & 454                  & 11.0\%                      & 7.0\%                       & 3.7\%                       & 185                    & 31                     \\
                                               & devexpress/devextreme            & 31               & 31                   & 10.6\%                      & 9.5\%                       & 3.9\%                       & 30                     & 5                      \\
                                               & home-assistant/home-assistant.io & 53               & 65                   & 7.6\%                       & 6.5\%                       & 3.7\%                       & 60                     & 7                      \\
                                               & nixos/nixpkgs                    & 2,054            & 2,324                & 6.9\%                       & 5.3\%                       & 0.0\%                       & 180                    & --                     \\
                                               & automattic/wp-calypso            & 293              & 349                  & 2.9\%                       & 2.5\%                       & 1.8\%                       & 270                    & 7                      \\
            \bottomrule
        \end{tabular}
    }
\end{table}

For easier comparison, we group the projects based on the average monthly ratio of open PRs intervened by Stale bot (i.e., \%~Intervened~PRs) into three levels of usage: \begin{inparaenum} \item six projects have a high usage level where Stale bot intervened in more than 50\% of monthly open PRs; \item five projects have a moderate usage level where Stale bot intervened in between 25\% and 50\% of monthly open PRs; and \item nine projects have a low usage level where Stale bot intervened in less than 25\% of monthly open PRs. \end{inparaenum} In the following, we discuss our findings in more detail.

\noindentparagraph{\emph{\textbf{The usage level of Stale bot widely varies among the projects.}}} On average, Stale bot intervened between 2.9\% and 70.8\% of open PRs, warned between 2.5\% and 65.5\% of open PRs, and closed between 0\% and 37.3\% of open PRs each month in the projects. For example, in the nixos/nixpkgs project, which has both the largest PR backlog upon the adoption and the highest average monthly number of open PRs after the adoption among the studied projects, Stale bot only intervened in 6.9\% of open PRs on average each month, and was not even configured to close any of them. In contrast, the solana-labs/solana project has the second lowest average monthly number of open PRs after the adoption among the studied projects, yet Stale bot warned 65.5\% of open PRs and closed 30.5\% of them on average each month. This wide range of activities indicates that while some projects rely conservatively on Stale bot, others rely on it aggressively to deal with their unresolved PRs.

\noindentparagraph{\emph{\textbf{The projects with a larger backlog of unresolved PRs have typically not relied more aggressively on Stale bot.}}} The average monthly ratio of open PRs intervened by Stale bot is not significantly associated with the size of the PR backlog upon the adoption (Spearman's $\rho=-0.31$). Similarly, it is not significantly associated with the average monthly number of open PRs after the adoption (Spearman's $\rho=-0.38$). These results suggest that the higher activity of Stale bot in a project is not usually due to more accumulated unresolved PRs. Indeed, there is a significant negative correlation between the average monthly ratio of open PRs intervened by Stale bot and the average configured number of days to mark a PR as stale (Spearman's $\rho=-0.79$), suggesting that the higher activity of Stale bot in a project is usually due to a more aggressive configuration (i.e., fewer days of inactivity before a PR is marked as stale). In other words, the differences in the activity of Stale bot in different projects tend to be due to its configuration rather than the PR backlog size in a project.

\bigskip
\begin{tcolorbox}
    \paragraph{\emph{\textbf{Answer to RQ\textsubscript{1}}}} We find that the usage level of Stale bot widely varies among the studied projects. On average each month, Stale bot intervened in less than 25\% of open PRs in nine projects, between 25\% and 50\% of open PRs in five projects, and more than 50\% of open PRs in six projects. Unexpectedly, the projects with a larger backlog of unresolved PRs have typically not relied more aggressively on Stale bot to deal with their unresolved PRs.
\end{tcolorbox}
\medskip

\section{RQ\texorpdfstring{\textsubscript{2}}{2}: \rqii}
\label{sec:rqii}
In the previous research question, we found that the studied projects rely on Stale bot to deal with their accumulated backlog of unresolved PRs. As our second research question, we aim to understand if and how adopting Stale bot improves the efficiency and effectiveness of the pull-based development workflow in large open-source projects. In the following, we first explain our approach and then discuss our findings to answer this research question.

\subsection{Approach}
To investigate the impact of adopting Stale bot, we first need to quantify the performance of the pull-based development workflow in the studied projects. For this purpose, we consult similar studies that investigated the effect of an intervention on the pull-based development of open-source projects \citep{wessel_quality_2022, wessel_github_2022}. As shown in \Cref{tab:indicators}, we measure 13 performance indicators covering six dimensions: \begin{inparaenum} \item resolved PRs, \item review latency, \item resolution time, \item review discussion, \item PR updates, and \item contributor retention. \end{inparaenum}

\begin{table}
    \caption{Overview of the indicators measured to quantify the performance of the pull-based development workflow in the studied projects.}
    \label{tab:indicators}
    \resizebox{\textwidth}{!}{%
        \begin{tabular}{@{}l|p{2.8in}|l|l@{}}
            \toprule

            \textbf{Dimension}                              & \textbf{Rationale}                                                                                                                      & \textbf{Indicator}            & \textbf{Description}                                                                             \\
            \midrule
            \multirow{2}{*}{\textbf{Resolved PRs}}          & \multirow{2}{2.8in}{Stale bot closes inactive PRs and frees maintainers from manually tracking inactive PRs.}                           & merged\_pulls                 & Number of merged PRs within a month                                                              \\
                                                            &                                                                                                                                         & closed\_pulls                 & Number of closed PRs within a month                                                              \\
            \midrule
            \multirow{4.5}{*}{\textbf{Review Latency}}      & \multirow{4.5}{2.8in}{Stale bot frees maintainers from triaging inactive PRs by automatically tracking, warning, and closing such PRs.} & first\_latency\_m             & Average first response latency (excluding Stale bot) of merged PRs within a month in hours       \\
                                                            &                                                                                                                                         & first\_latency\_c             & Average first response latency (excluding Stale bot) of closed PRs within a month in hours       \\
            \cmidrule{3-4}
                                                            &                                                                                                                                         & mean\_latency\_m              & Average of the mean response latency (excluding Stale bot) of merged PRs within a month in hours \\
                                                            &                                                                                                                                         & mean\_latency\_c              & Average of the mean response latency (excluding Stale bot) of closed PRs within a month in hours \\
            \midrule
            \multirow{2}{*}{\textbf{Resolution Time}}       & \multirow{2}{2.8in}{Stale bot prevents PRs from staying indefinitely open without any progress by closing inactive PRs.}                & resolution\_time\_m           & Average resolution time of merged PRs within a month in hours                                    \\
                                                            &                                                                                                                                         & resolution\_time\_c           & Average resolution time of closed PRs within a month in hours                                    \\
            \midrule
            \multirow{2}{*}{\textbf{Review Discussion}}     & \multirow{2}{2.8in}{Stale bot attracts the attention of participants by warning about the lack of activity.}                            & comments\_m                   & Average number of comments (excluding Stale bot) in merged PRs within a month                    \\
                                                            &                                                                                                                                         & comments\_c                   & Average number of comments (excluding Stale bot) in closed PRs within a month                    \\
            \midrule
            \multirow{2}{*}{\textbf{PR Updates}}            & \multirow{2}{2.8in}{Stale bot prevents PRs from getting too outdated by warning about the lack of activity.}                            & commits\_m                    & Average number of commits in merged PRs within a month                                           \\
                                                            &                                                                                                                                         & commits\_c                    & Average number of commits in closed PRs within a month                                           \\
            \midrule
            \multirow{2}{*}{\textbf{Contributor Retention}} & \multirow{2}{2.8in}{Stale bot is frequently reported to frustrate the community by attempting to close PRs.}                            & \multirow{2}{*}{contributors} & \multirow{2}{*}{Number of active contributors within a month}                                    \\
                                                            &                                                                                                                                         &                               &                                                                                                  \\
            \bottomrule
        \end{tabular}
    }
\end{table}

Similar to previous studies \citep{wessel_quality_2022, wessel_github_2022}, we differentiate between the indicators of merged PRs (ending with \_m) and closed PRs (ending with \_c) to take into account the inherent differences in the characteristics of accepted and rejected PRs. To identify merged PRs, we resort to heuristics \citep{khatoonabadi_gap2wss_2021} similar to \citep{kalliamvakou_-depth_2016} because accepted PRs are not always merged using the standard methods provided through the GitHub interface. Accordingly, besides PRs with an explicit merged status (i.e., merged using the GitHub interface), we consider closed PRs with a commit inside the project that references them (e.g., ``Close \#123'') as merged. For each month before and after the adoption, we then measure the indicators of each project by aggregating the data about PRs that have been merged or closed in that month's timeframe.

To estimate the potential impact of Stale bot on the indicators of the projects, we rely on interrupted time-series analysis (ITS) \citep{wagner_segmented_2002}. This method is a well-established quasi-experiment previously used in the software engineering literature to study the impact of an intervention (e.g., \citep{zhao_impact_2017, wessel_quality_2022, wessel_github_2022}). To estimate the longitudinal impact of an intervention (i.e., the adoption of Stale bot in our case), ITS compares a period before and after the intervention assuming the trend would be retained if the intervention had not occurred. To perform ITS in our study, we use the following linear regression model:
\begin{equation}
    \label{eq:its}
    Y_t = \beta_0 + \beta_1 \times \mathit{time}_t + \beta_2 \times \mathit{adoption}_t + \beta_3 \times \mathit{time\_since\_adoption}_t + \beta_4 \times \mathit{controls}
\end{equation}
where $Y_t$ represents the value of indicator $Y$ at time $t$; \textit{time} represents the number of months passed since the start of the observation period at time $t$ (encoded from 1 to 24 covering one year before and one year after the adoption); \textit{adoption} represents whether Stale bot has been adopted at time $t$ (encoded as 0 before the adoption and as 1 after the adoption); \textit{time\_since\_adoption} represents the number of months passed since the adoption of Stale bot at time $t$ (encoded as 1 to 12 covering one year after the adoption and as 0 before the adoption); and \textit{controls} includes a set of variables to control for the inherent differences among the studied projects. These control variables include: \begin{inparaenum} \item the age of the project at the adoption time in months (denoted as \textit{age\_at\_adoption}) as a proxy for the level of maturity in the project, \item the number of PRs at the adoption time in the project (denoted as \textit{pulls\_at\_adoption}) as a proxy for the level of activity in the project, \item the number of contributors at the adoption time in the project (denoted as \textit{contributors\_at\_adoption}) as a proxy for the size of the project community, and \item the number of maintainers at the adoption time in the project (denoted as \textit{maintainers\_at\_adoption}) as a proxy for the size of the project core team. \end{inparaenum}

To implement the ITS regression in \Cref{eq:its}, we build linear mixed-effects models \citep{west_linear_2014, galecki_linear_2013} using the \texttt{lmerTest} package \citep{kuznetsova_lmertest_2017}. To do so, we consider \textit{age\_at\_adoption}, \textit{pulls\_at\_adoption}, \textit{contributors\_at\_adoption}, and \textit{maintainers\_at\_adoption} as fixed effects and the name of the project (denoted as \textit{project\_name}) as the random intercept. While both these fixed-effect and random-effect variables aim to capture project-to-project variability, the random intercept allows for capturing unmeasured variability between the projects by assigning a different intercept to each project. We also log-transform all the variables with a skewed distribution to better satisfy the assumptions of linear mixed-effects models, such as linearity and homoscedasticity \citep{west_linear_2014}.

To evaluate the goodness of fit of our models, we measure the marginal and conditional coefficients of determination ($R^2$) \citep{nakagawa_coefficient_2017} using the \texttt{performance} package \citep{ludecke_performance_2021}. While the marginal $R^2$ describes the proportion of the total variance explained only by the fixed effects, the conditional $R^2$ describes the proportion of the total variance explained by both the fixed and the random effects. To estimate the statistical significance of the coefficients in our models, we rely on the Satterthwaite's method \citep{hrong-tai_fai_approximate_1996} calculated using the \texttt{lmerTest} package \citep{kuznetsova_lmertest_2017}. We consider the traditional confidence level of 95\% (i.e., $\alpha = 0.05$) \citep{chavalarias_evolution_2016} to identify significant variables. Additionally, to estimate the effect size of each variable, we measure its sum of squares based on type III analysis of variance (a.k.a. ANOVA) \citep{goodnight_tests_1980} using the \texttt{lmerTest} package \citep{kuznetsova_lmertest_2017}. The results describe the fraction of the total variance explained by the model that can be attributed to each variable.

To determine if the adoption of Stale bot has a potential impact on an indicator, we check the statistical significance of \textit{time}, \textit{adoption}, and \textit{time\_since\_adoption} in the corresponding model. Specifically, a significant \textit{time} shows the existing trend before the adoption, a significant \textit{adoption} shows the change in level at the adoption time, and a significant \textit{time\_since\_adoption} shows the change in the trend after the adoption. Accordingly, the sum of \textit{time} and \textit{time\_since\_adoption} represents the new trend after the adoption. To also estimate the effect size of the adoption when a significant change is observed, we rely on counterfactuals \citep{wagner_segmented_2002}: what would have happened if the adoption had never occurred and thus the trend before the adoption had continued unchanged. For this purpose, we assume that there has been neither a change in the level (i.e., \textit{adoption} = 0) nor the trend (i.e., \textit{time\_since\_adoption} = 0) and measure the change percentage compared to the actual predicted values of the model.

\subsection{Findings}
The results of the models to estimate the impact of Stale bot on different performance indicators are presented in \Cref{appendix:models}. For easier comparison, \Cref{fig:impact} visualizes how the values of the impacted indicators vary each month during our observation period, along with the average of the model predictions (the solid red lines) and the average of the counterfactual predictions for all the studied projects (the dashed red lines). The variation plots for the remaining indicators can also be found in \Cref{appendix:variation}. We observe that the projects closed more PRs within the first few months of adopting Stale bot, but overall closed and merged fewer PRs afterward. The adoption of Stale bot is also associated with faster first reviews in merged PRs, faster resolutions in closed PRs, slightly fewer updates in merged PRs, and considerably fewer active contributors in the projects. In the following, we discuss our findings in more detail.

\begin{figure}
    \begin{subfigure}{0.435\textwidth}
        \includegraphics[width=\textwidth]{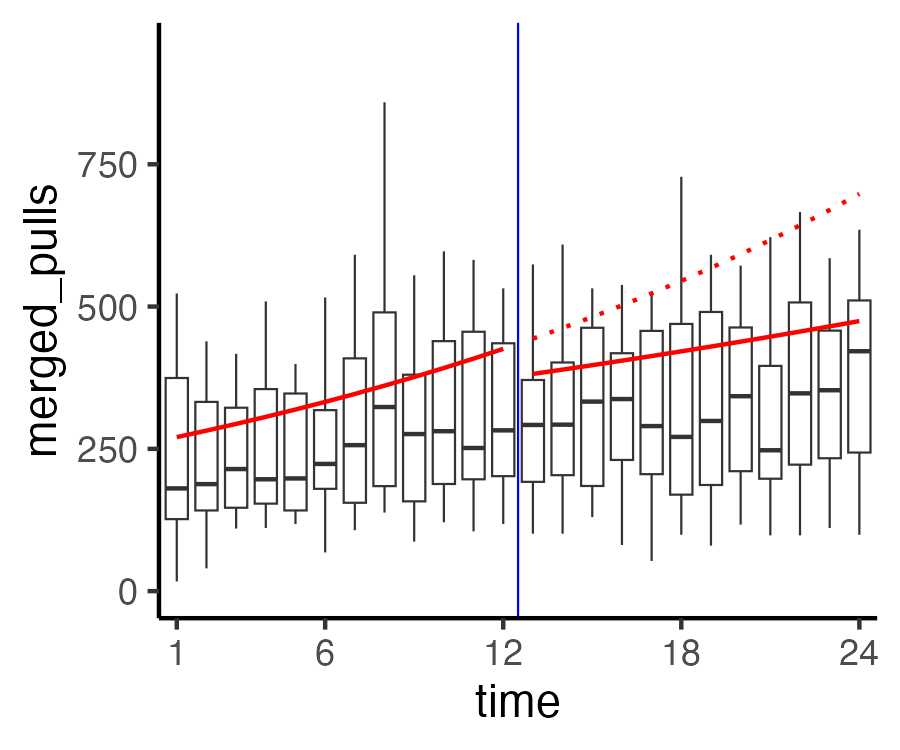}
        \caption{}
        \label{fig:merged_pulls}
    \end{subfigure}
    \begin{subfigure}{0.435\textwidth}
        \includegraphics[width=\textwidth]{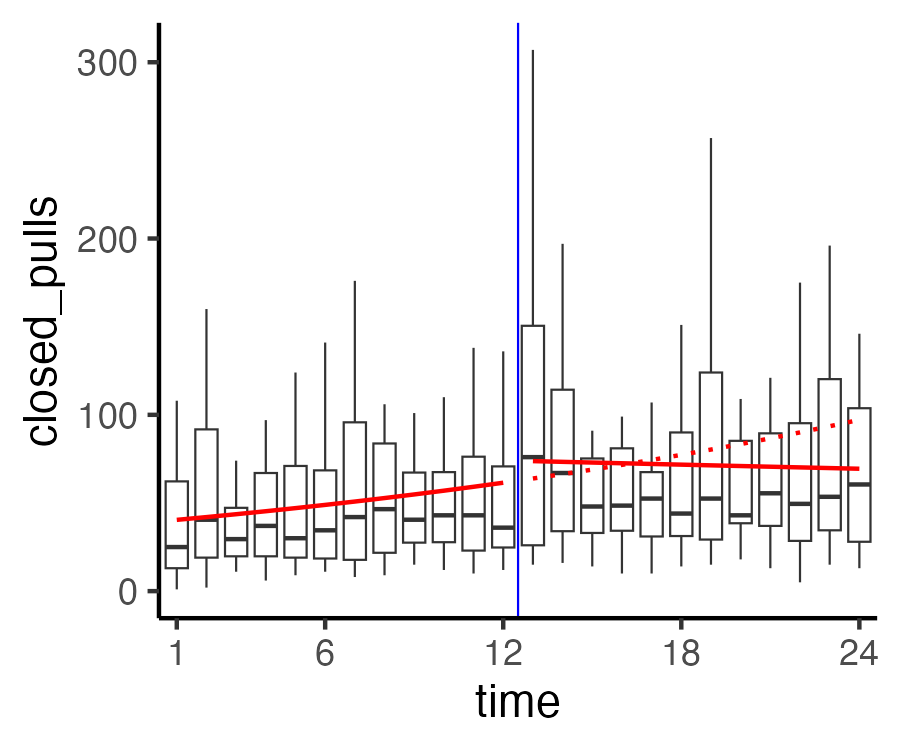}
        \caption{}
        \label{fig:closed_pulls}
    \end{subfigure}
    \begin{subfigure}{0.435\textwidth}
        \includegraphics[width=\textwidth]{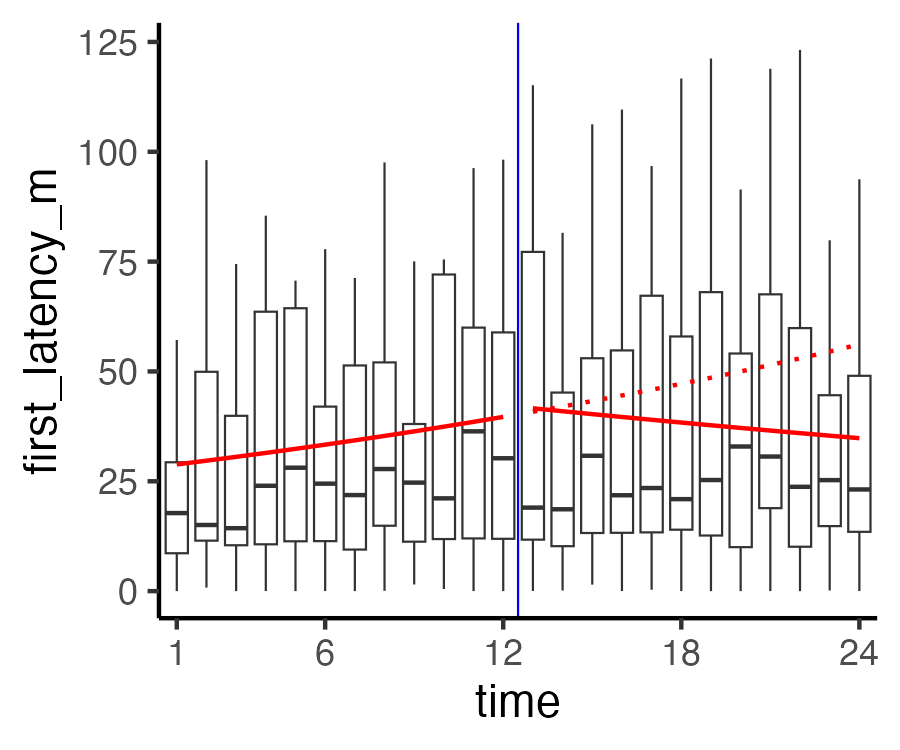}
        \caption{}
        \label{fig:first_latency_m}
    \end{subfigure}
    \begin{subfigure}{0.435\textwidth}
        \includegraphics[width=\textwidth]{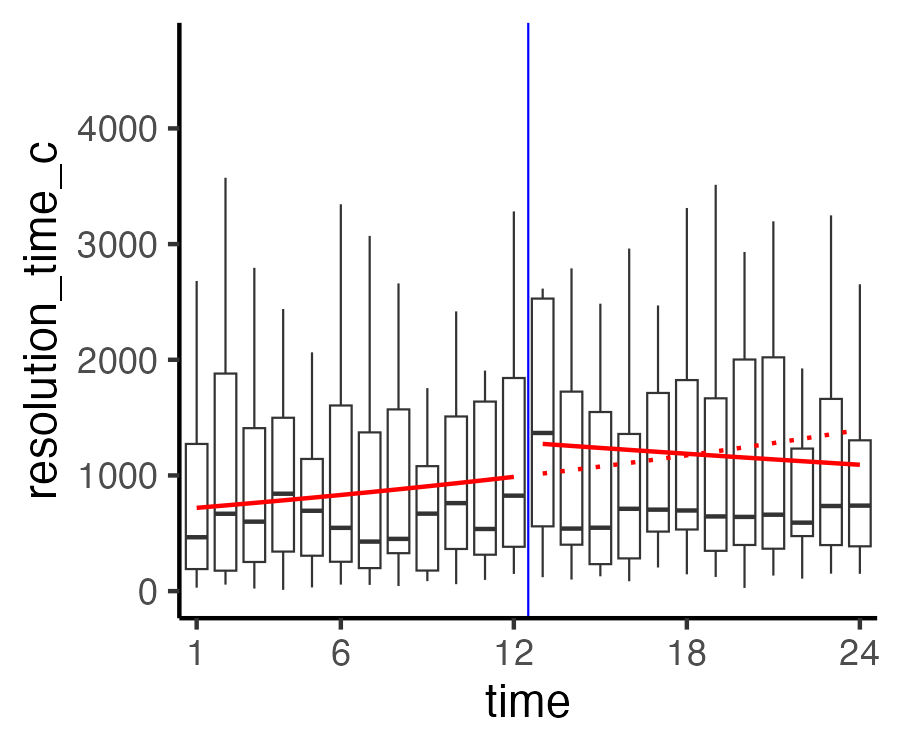}
        \caption{}
        \label{fig:resolution_time_c}
    \end{subfigure}
    \begin{subfigure}{0.435\textwidth}
        \includegraphics[width=\textwidth]{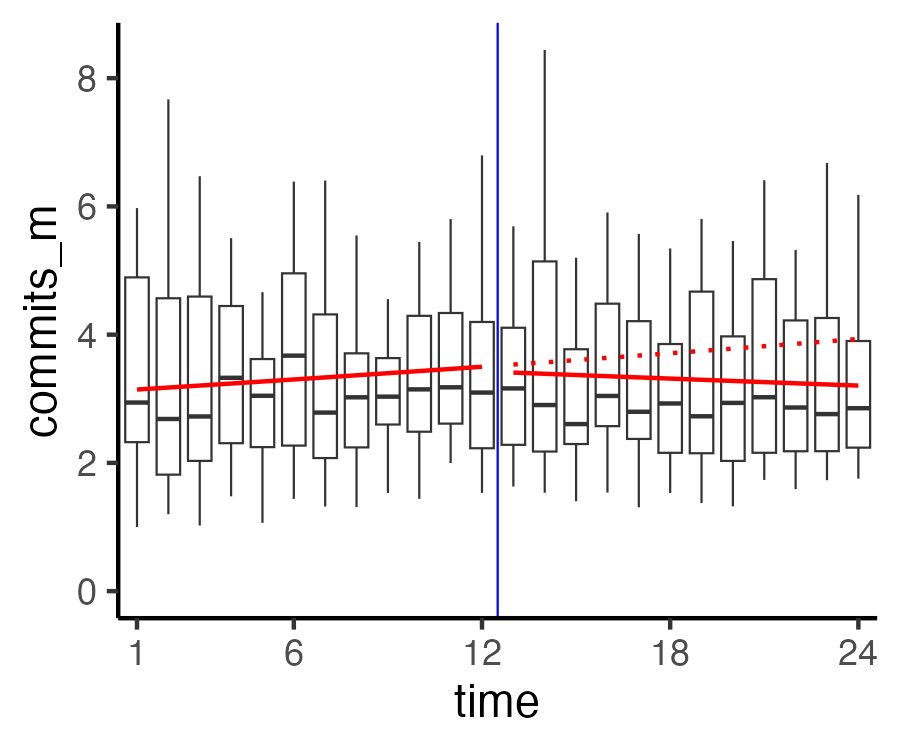}
        \caption{}
        \label{fig:commits_m}
    \end{subfigure}
    \begin{subfigure}{0.435\textwidth}
        \includegraphics[width=\textwidth]{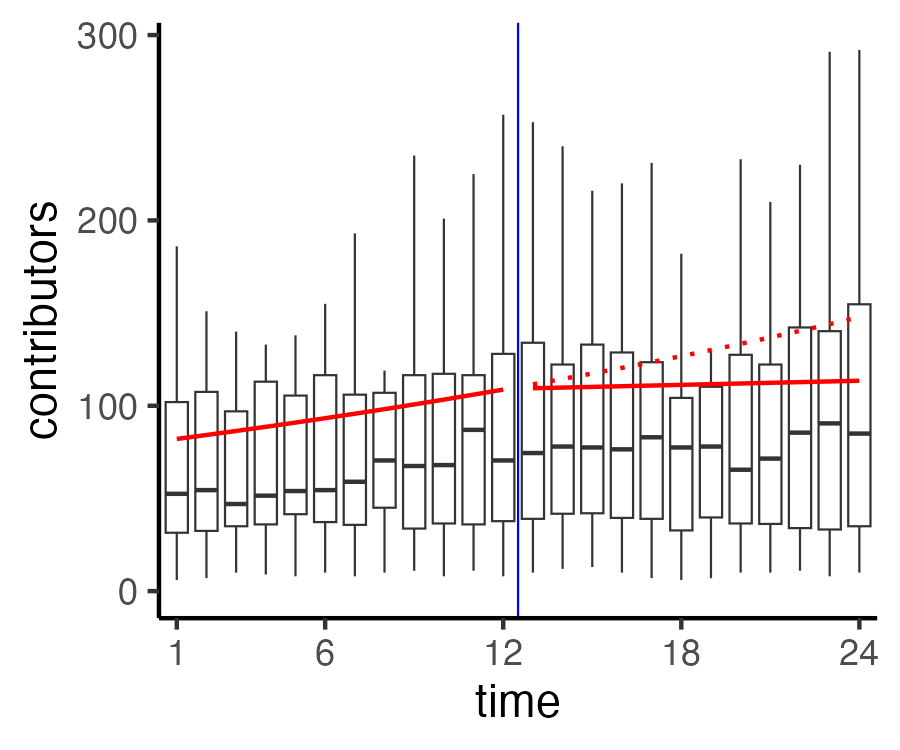}
        \caption{}
        \label{fig:contributors}
    \end{subfigure}
    \caption{Variation in (a) the number of merged PRs, (b) the number of closed PRs, (c) the first response latency of merged PRs, (d) the resolution time of closed PRs, (e) the number of commits in merged PRs, and (f) the number of active contributors each month during our observation period. The \textcolor{blue}{blue lines} show the adoption time, the \textcolor{red}{solid red lines} show the average predictions, and the \textcolor{red}{dashed red lines} show the average counterfactual predictions of the model for all the studied projects. Note that we dropped the outliers in the plots and used the exponential function to convert the log-transformed output of the models back to real values.}
    \label{fig:impact}
\end{figure}

\noindentparagraph{\emph{\textbf{While more PRs are closed within the first few months of adopting Stale bot, overall fewer PRs are closed and merged afterward.}}} As shown in \Cref{fig:closed_pulls}, the number of closed PRs experienced an increase in the level but a decrease in the slope that reversed the increasing trend before the adoption. Specifically, our predictions indicate that 15\% more PRs were closed in the first month of adoption but overall 10\% fewer PRs were closed by the end of the first year of adoption. The short-term increase in the number of monthly closed PRs is expected as Stale bot closes accumulated inactive PRs immediately upon its adoption. From \Cref{fig:merged_pulls}, we also observe a decrease in both the level and the slope of the number of merged PRs that decelerated the increasing trend before the adoption. While more PRs are still merged each month, our predictions indicate that overall 24\% fewer PRs were merged by the end of the first year of adoption.

\noindentparagraph{\emph{\textbf{Merged PRs tend to have faster first reviews after the adoption of Stale bot.}}} As shown in \Cref{fig:first_latency_m}, the first review latency of merged PRs experienced a decrease in the slope that reversed the increasing trend before the adoption. Specifically, our predictions indicate that merged PRs had an overall 21\% lower first review latency during the first year of adoption. However, the first review latency of closed PRs and the mean review latency of both closed PRs and merged PRs have not significantly changed after the adoption. Still, closed PRs tend to take longer to receive a first review over time, and this trend has not significantly changed after the adoption. The results suggest that the maintainers are focusing on PRs that are more likely to get merged, but at the cost of leaving the remaining PRs to linger until Stale bot closes them after a period of inactivity.

\noindentparagraph{\emph{\textbf{Closed PRs tend to have faster resolutions after the adoption of Stale bot.}}} As shown in \Cref{fig:resolution_time_c}, the resolution time experienced a decrease in the slope across closed PRs that reversed the increasing trend before the adoption, but has not significantly changed across merged PRs. For example, our predictions indicate that closed PRs had 22\% lower resolution time in the last month of the first year of adoption. The decrease in the resolution time of closed PRs is expected as the auto-close function of Stale bot does not allow PRs to stay open indefinitely without any activity.

\noindentparagraph{\emph{\textbf{The amount of discussions in PRs is not affected by the adoption of Stale bot.}}} We find that the number of discussion comments has not significantly changed after the adoption of Stale bot across both merged PRs and closed PRs. This observation comes as a surprise since we expected Stale bot to encourage more communication during the review process of PRs by notifying the participants about the lack of activity and by reducing the workload of the maintainers.

\noindentparagraph{\emph{\textbf{Merged PRs tend to have slightly fewer updates after the adoption of Stale bot.}}} As shown in \Cref{fig:commits_m}, the total number of commits across merged PRs experienced a decrease in the slope but has not significantly changed across closed PRs. Specifically, our predictions indicate that merged PRs overall contained 11\% fewer commits during the first year of adoption. Still, closed PRs tend to have more commits over time, and this trend has not significantly changed after the adoption.

\noindentparagraph{\emph{\textbf{The adoption of Stale bot is associated with a considerable decrease in the number of active contributors.}}} As shown in \Cref{fig:contributors}, the number of monthly active contributors experienced a decrease in the slope that significantly decelerated the increasing trend before the adoption. Specifically, our predictions indicate that overall 14\% fewer contributors were active each month during the first year of adoption. The decreased number of active contributors may reflect the frustration of the community regarding the usage of Stale bot in open-source projects.

\bigskip
\begin{tcolorbox}
    \paragraph{\emph{\textbf{Answer to RQ\textsubscript{2}}}} We find that the studied projects closed more PRs within the first few months of adopting Stale bot, but overall closed and merged fewer PRs afterward. The adoption of Stale bot is also associated with faster first reviews in merged PRs, faster resolutions in closed PRs, slightly fewer updates in merged PRs, and considerably fewer active contributors in the projects.
\end{tcolorbox}
\medskip

\section{RQ\texorpdfstring{\textsubscript{3}}{3}: \rqiii}
\label{sec:rqiii}
In the previous research question, we found that the adoption of Stale bot is associated with both positive and negative changes in the pull-based development workflow of the studied projects. As our last research question, we aim to understand what characteristics of PRs, their contributors, and their review processes are associated with a higher probability of getting intervened by Stale bot in large open-source projects. In the following, we first explain our approach and then discuss our findings to answer this research question.

\subsection{Approach}
To investigate the kinds of PRs intervened by Stale bot, we first need to characterize the PRs, their contributors, and their review processes. For this purpose, we consult the pull-based development literature \citep{zhang_pull_2023, zhang_pull_2022, zhang_shoulders_2020, gousios_dataset_2014}. As shown in \Cref{tab:factors}, we measure 16 factors covering three dimensions: \begin{inparaenum} \item PR characteristics, \item contributor characteristics, and \item review process characteristics \end{inparaenum}. After measuring these factors, we perform statistical analyses to determine how each factor differs in PRs that are intervened by Stale bot. To have a fair comparison between intervened and not intervened PRs, we filter our dataset to include only PRs that are closed or merged within the first year of adoption (i.e., our observation period). After this step, our dataset includes a total of 126,378 PRs, of which 6,833 PRs ($\sim$5.4\%) have been intervened by Stale bot.

\begin{table}
    \caption{Overview of the factors measured to characterize PRs, their contributors, and their review processes.}
    \label{tab:factors}
    \resizebox{\textwidth}{!}{%
        \begin{tabular}{@{}l|l|l@{}}
            \toprule

            \textbf{Dimension}                       & \textbf{Factor}          & \textbf{Description}                                                                           \\
            \midrule
            \multirow{7}{*}{\textbf{Pull Request}}   & description              & Number of words in the title and description of the PR                                         \\
                                                     & initial\_commits         & Number of commits at the submission time of the PR                                             \\
                                                     & followup\_commits        & Number of commits after the submission time of the PR                                          \\
                                                     & initial\_changed\_lines  & Number of changed lines at the submission time of the PR                                       \\
                                                     & followup\_changed\_lines & Number of changed lines after the submission time of the PR                                    \\
                                                     & initial\_changed\_files  & Number of changed files at the submission time of the PR                                       \\
                                                     & followup\_changed\_files & Number of changed files after the submission time of the PR                                    \\
            \midrule
            \multirow{3}{*}{\textbf{Contributor}}    & submitted\_pulls         & Number of previously submitted PRs by the contributor of the PR in the project                 \\
                                                     & acceptance\_rate         & Ratio of the previously merged PRs of the contributor of the PR in the project                 \\
                                                     & contribution\_period     & Number of months since the first submitted PR of the contributor of the PR in the project      \\
            \midrule
            \multirow{6}{*}{\textbf{Review Process}} & participants             & Number of participants (excluding Stale bot) during the lifecycle of the PR                    \\
                                                     & participant\_comments    & Number of comments from the participants (excluding Stale bot) during the lifecycle of the PR  \\
                                                     & contributor\_comments    & Number of comments from the contributor during the lifecycle of the PR                         \\
                                                     & first\_latency           & Number of hours till the first response of the participants (excluding Stale bot) in the PR    \\
                                                     & mean\_latency            & Mean number of hours between the responses of the participants (excluding Stale bot) in the PR \\
                                                     & resolution\_time         & Number of hours between the submission and resolution times of the PR                          \\
            \bottomrule
        \end{tabular}
    }
\end{table}

Then, we compare the distribution of the measured factors between intervened and not intervened PRs by generating violin plots \citep{hintze_violin_1998} for each project using the \texttt{ggstatsplot} package \citep{patil_visualizations_2021}. The generated plots for each factor showing their median values (denoted by $M$), interquartile ranges (the box inside the violin), and probability densities (the width of the violin at each value) are presented in \Cref{appendix:stats}. To test the statistical difference between the factors of intervened and not intervened PRs, we apply the Mann–Whitney $U$ test \citep{mann_test_1947} as a nonparametric test that does not require the distribution of the factors to be normal. To perform this test, we use the \texttt{stats} package \citep{r_core_team_r_2023} and add the results to the generated plots. We consider the traditional confidence level of 95\% (i.e., $\alpha = 0.05$) \citep{chavalarias_evolution_2016} to identify statistically significant factors.

While statistical significance verifies whether a difference exists between the factors of intervened and not intervened PRs, we also need to test their practical difference \citep{kirk_practical_1996}. For this purpose, we use Cliff's delta \citep{cliff_dominance_1993} to estimate their magnitude of difference (i.e., effect size). The value of Cliff's delta (denoted by $d$) ranges from $-1$ to $+1$, where a positive $d$ implies that the values of the factor in intervened PRs are usually greater than those of not intervened PRs, while a negative $d$ implies the opposite. To calculate this statistic, we use the \texttt{effectsize} package \citep{ben-shachar_effectsize_2020} and add the results to the generated plots. Finally, for easier comparison, we convert the $d$ values to qualitative magnitudes according to the following thresholds as suggested by \citet{hess_robust_2004}:

\begin{equation*}
    \text{Effect size}=
    \begin{cases}
        \text{Negligible}, & \text{if         $\lvert d \rvert \leq 0.147$} \\
        \text{Small},      & \text{if $0.147 < \lvert d \rvert \leq 0.33$}  \\
        \text{Medium},     & \text{if $0.33  < \lvert d \rvert \leq 0.474$} \\
        \text{Large},      & \text{if $0.474 < \lvert d \rvert \leq 1$}     \\
    \end{cases}
\end{equation*}

\subsection{Findings}
\Cref{tab:difference} summarizes the significant differences in the characteristics of PRs with and without intervention from Stale bot across the studied projects. For each factor, the table shows the number of projects in which intervened PRs tend to have significantly higher values compared to not intervened PRs, the number of projects in which intervened PRs and not intervened PRs are not significantly different, and the number of projects in which intervened PRs tend to have significantly lower values compared to not intervened PRs. We consider a factor significant if its difference between intervened and not intervened PRs is both statistically significant (i.e., $p < 0.05$) and practically significant (i.e., the effect size is small, medium, or large). We observe the most common differences in the characteristics of contributors and the review processes of PRs. Specifically, the largest differences are in the resolution time of PRs (higher in all the projects), the number of prior PRs by contributors (lower in all the projects), the mean response latency of PRs (higher in 19 projects), the acceptance rate of contributors (lower in 19 projects), and the contribution period of contributors (lower in 18 projects). In the following, we discuss our findings in more detail.

\begin{table}
    \caption{Differences in the characteristics of PRs with and without intervention from Stale bot across the studied projects.}
    \label{tab:difference}
    \resizebox{\textwidth}{!}{%
        \begin{tabular}{@{}l|l|ccc|c|ccc@{}}
            \toprule
            \multirow{2}{*}{\textbf{Dimension}}      & \multirow{2}{*}{\textbf{Factor}} & \multicolumn{3}{c|}{\textbf{Higher in Intervened PRs}} & \multirow{2}{*}{\textbf{No Difference}} & \multicolumn{3}{c}{\textbf{Lower in Intervened PRs}} \\
                                                     &                                  & \textbf{Small} & \textbf{Medium} & \textbf{Large}      &                                         & \textbf{Small} & \textbf{Medium} & \textbf{Large}    \\
            \midrule
            \multirow{7}{*}{\textbf{Pull Request}}   & description                      & 4              & 1               & 1                   & 10                                      & 3              & --              & 1                 \\
                                                     & initial\_commits                 & --             & --              & --                  & 12                                      & 7              & --              & 1                 \\
                                                     & followup\_commits                & 8              & 4               & 2                   & 6                                       & --             & --              & --                \\
                                                     & initial\_changed\_lines          & 1              & --              & --                  & 16                                      & 2              & --              & 1                 \\
                                                     & followup\_changed\_lines         & 8              & 4               & 2                   & 6                                       & --             & --              & --                \\
                                                     & initial\_changed\_files          & --             & --              & --                  & 13                                      & 6              & --              & 1                 \\
                                                     & followup\_changed\_files         & 9              & 2               & --                  & 9                                       & --             & --              & --                \\
            \midrule
            \multirow{3}{*}{\textbf{Contributor}}    & submitted\_pulls                 & --             & --              & --                  & --                                      & 3              & 14              & 3                 \\
                                                     & acceptance\_rate                 & --             & --              & --                  & 1                                       & 6              & 11              & 2                 \\
                                                     & contribution\_period             & --             & --              & --                  & 2                                       & 12             & 5               & 1                 \\
            \midrule
            \multirow{6}{*}{\textbf{Review Process}} & participants                     & 5              & --              & 8                   & 4                                       & 2              & 1               & --                \\
                                                     & participant\_comments            & 4              & 6               & 4                   & 3                                       & --             & 2               & 1                 \\
                                                     & contributor\_comments            & 5              & 6               & 3                   & 6                                       & --             & --              & --                \\
                                                     & first\_latency                   & 2              & 5               & 6                   & 4                                       & 2              & --              & 1                 \\
                                                     & mean\_latency                    & --             & 2               & 17                  & 1                                       & --             & --              & --                \\
                                                     & resolution\_time                 & --             & --              & 20                  & --                                      & --             & --              & --                \\
            \bottomrule
        \end{tabular}
    }
\end{table}

\begin{figure}
    \begin{subfigure}{0.325\textwidth}
        \includegraphics[width=\textwidth]{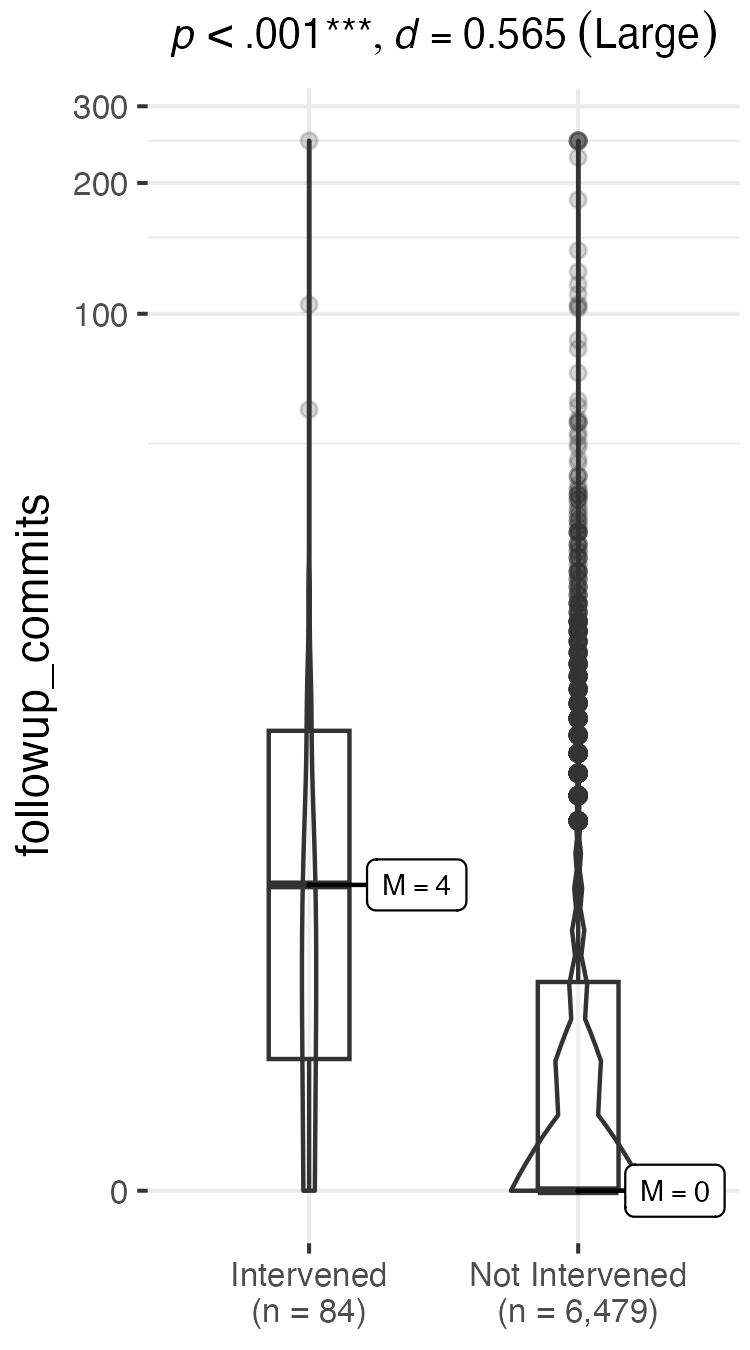}
        \caption{}
    \end{subfigure}
    \begin{subfigure}{0.325\textwidth}
        \includegraphics[width=\textwidth]{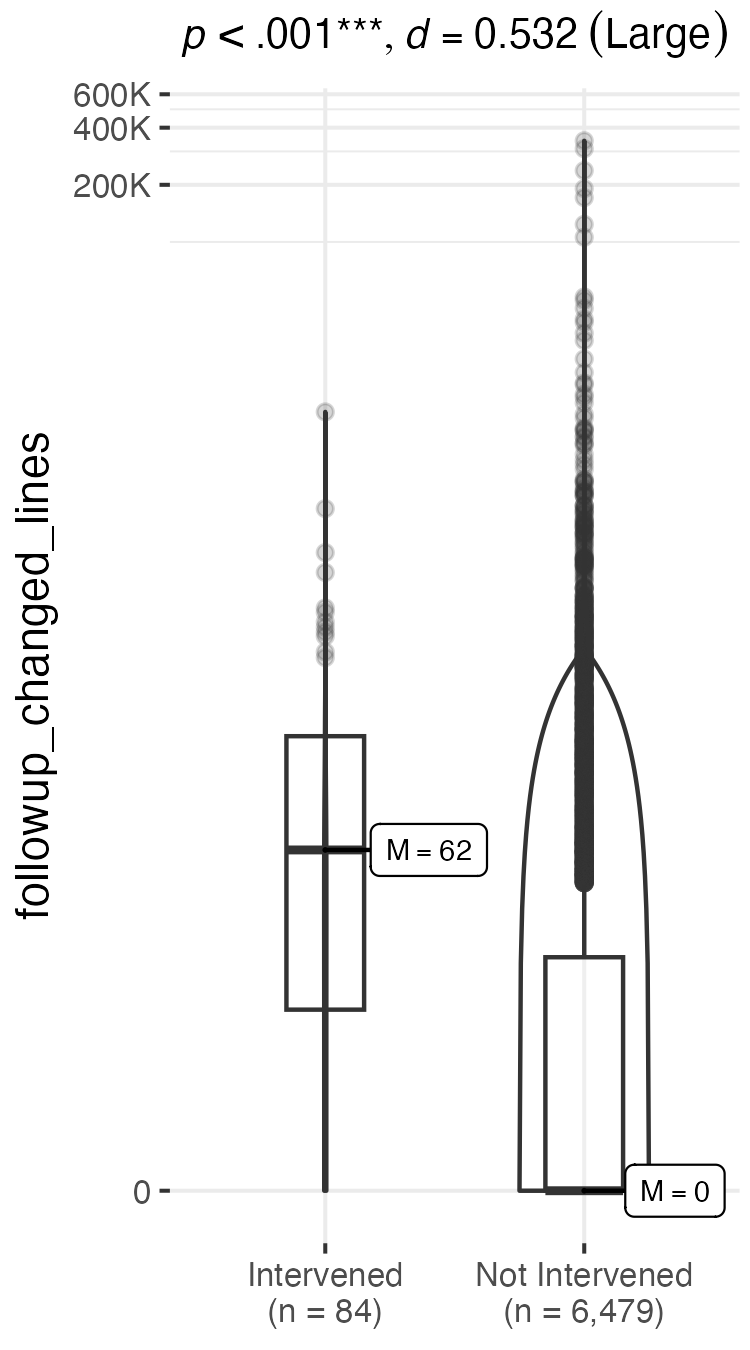}
        \caption{}
    \end{subfigure}
    \begin{subfigure}{0.325\textwidth}
        \includegraphics[width=\textwidth]{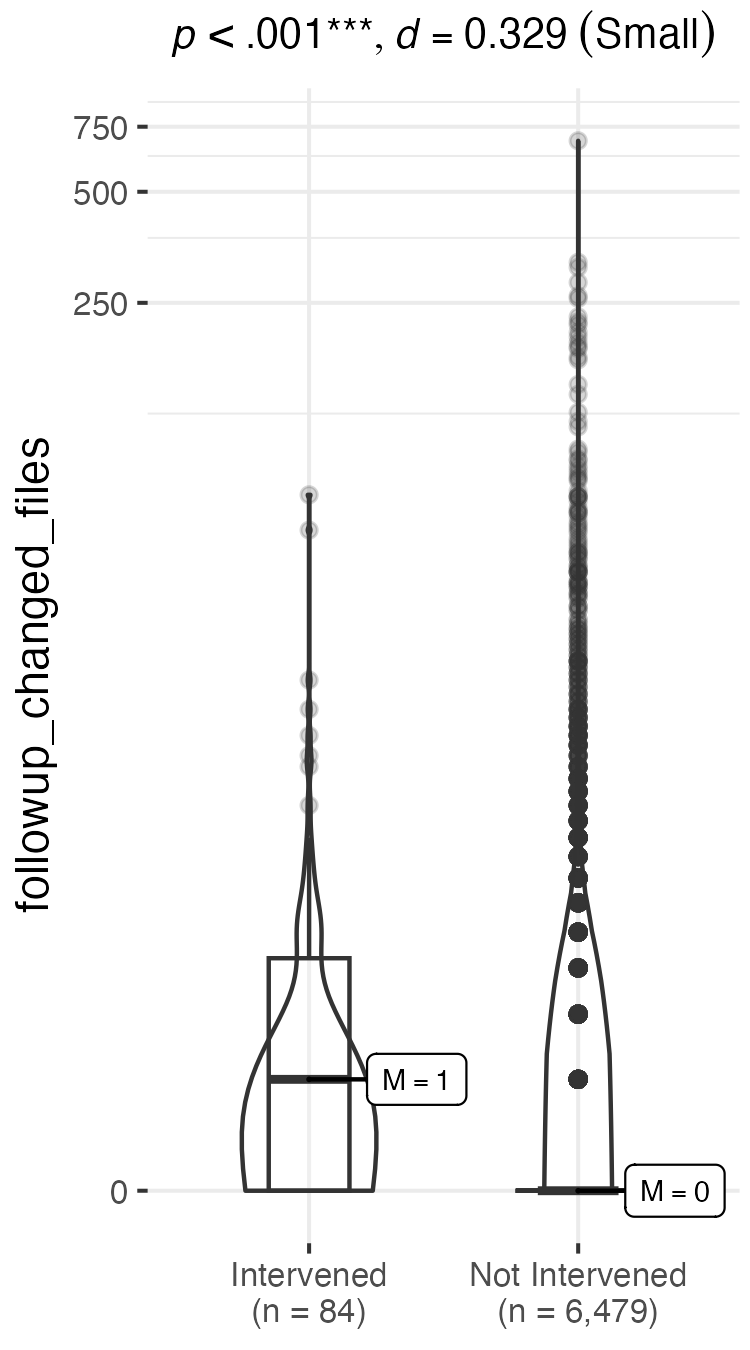}
        \caption{}
    \end{subfigure}
    \caption{Differences between PRs with and without intervention from Stale bot in the cleverraven/cataclysm-dda project regarding (a) the number of follow-up commits, (b) the number of follow-up changed lines, and (c) the number of follow-up changed files after the submission.}
    \label{fig:pr_stats_cleverraven}
\end{figure}

\noindentparagraph{\emph{\textbf{PRs intervened by Stale bot tend to be more complex.}}} As shown in the PR dimension of \Cref{tab:difference}, intervened PRs tend to significantly include fewer commits (8 projects), contain smaller changes (3 projects), and touch fewer files (7 projects) upon submission compared to not intervened PRs. However, after the submission, intervened PRs tend to significantly include more commits (14 projects), contain larger changes (14 projects), and touch more files (11 projects). For example, the intervened PRs in the cleverraven/cataclysm-dda project on median have 4 more commits, 62 more changed lines, and 1 more changed file after their submission (see \Cref{fig:pr_stats_cleverraven}). Regarding the description length, we also observe that 6 projects tend to have lengthier descriptions and 4 projects tend to have shorter descriptions in intervened PRs. While the description length of PRs rarely changes during the review process of a PR, the size of changes (i.e., the number of commits, changed lines, or changed files) is likely to increase as the PR gets iteratively reviewed and updated. The results indicate that PRs mostly intervened by Stale bot have received considerable effort from their contributors to make it ready to get merged into the project.

\begin{figure}
    \begin{subfigure}{0.325\textwidth}
        \includegraphics[width=\textwidth]{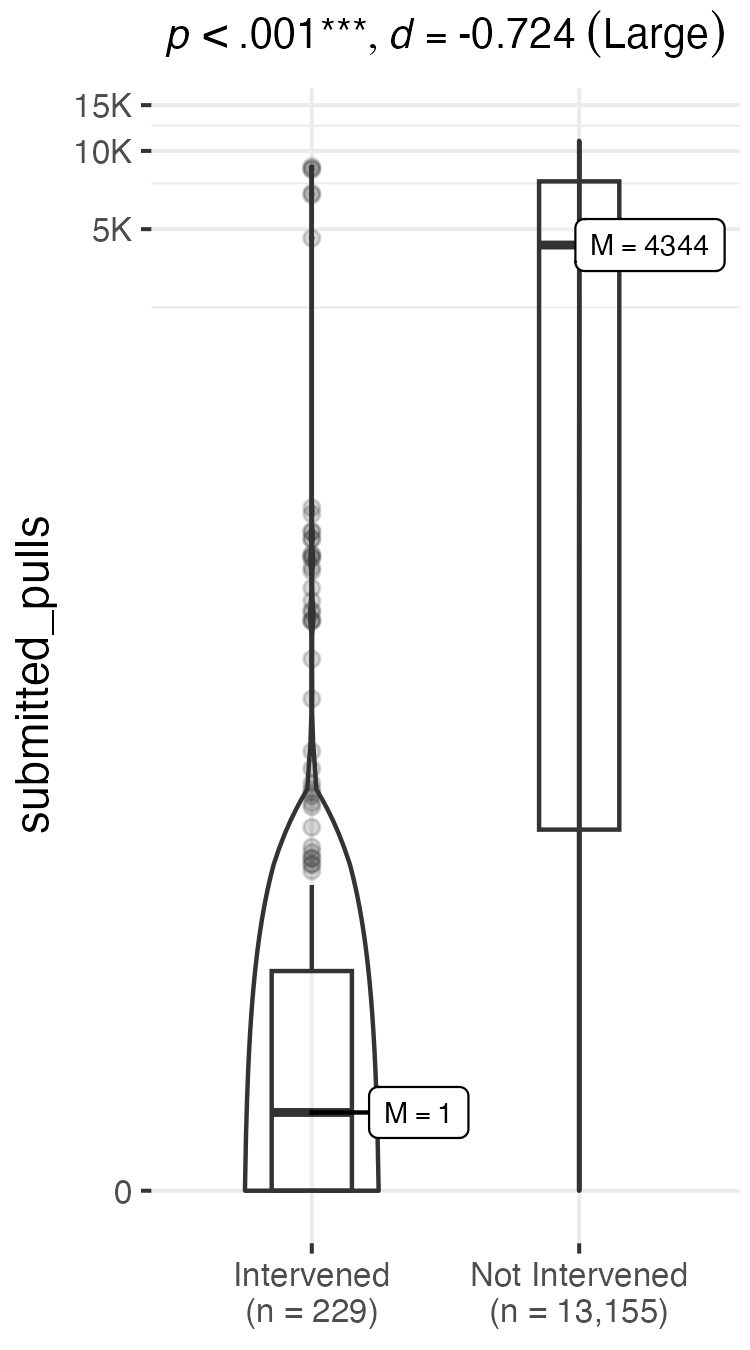}
        \caption{}
    \end{subfigure}
    \begin{subfigure}{0.325\textwidth}
        \includegraphics[width=\textwidth]{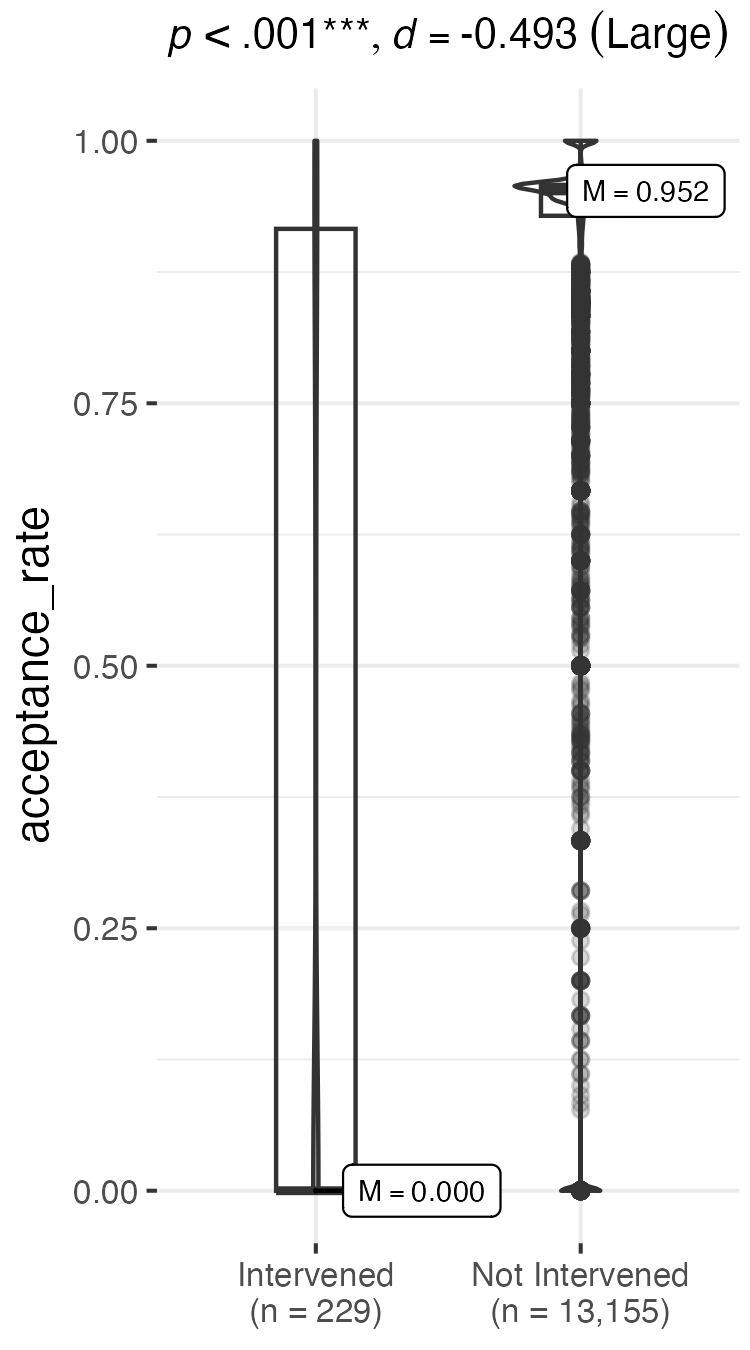}
        \caption{}
    \end{subfigure}
    \begin{subfigure}{0.325\textwidth}
        \includegraphics[width=\textwidth]{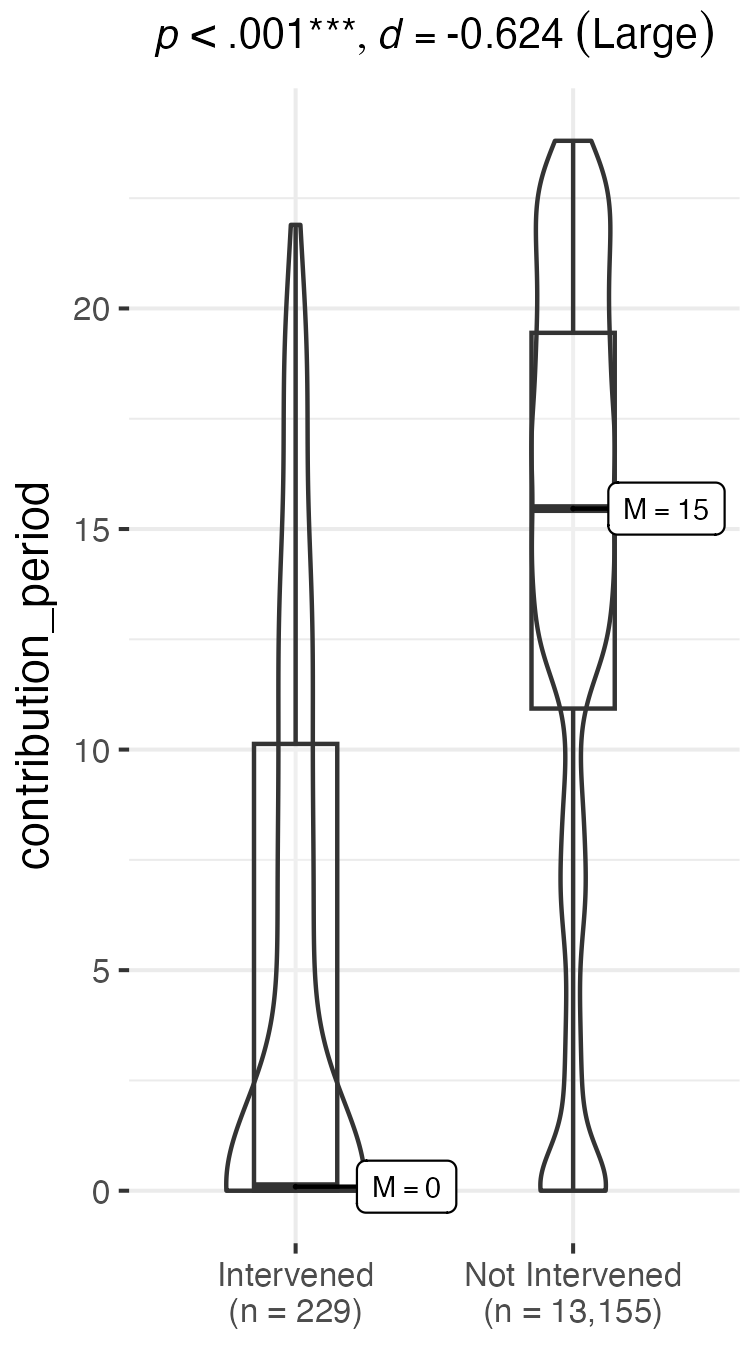}
        \caption{}
    \end{subfigure}
    \caption{Differences between the contributors of PRs with and without intervention from Stale bot in the homebrew/homebrew-core project regarding (a) the number of prior PRs, (b) the acceptance rate, and (c) the contribution period.}
    \label{fig:contributor_stats_homebrew}
\end{figure}

\noindentparagraph{\emph{\textbf{The contributors of PRs intervened by Stale bot tend to be less experienced.}}} As shown in the contributor dimension of \Cref{tab:difference}, the contributors of intervened PRs tend to significantly have previously submitted fewer PRs (all the projects), have a lower acceptance rate (19 projects), and have a lower contribution period (18 projects) compared to the contributors of not intervened PRs. For example, the contributors of intervened PRs in the homebrew/homebrew-core project on median have previously submitted 4,343 fewer PRs, have a 0.95 lower rate of acceptance, and have contributed to the project for 15 fewer months (see \Cref{fig:contributor_stats_homebrew}). However, novice contributors are the ones who face the most barriers and thus need the most guidance from the maintainers \citep{steinmacher_social_2015, steinmacher_preliminary_2014, steinmacher_why_2013}. The lack of responsiveness from the reviewers is also cited as a major reason why contributors (especially novice or casual contributors) leave the review processes of PRs unfinished and even stop further contributing to the project \citep{khatoonabadi_wasted_2023, li_are_2022, wang_why_2019, steinmacher_why_2013}. Therefore, the results indicate that while Stale bot can help projects automatically deal with their inactive PRs, it may also lead to lower engagement of the community and eventually contributor abandonment, especially if the reviewers' input is required to continue the review process.

\begin{figure}
    \begin{subfigure}{0.325\textwidth}
        \includegraphics[width=\textwidth]{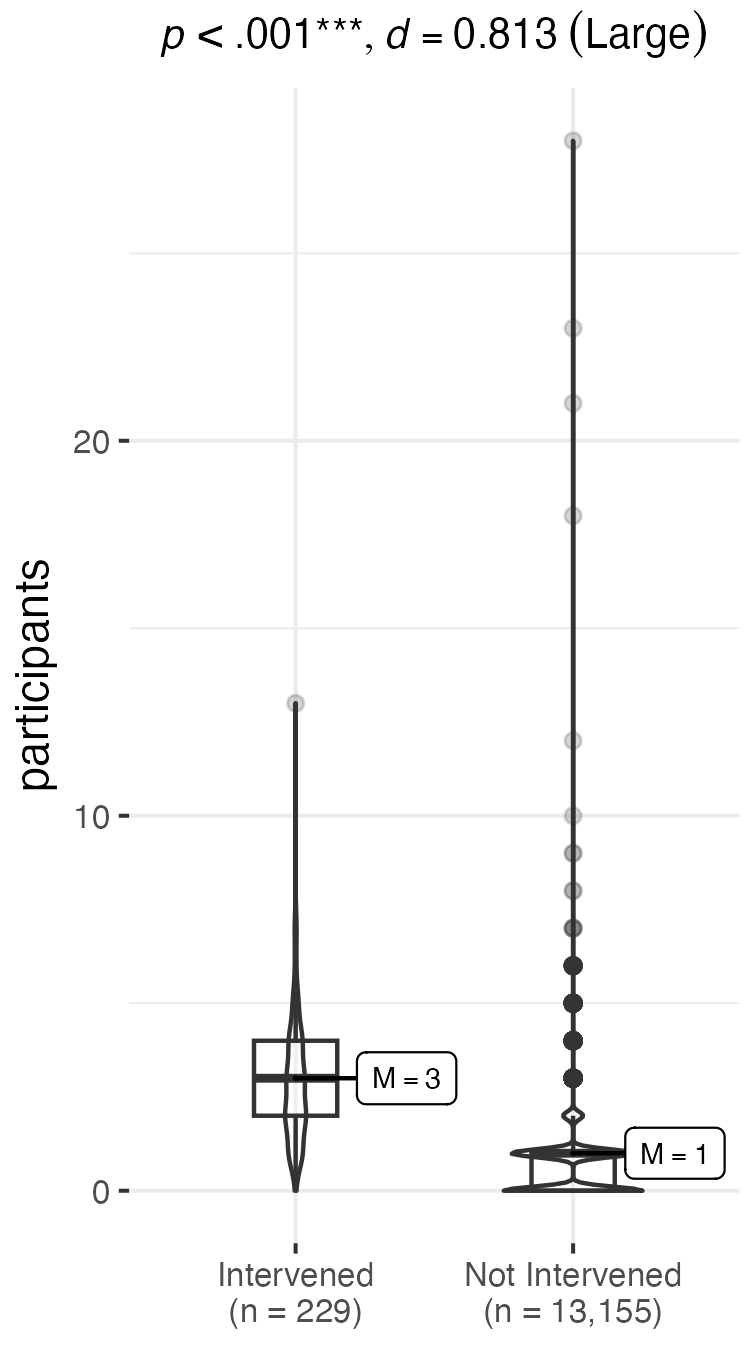}
        \caption{}
    \end{subfigure}
    \begin{subfigure}{0.325\textwidth}
        \includegraphics[width=\textwidth]{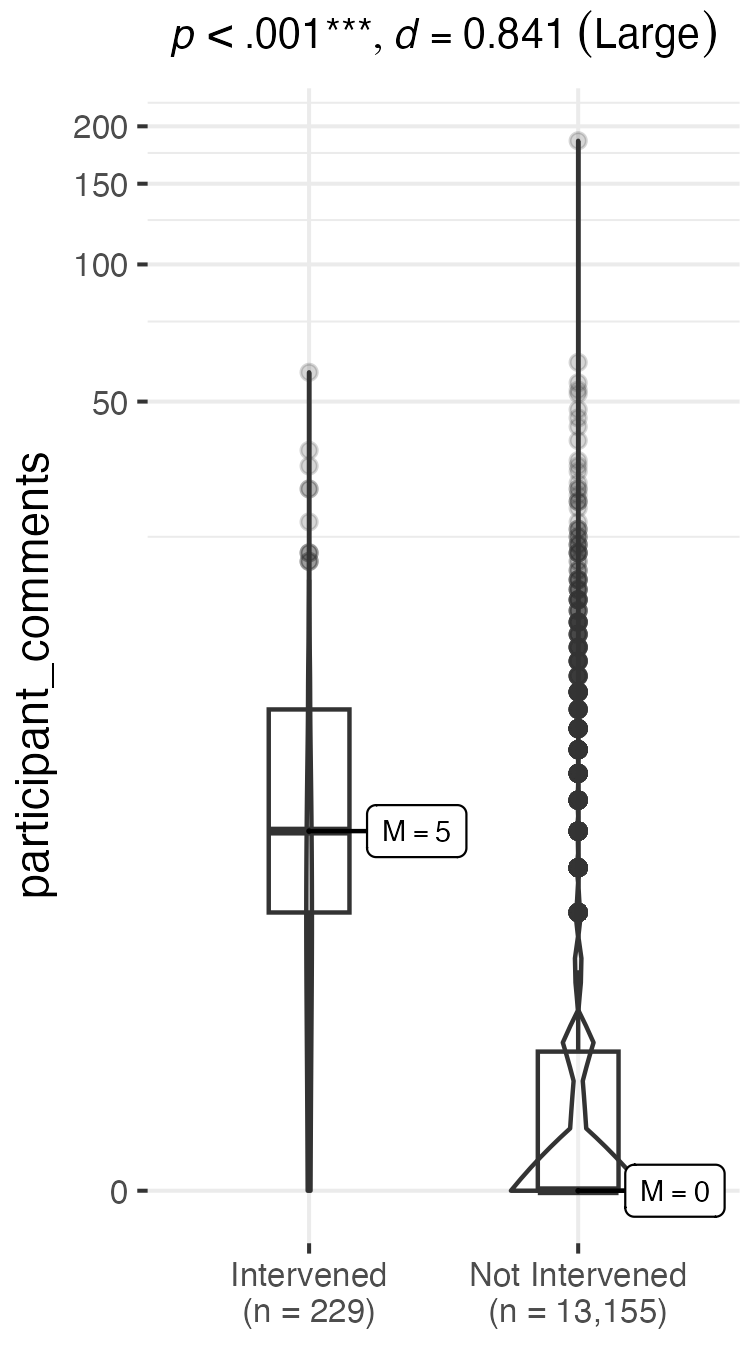}
        \caption{}
    \end{subfigure}
    \begin{subfigure}{0.325\textwidth}
        \includegraphics[width=\textwidth]{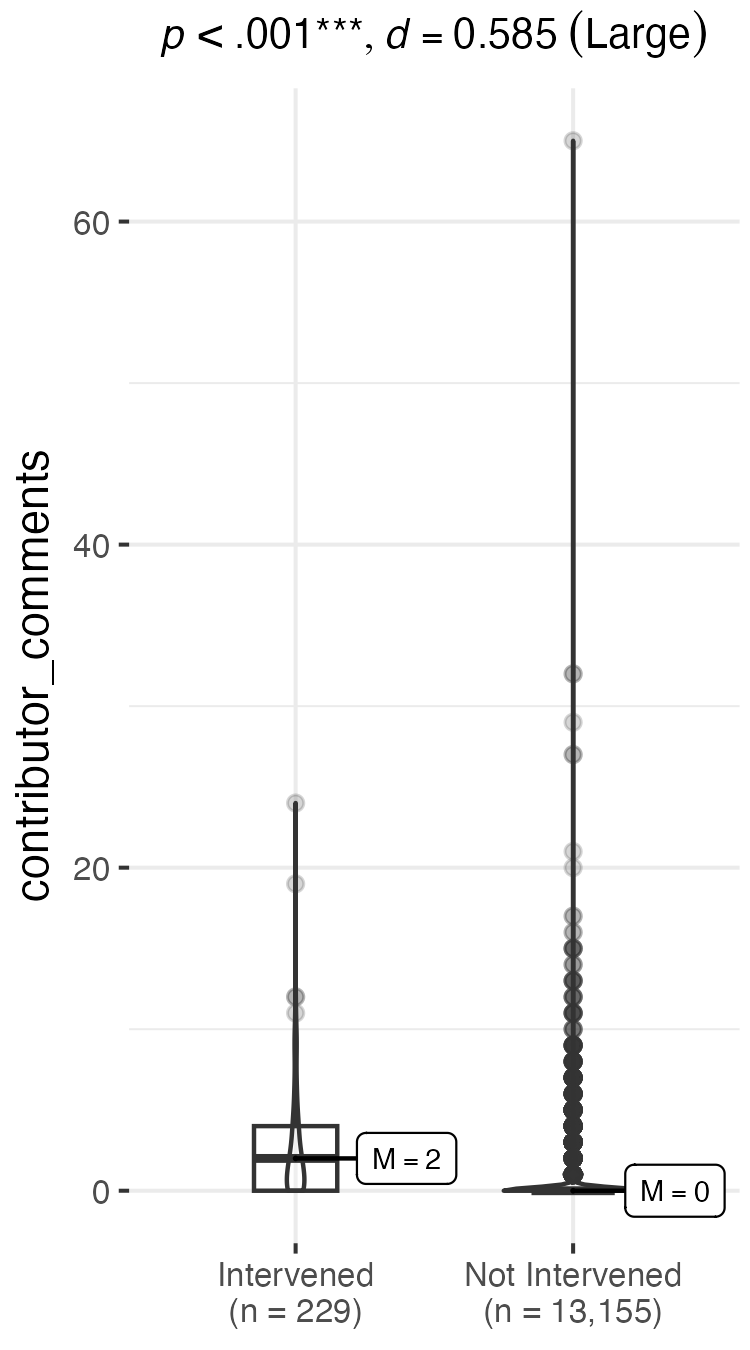}
        \caption{}
    \end{subfigure}
    \begin{subfigure}{0.325\textwidth}
        \includegraphics[width=\textwidth]{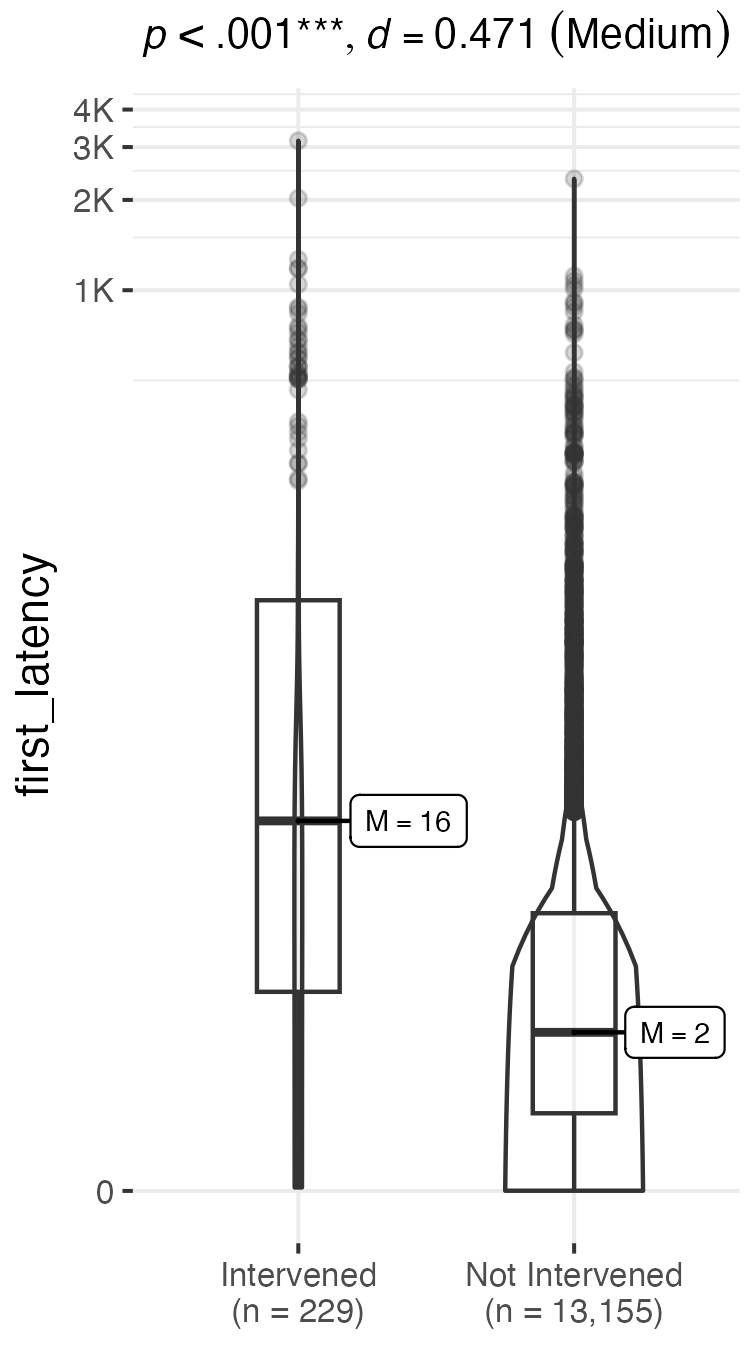}
        \caption{}
    \end{subfigure}
    \begin{subfigure}{0.325\textwidth}
        \includegraphics[width=\textwidth]{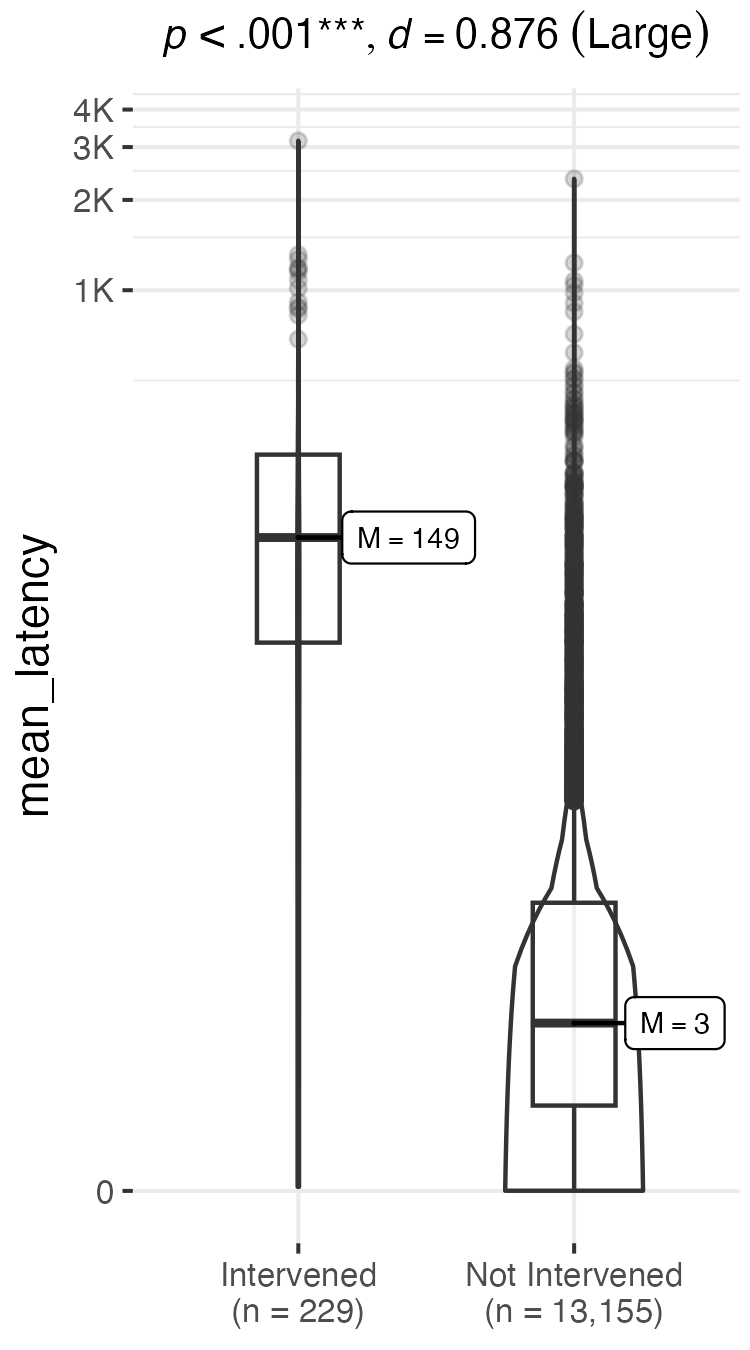}
        \caption{}
    \end{subfigure}
    \begin{subfigure}{0.325\textwidth}
        \includegraphics[width=\textwidth]{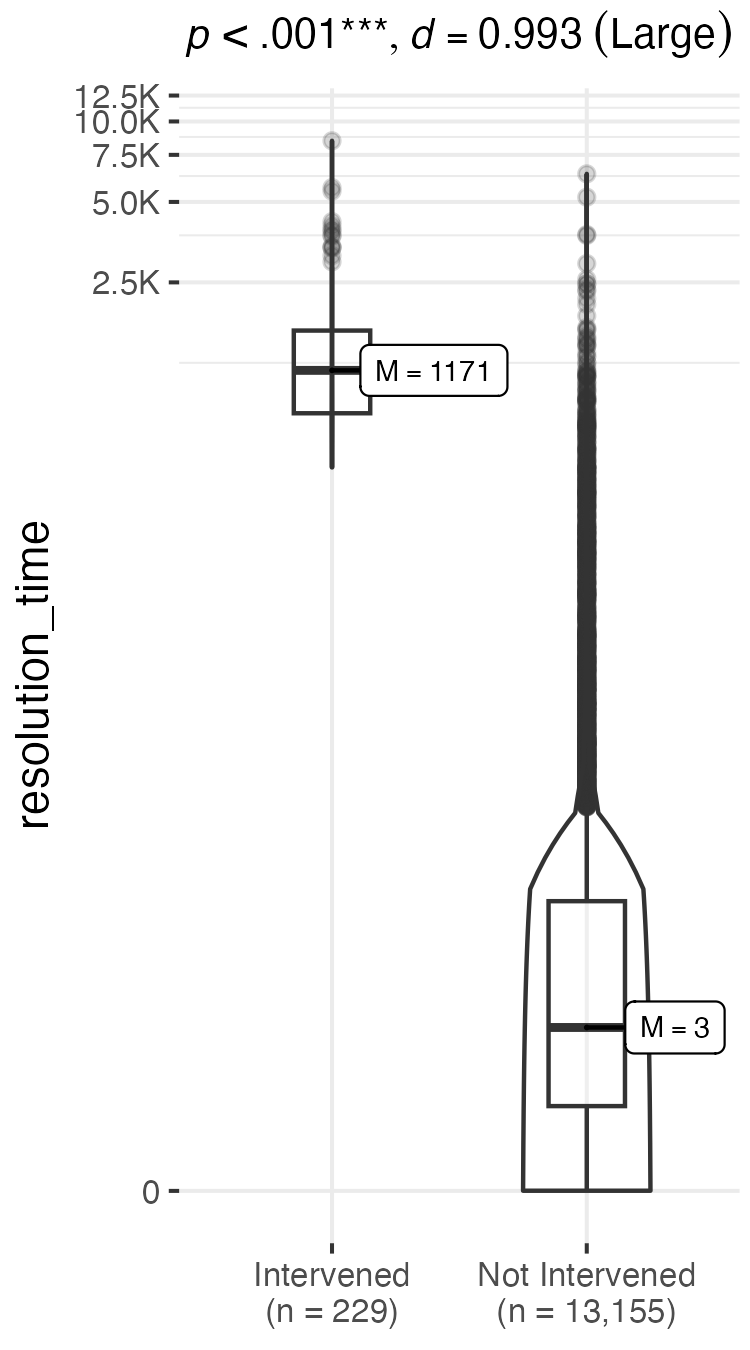}
        \caption{}
    \end{subfigure}
    \caption{Differences between the review processes of PRs with and without intervention from Stale bot in the homebrew/homebrew-core project regarding (a) the number of participants, (b) the number of comments from the participants, (c) the number of comments from the contributors, (d) the first review latency, (e) the mean review latency, and (f) the resolution time.}
    \label{fig:review_stats_homebrew}
\end{figure}

\noindentparagraph{\emph{\textbf{The review processes of PRs intervened by Stale bot tend to be lengthier.}}} As shown in the review process dimension of \Cref{tab:difference}, the review processes of intervened PRs tend to significantly involve more participants (13 projects), receive more comments from the participants (14 projects), receive more comments from the contributors (14 projects), take longer to receive their first review (13 projects), take longer to receive a follow-up review (19 projects), and take longer to get resolved (all the projects) compared to the review processes of not intervened PRs. For example, the review processes of intervened PRs in the homebrew/homebrew-core project on median involve 2 more participants, receive 5 more comments from the participants, receive 2 more comments from the contributors, takes 14 more hours to receive their first review, takes 146 more hours to receive a follow-up review, and takes 1,168 more hours to get resolved (see \Cref{fig:review_stats_homebrew}). While a longer resolution time in intervened PRs is expected as Stale bot intervenes in PRs that have been idle for a while, a longer time to receive reviews (both first and follow-up) from the participants is rather interesting.

\bigskip
\begin{tcolorbox}
    \paragraph{\emph{\textbf{Answer to RQ\textsubscript{3}.}}} We find that Stale bot tends to intervene more in complex PRs, PRs from novice contributors, and PRs with lengthy review processes. Specifically, besides the resolution time of PRs, the largest differences are observed in the number of prior PRs by contributors, the mean response latency of PRs, the acceptance rate of contributors, and the contribution period of contributors.
\end{tcolorbox}
\medskip

\section{Implications}
\label{implications}
In the following, we combine our findings to further discuss the implications of our study.

\noindentparagraph{\emph{\textbf{Stale bot can help projects deal with a backlog of unresolved PRs.}}} Almost all the studied projects relied on Stale bot to automatically close their accumulated PRs after a period of inactivity. In fact, the projects closed more PRs within the first few months of adopting Stale bot, yet closed and even merged considerably fewer PRs afterward. These findings suggest that adopting Stale bot could be an effective strategy for quickly dealing with a backlog of unresolved PRs in the short term. However, projects should be aware that the rule-based nature of Stale bot could wrongly close PRs that may still be under progress despite being inactive for some time \citep{wessel_inconvenient_2020}. Moreover, the automatic closure of PRs by Stale bot is known to raise the most negative reactions from the contributors and participants, especially when they perceive the closure as erroneous or unjustified \citep{farah_exploratory_2022}.

\noindentparagraph{\emph{\textbf{Stale bot can help projects improve the review process of PRs.}}} After the adoption of Stale bot, PRs that ended up being merged received faster reviews after their submission, and PRs that ended up being closed were also resolved much faster in the studied projects. However, PRs that end up being closed are increasingly taking longer to receive a review from the reviewers. The adoption of Stale bot did not improve this trend or encourage more communication during the review process of PRs. These findings suggest that maintainers are focusing on PRs that are more likely to get merged, but at the cost of leaving the remaining PRs to linger until Stale bot closes them after a period of inactivity. However, projects should be mindful that such a strategy will likely result in fewer merged PRs over time.

\noindentparagraph{\emph{\textbf{Stale bot can negatively affect the contributors of projects.}}} The studied projects experienced a considerable decrease in their number of active contributors after the adoption of Stale bot. While Stale bot can help projects automatically deal with their inactive PRs, it is also more likely to intervene in PRs submitted by less experienced contributors. Projects should ensure that PRs receive timely reviews and responses from their reviewers, particularly if their input is needed to continue the review process. This is particularly important as the lack of responsiveness from reviewers is cited as a major reason why contributors (especially novice or casual contributors) leave the review process of PRs unfinished and even stop further contributing to a project \citep{khatoonabadi_wasted_2023, li_are_2022, wang_why_2019, steinmacher_why_2013}. Therefore, implicitly ignoring unresolved PRs till Stale bot eventually closes them may lead to decreased community engagement and an increased probability of contributor abandonment.

\section{Limitations}
\label{sec:limitations}
In the following, we discuss threats to the internal, construct, and external validity \citep{kitchenham_evidence-based_2015} of our study and explain the measures taken to mitigate them.

\noindentparagraph{\emph{\textbf{Construct Validity.}}} Construct validity is concerned with how accurately we quantified the helpfulness of Stale bot for pull-based development. We measured 13 different performance indicators covering various dimensions, including resolved PRs, review latency, resolution time, review discussion, PR updates, and contributor retention. To identify these indicators, we consulted similar studies that investigated the effects of interventions on the pull-based development of projects \citep{wessel_quality_2022, wessel_github_2022} and also drew from our previous experience studying abandoned PRs \citep{khatoonabadi_wasted_2023}. Nevertheless, projects may experience changes in aspects that we did not consider or that are difficult to quantify, such as code quality. \citet{forsgren_space_2021} describes various facets of productivity for individual developers and software teams, which future studies can consult for a more comprehensive investigation of the helpfulness of Stale bot.

\noindentparagraph{\emph{\textbf{Internal Validity.}}} Internal validity is concerned with whether the adoption of Stale bot has actually caused the observed effects. Although we applied interrupted time-series analysis as a well-established quasi-experiment to understand the impact of adopting Stale bot, we cannot conclusively claim that Stale bot caused the observed changes. It is possible that the adoption of Stale bot in a project occurred simultaneously with other events that could explain the observed changes, such as the introduction of new tools or practices. For example, less than a month before adopting Stale bot, the devexpress/devextreme project modified its continuous integration (CI) workflow to also run on submitted PRs in addition to pushed commits \citep{ziborov_commit_2020}. To mitigate this threat, we utilized mixed-effects models for implementing the interrupted time-series analysis. This approach allows us to identify changes that are commonly observed across the majority of projects rather than changes that are specific to only a few projects. In other words, it is unlikely that the majority of the studied projects experienced significant changes to their pull-based development workflow during the same month they adopted Stale bot.

\noindentparagraph{\emph{\textbf{External Validity.}}} External validity is concerned with how well our findings may be generalized to other open-source projects. In this study, we aimed to evaluate the helpfulness of Stale bot in open-source projects, specifically concerning their pull-based development. To conduct our study, we selected 20 large and popular open-source projects with a rich history of using Stale bot in their pull-based development workflow. While we believe these projects are more likely to benefit from adopting Stale bot, we recognize that they cannot represent the entire open-source ecosystem. In other words, the selected projects may not be representative of other open-source projects with different characteristics, such as their size, maturity, popularity, workload, culture, and development practices. Therefore, the findings of this study may not apply to all open-source projects. To acquire more broadly applicable insights, future research can replicate this study using a more diverse selection of projects.

\section{Conclusion}
\label{sec:conclusion}
PRs that are neither progressed nor resolved accumulate over time, clutter the list of PRs, and eventually make it difficult for the maintainers to manage and prioritize unresolved PRs. To automatically track such inactive PRs, follow up on their progress, and close them if needed, Stale bot was introduced by GitHub. Despite its increasing adoption, there are ongoing debates on whether using Stale bot alleviates or exacerbates the problem of inactive PRs. To better understand if and how Stale bot helps open-source projects in their pull-based development workflow, we conducted an empirical study of 20 large and popular open-source projects. First, we analyzed the configuration and activity of Stale bot to understand the extent to which the projects relied on Stale bot to automatically deal with their unresolved PRs. Then, we applied interrupted time-series analysis as a well-established quasi-experiment to understand if and how adopting Stale bot improved the efficiency and effectiveness of the pull-based development workflow in the projects. Next, we analyzed the characteristics of PRs, their contributors, and their review processes to understand the factors that were associated with a higher probability of getting intervened by Stale bot in the projects. Finally, we combined our observations to discuss the potential benefits and drawbacks of employing Stale bot within a pull-based development workflow. In summary, we found that Stale bot can help projects deal with a backlog of unresolved PRs and also improve the review process of PRs. However, the adoption of Stale bot can negatively affect the contributors (especially novice or casual contributors) in a project.

\bibliographystyle{ACM-Reference-Format}
\bibliography{references}


\begin{thebibliography}{77}


\ifx \showCODEN    \undefined \def \showCODEN     #1{\unskip}     \fi
\ifx \showDOI      \undefined \def \showDOI       #1{#1}\fi
\ifx \showISBNx    \undefined \def \showISBNx     #1{\unskip}     \fi
\ifx \showISBNxiii \undefined \def \showISBNxiii  #1{\unskip}     \fi
\ifx \showISSN     \undefined \def \showISSN      #1{\unskip}     \fi
\ifx \showLCCN     \undefined \def \showLCCN      #1{\unskip}     \fi
\ifx \shownote     \undefined \def \shownote      #1{#1}          \fi
\ifx \showarticletitle \undefined \def \showarticletitle #1{#1}   \fi
\ifx \showURL      \undefined \def \showURL       {\relax}        \fi
\providecommand\bibfield[2]{#2}
\providecommand\bibinfo[2]{#2}
\providecommand\natexlab[1]{#1}
\providecommand\showeprint[2][]{arXiv:#2}

\bibitem[Ben-Shachar et~al\mbox{.}(2020)]%
        {ben-shachar_effectsize_2020}
\bibfield{author}{\bibinfo{person}{Mattan~S. Ben-Shachar},
  \bibinfo{person}{Daniel Lüdecke}, {and} \bibinfo{person}{Dominique
  Makowski}.} \bibinfo{year}{2020}\natexlab{}.
\newblock \showarticletitle{effectsize: estimation of effect size indices and
  standardized parameters}.
\newblock \bibinfo{journal}{\emph{Journal of Open Source Software}}
  \bibinfo{volume}{5}, \bibinfo{number}{56} (\bibinfo{year}{2020}),
  \bibinfo{pages}{1--7}.
\newblock
\urldef\tempurl%
\url{https://doi.org/10.21105/joss.02815}
\showDOI{\tempurl}


\bibitem[Cassee et~al\mbox{.}(2020)]%
        {cassee_silent_2020}
\bibfield{author}{\bibinfo{person}{Nathan Cassee}, \bibinfo{person}{Bogdan
  Vasilescu}, {and} \bibinfo{person}{Alexander Serebrenik}.}
  \bibinfo{year}{2020}\natexlab{}.
\newblock \showarticletitle{The silent helper: the impact of continuous
  integration on code reviews}. In \bibinfo{booktitle}{\emph{Proceedings of the
  27th {International} {Conference} on {Software} {Analysis}, {Evolution}, and
  {Reengineering} ({SANER} 2020)}}. \bibinfo{pages}{423--434}.
\newblock
\urldef\tempurl%
\url{https://doi.org/10.1109/SANER48275.2020.9054818}
\showDOI{\tempurl}


\bibitem[Chavalarias et~al\mbox{.}(2016)]%
        {chavalarias_evolution_2016}
\bibfield{author}{\bibinfo{person}{David Chavalarias},
  \bibinfo{person}{Joshua~David Wallach}, \bibinfo{person}{Alvin Ho~Ting Li},
  {and} \bibinfo{person}{John P.~A. Ioannidis}.}
  \bibinfo{year}{2016}\natexlab{}.
\newblock \showarticletitle{Evolution of reporting \textit{{P}} values in the
  biomedical literature, 1990-2015}.
\newblock \bibinfo{journal}{\emph{JAMA}} \bibinfo{volume}{315},
  \bibinfo{number}{11} (\bibinfo{year}{2016}), \bibinfo{pages}{1141--1148}.
\newblock
\urldef\tempurl%
\url{https://doi.org/10.1001/jama.2016.1952}
\showDOI{\tempurl}


\bibitem[Cliff(1993)]%
        {cliff_dominance_1993}
\bibfield{author}{\bibinfo{person}{Norman Cliff}.}
  \bibinfo{year}{1993}\natexlab{}.
\newblock \showarticletitle{Dominance statistics: ordinal analyses to answer
  ordinal questions}.
\newblock \bibinfo{journal}{\emph{Psychological Bulletin}}
  \bibinfo{volume}{114}, \bibinfo{number}{3} (\bibinfo{year}{1993}),
  \bibinfo{pages}{494--509}.
\newblock
\urldef\tempurl%
\url{https://doi.org/10.1037/0033-2909.114.3.494}
\showDOI{\tempurl}


\bibitem[Creswell and Creswell(2017)]%
        {creswell_research_2017}
\bibfield{author}{\bibinfo{person}{John~W. Creswell} {and}
  \bibinfo{person}{J.~David Creswell}.} \bibinfo{year}{2017}\natexlab{}.
\newblock \bibinfo{booktitle}{\emph{Research {Design}: {Qualitative},
  {Quantitative}, and {Mixed} {Methods} {Approaches}} (\bibinfo{edition}{5th}
  ed.)}.
\newblock \bibinfo{publisher}{SAGE Publications, Inc}.
\newblock
\urldef\tempurl%
\url{https://us.sagepub.com/en-us/nam/research-design/book255675}
\showURL{%
\tempurl}


\bibitem[DeVault(2021)]%
        {devault_github_2021}
\bibfield{author}{\bibinfo{person}{Drew DeVault}.}
  \bibinfo{year}{2021}\natexlab{}.
\newblock \bibinfo{title}{{GitHub} {Stale} bot considered harmful}.
\newblock
\newblock
\urldef\tempurl%
\url{https://drewdevault.com/2021/10/26/stalebot.html}
\showURL{%
\tempurl}


\bibitem[Farah et~al\mbox{.}(2022)]%
        {farah_exploratory_2022}
\bibfield{author}{\bibinfo{person}{Juan~Carlos Farah}, \bibinfo{person}{Basile
  Spaenlehauer}, \bibinfo{person}{Xinyang Lu}, \bibinfo{person}{Sandy Ingram},
  {and} \bibinfo{person}{Denis Gillet}.} \bibinfo{year}{2022}\natexlab{}.
\newblock \showarticletitle{An exploratory study of reactions to bot comments
  on {GitHub}}. In \bibinfo{booktitle}{\emph{Proceedings of the 4th
  {International} {Workshop} on {Bots} in {Software} {Engineering} ({BotSE}
  2022)}}. \bibinfo{pages}{18--22}.
\newblock
\urldef\tempurl%
\url{https://doi.org/10.1145/3528228.3528409}
\showDOI{\tempurl}


\bibitem[Forsgren et~al\mbox{.}(2021)]%
        {forsgren_space_2021}
\bibfield{author}{\bibinfo{person}{Nicole Forsgren},
  \bibinfo{person}{Margaret-Anne Storey}, \bibinfo{person}{Chandra Maddila},
  \bibinfo{person}{Thomas Zimmermann}, \bibinfo{person}{Brian Houck}, {and}
  \bibinfo{person}{Jenna Butler}.} \bibinfo{year}{2021}\natexlab{}.
\newblock \showarticletitle{The {SPACE} of developer productivity: there's more
  to it than you think}.
\newblock \bibinfo{journal}{\emph{Queue}} \bibinfo{volume}{19},
  \bibinfo{number}{1} (\bibinfo{year}{2021}), \bibinfo{pages}{20--48}.
\newblock
\urldef\tempurl%
\url{https://doi.org/10.1145/3454122.3454124}
\showDOI{\tempurl}


\bibitem[Furtado et~al\mbox{.}(2021)]%
        {furtado_how_2021}
\bibfield{author}{\bibinfo{person}{Leonardo~B. Furtado}, \bibinfo{person}{Bruno
  Cartaxo}, \bibinfo{person}{Christoph Treude}, {and} \bibinfo{person}{Gustavo
  Pinto}.} \bibinfo{year}{2021}\natexlab{}.
\newblock \showarticletitle{How successful are open source contributions from
  countries with different levels of human development?}
\newblock \bibinfo{journal}{\emph{IEEE Software}} \bibinfo{volume}{38},
  \bibinfo{number}{2} (\bibinfo{year}{2021}), \bibinfo{pages}{58--63}.
\newblock
\urldef\tempurl%
\url{https://doi.org/10.1109/MS.2020.3044020}
\showDOI{\tempurl}


\bibitem[Gałecki and Burzykowski(2013)]%
        {galecki_linear_2013}
\bibfield{author}{\bibinfo{person}{Andrzej Gałecki} {and}
  \bibinfo{person}{Tomasz Burzykowski}.} \bibinfo{year}{2013}\natexlab{}.
\newblock \bibinfo{booktitle}{\emph{Linear {Mixed}-{Effects} {Models} {Using}
  {R}: {A} {Step}-by-{Step} {Approach}}}.
\newblock \bibinfo{publisher}{Springer}.
\newblock
\urldef\tempurl%
\url{https://doi.org/10.1007/978-1-4614-3900-4}
\showDOI{\tempurl}


\bibitem[{GitHub}(2023a)]%
        {github_events_2023}
\bibfield{author}{\bibinfo{person}{{GitHub}}.}
  \bibinfo{year}{2023}\natexlab{a}.
\newblock \bibinfo{title}{Events - {GitHub} {Docs}}.
\newblock
\newblock
\urldef\tempurl%
\url{https://docs.github.com/en/rest/activity/events}
\showURL{%
\tempurl}


\bibitem[{GitHub}(2023b)]%
        {github_issues_2023}
\bibfield{author}{\bibinfo{person}{{GitHub}}.}
  \bibinfo{year}{2023}\natexlab{b}.
\newblock \bibinfo{title}{Issues - {GitHub} {Docs}}.
\newblock
\newblock
\urldef\tempurl%
\url{https://docs.github.com/en/rest/reference/issues}
\showURL{%
\tempurl}


\bibitem[{GitHub}(2023c)]%
        {github_pulls_2023}
\bibfield{author}{\bibinfo{person}{{GitHub}}.}
  \bibinfo{year}{2023}\natexlab{c}.
\newblock \bibinfo{title}{Pulls - {GitHub} {Docs}}.
\newblock
\newblock
\urldef\tempurl%
\url{https://docs.github.com/en/rest/reference/pulls}
\showURL{%
\tempurl}


\bibitem[{GitHub}(2023d)]%
        {github_rest_2023}
\bibfield{author}{\bibinfo{person}{{GitHub}}.}
  \bibinfo{year}{2023}\natexlab{d}.
\newblock \bibinfo{title}{{REST} {API} - {GitHub} {Docs}}.
\newblock
\newblock
\urldef\tempurl%
\url{https://docs.github.com/en/rest}
\showURL{%
\tempurl}


\bibitem[{GitHub}(2023e)]%
        {github_stale_2023}
\bibfield{author}{\bibinfo{person}{{GitHub}}.}
  \bibinfo{year}{2023}\natexlab{e}.
\newblock \bibinfo{title}{Stale - {GitHub} {Marketplace}}.
\newblock
\newblock
\urldef\tempurl%
\url{https://github.com/marketplace/stale}
\showURL{%
\tempurl}


\bibitem[{GitHub}(2023f)]%
        {github_timeline_2023}
\bibfield{author}{\bibinfo{person}{{GitHub}}.}
  \bibinfo{year}{2023}\natexlab{f}.
\newblock \bibinfo{title}{Timeline - {GitHub} {Docs}}.
\newblock
\newblock
\urldef\tempurl%
\url{https://docs.github.com/en/rest/issues/timeline}
\showURL{%
\tempurl}


\bibitem[Goodnight(1980)]%
        {goodnight_tests_1980}
\bibfield{author}{\bibinfo{person}{James~Howard Goodnight}.}
  \bibinfo{year}{1980}\natexlab{}.
\newblock \showarticletitle{Tests of hypotheses in fixed effects linear
  models}.
\newblock \bibinfo{journal}{\emph{Communications in Statistics - Theory and
  Methods}} \bibinfo{volume}{9}, \bibinfo{number}{2} (\bibinfo{year}{1980}),
  \bibinfo{pages}{167--180}.
\newblock
\urldef\tempurl%
\url{https://doi.org/10.1080/03610928008827869}
\showDOI{\tempurl}


\bibitem[{Google}(2023)]%
        {google_bigquery_2023}
\bibfield{author}{\bibinfo{person}{{Google}}.} \bibinfo{year}{2023}\natexlab{}.
\newblock \bibinfo{title}{{BigQuery}: cloud data warehouse}.
\newblock
\newblock
\urldef\tempurl%
\url{https://cloud.google.com/bigquery}
\showURL{%
\tempurl}


\bibitem[Gousios et~al\mbox{.}(2014)]%
        {gousios_exploratory_2014}
\bibfield{author}{\bibinfo{person}{Georgios Gousios}, \bibinfo{person}{Martin
  Pinzger}, {and} \bibinfo{person}{Arie van Deursen}.}
  \bibinfo{year}{2014}\natexlab{}.
\newblock \showarticletitle{An exploratory study of the pull-based software
  development model}. In \bibinfo{booktitle}{\emph{Proceedings of the 36th
  {International} {Conference} on {Software} {Engineering} ({ICSE} 2014)}}.
  \bibinfo{pages}{345--355}.
\newblock
\urldef\tempurl%
\url{https://doi.org/10.1145/2568225.2568260}
\showDOI{\tempurl}


\bibitem[Gousios and Zaidman(2014)]%
        {gousios_dataset_2014}
\bibfield{author}{\bibinfo{person}{Georgios Gousios} {and}
  \bibinfo{person}{Andy Zaidman}.} \bibinfo{year}{2014}\natexlab{}.
\newblock \showarticletitle{A dataset for pull-based development research}. In
  \bibinfo{booktitle}{\emph{Proceedings of the 11th {Working} {Conference} on
  {Mining} {Software} {Repositories} ({MSR} 2014)}}. \bibinfo{pages}{368--371}.
\newblock
\urldef\tempurl%
\url{https://doi.org/10.1145/2597073.2597122}
\showDOI{\tempurl}


\bibitem[Gousios et~al\mbox{.}(2015)]%
        {gousios_work_2015}
\bibfield{author}{\bibinfo{person}{Georgios Gousios}, \bibinfo{person}{Andy
  Zaidman}, \bibinfo{person}{Margaret-Anne Storey}, {and} \bibinfo{person}{Arie
  van Deursen}.} \bibinfo{year}{2015}\natexlab{}.
\newblock \showarticletitle{Work practices and challenges in pull-based
  development: the integrator's perspective}. In
  \bibinfo{booktitle}{\emph{Proceedings of the {IEEE}/{ACM} 37th
  {International} {Conference} on {Software} {Engineering} ({ICSE} 2015)}}.
  \bibinfo{pages}{358--368}.
\newblock
\urldef\tempurl%
\url{https://doi.org/10.1109/ICSE.2015.55}
\showDOI{\tempurl}


\bibitem[Grigorik(2023)]%
        {grigorik_gh_2023}
\bibfield{author}{\bibinfo{person}{Ilya Grigorik}.}
  \bibinfo{year}{2023}\natexlab{}.
\newblock \bibinfo{title}{{GH} {Archive}: {A} project to record the public
  {GitHub} timeline, archive it, and make it easily accessible for further
  analysis}.
\newblock
\newblock
\urldef\tempurl%
\url{https://www.gharchive.org}
\showURL{%
\tempurl}


\bibitem[Hess and Kromrey(2004)]%
        {hess_robust_2004}
\bibfield{author}{\bibinfo{person}{Melinda~R. Hess} {and}
  \bibinfo{person}{Jeffrey~D. Kromrey}.} \bibinfo{year}{2004}\natexlab{}.
\newblock \showarticletitle{Robust confidence intervals for effect sizes: a
  comparative study of {Cohen}'s d and {Cliff}'s delta under non-normality and
  heterogeneous variances}. In \bibinfo{booktitle}{\emph{Presented at the
  {Annual} {Meeting} of the {American} {Educational} {Research} {Association}
  ({AERA} 2004)}}. \bibinfo{pages}{1--30}.
\newblock
\urldef\tempurl%
\url{https://citeseerx.ist.psu.edu/viewdoc/summary?doi=10.1.1.487.8299}
\showURL{%
\tempurl}


\bibitem[Hintze and Nelson(1998)]%
        {hintze_violin_1998}
\bibfield{author}{\bibinfo{person}{Jerry~L. Hintze} {and}
  \bibinfo{person}{Ray~D. Nelson}.} \bibinfo{year}{1998}\natexlab{}.
\newblock \showarticletitle{Violin plots: a box plot-density trace synergism}.
\newblock \bibinfo{journal}{\emph{The American Statistician}}
  \bibinfo{volume}{52}, \bibinfo{number}{2} (\bibinfo{year}{1998}),
  \bibinfo{pages}{181--184}.
\newblock
\urldef\tempurl%
\url{https://doi.org/10.1080/00031305.1998.10480559}
\showDOI{\tempurl}


\bibitem[Hrong-Tai~Fai and Cornelius(1996)]%
        {hrong-tai_fai_approximate_1996}
\bibfield{author}{\bibinfo{person}{Alex Hrong-Tai~Fai} {and}
  \bibinfo{person}{Paul~L. Cornelius}.} \bibinfo{year}{1996}\natexlab{}.
\newblock \showarticletitle{Approximate {F}-tests of multiple degree of freedom
  hypotheses in generalized least squares analyses of unbalanced split-plot
  experiments}.
\newblock \bibinfo{journal}{\emph{Journal of Statistical Computation and
  Simulation}} \bibinfo{volume}{54}, \bibinfo{number}{4}
  (\bibinfo{year}{1996}), \bibinfo{pages}{363--378}.
\newblock
\urldef\tempurl%
\url{https://doi.org/10.1080/00949659608811740}
\showDOI{\tempurl}


\bibitem[Iyer et~al\mbox{.}(2021)]%
        {iyer_effects_2021}
\bibfield{author}{\bibinfo{person}{Rahul~N. Iyer}, \bibinfo{person}{S.~Alex
  Yun}, \bibinfo{person}{Meiyappan Nagappan}, {and} \bibinfo{person}{Jesse
  Hoey}.} \bibinfo{year}{2021}\natexlab{}.
\newblock \showarticletitle{Effects of personality traits on pull request
  acceptance}.
\newblock \bibinfo{journal}{\emph{IEEE Transactions on Software Engineering}}
  \bibinfo{volume}{47}, \bibinfo{number}{11} (\bibinfo{year}{2021}),
  \bibinfo{pages}{2632--2643}.
\newblock
\urldef\tempurl%
\url{https://doi.org/10.1109/TSE.2019.2960357}
\showDOI{\tempurl}


\bibitem[Jacques(2023)]%
        {jacques_pygithub_2023}
\bibfield{author}{\bibinfo{person}{Vincent Jacques}.}
  \bibinfo{year}{2023}\natexlab{}.
\newblock \bibinfo{title}{{PyGithub}: typed interactions with the {GitHub}
  {API} v3}.
\newblock
\newblock
\urldef\tempurl%
\url{https://github.com/PyGithub/PyGithub}
\showURL{%
\tempurl}


\bibitem[Kalliamvakou et~al\mbox{.}(2016)]%
        {kalliamvakou_-depth_2016}
\bibfield{author}{\bibinfo{person}{Eirini Kalliamvakou},
  \bibinfo{person}{Georgios Gousios}, \bibinfo{person}{Kelly Blincoe},
  \bibinfo{person}{Leif Singer}, \bibinfo{person}{Daniel~M. German}, {and}
  \bibinfo{person}{Daniela Damian}.} \bibinfo{year}{2016}\natexlab{}.
\newblock \showarticletitle{An in-depth study of the promises and perils of
  mining {GitHub}}.
\newblock \bibinfo{journal}{\emph{Empirical Software Engineering}}
  \bibinfo{volume}{21}, \bibinfo{number}{5} (\bibinfo{year}{2016}),
  \bibinfo{pages}{2035--2071}.
\newblock
\urldef\tempurl%
\url{https://doi.org/10.1007/s10664-015-9393-5}
\showDOI{\tempurl}


\bibitem[Keepers(2023)]%
        {keepers_stale_2023}
\bibfield{author}{\bibinfo{person}{Brandon Keepers}.}
  \bibinfo{year}{2023}\natexlab{}.
\newblock \bibinfo{title}{Stale - {GitHub} {Repository}}.
\newblock
\newblock
\urldef\tempurl%
\url{https://github.com/probot/stale}
\showURL{%
\tempurl}


\bibitem[Khatoonabadi et~al\mbox{.}(2023)]%
        {khatoonabadi_wasted_2023}
\bibfield{author}{\bibinfo{person}{SayedHassan Khatoonabadi},
  \bibinfo{person}{Diego~Elias Costa}, \bibinfo{person}{Rabe Abdalkareem},
  {and} \bibinfo{person}{Emad Shihab}.} \bibinfo{year}{2023}\natexlab{}.
\newblock \showarticletitle{On wasted contributions: understanding the dynamics
  of contributor-abandoned pull requests – a mixed-methods study of 10 large
  open-source projects}.
\newblock \bibinfo{journal}{\emph{ACM Transactions on Software Engineering and
  Methodology}} \bibinfo{volume}{32}, \bibinfo{number}{1}
  (\bibinfo{year}{2023}), \bibinfo{pages}{1--39}.
\newblock
\urldef\tempurl%
\url{https://doi.org/10.1145/3530785}
\showDOI{\tempurl}


\bibitem[Khatoonabadi et~al\mbox{.}(2021)]%
        {khatoonabadi_gap2wss_2021}
\bibfield{author}{\bibinfo{person}{SayedHassan Khatoonabadi},
  \bibinfo{person}{Shahriar Lotfi}, {and} \bibinfo{person}{Ayaz Isazadeh}.}
  \bibinfo{year}{2021}\natexlab{}.
\newblock \bibinfo{title}{{GAP2WSS}: a genetic algorithm based on the {Pareto}
  principle for web service selection}.
\newblock
\newblock
\urldef\tempurl%
\url{https://doi.org/10.48550/arXiv.2109.10430}
\showDOI{\tempurl}


\bibitem[Kirk(1996)]%
        {kirk_practical_1996}
\bibfield{author}{\bibinfo{person}{Roger~E. Kirk}.}
  \bibinfo{year}{1996}\natexlab{}.
\newblock \showarticletitle{Practical significance: a concept whose time has
  come}.
\newblock \bibinfo{journal}{\emph{Educational and Psychological Measurement}}
  \bibinfo{volume}{56}, \bibinfo{number}{5} (\bibinfo{year}{1996}),
  \bibinfo{pages}{746--759}.
\newblock
\urldef\tempurl%
\url{https://doi.org/10.1177/0013164496056005002}
\showDOI{\tempurl}


\bibitem[Kitchenham et~al\mbox{.}(2015)]%
        {kitchenham_evidence-based_2015}
\bibfield{author}{\bibinfo{person}{Barbara~Ann Kitchenham},
  \bibinfo{person}{David Budgen}, {and} \bibinfo{person}{Pearl Brereton}.}
  \bibinfo{year}{2015}\natexlab{}.
\newblock \bibinfo{booktitle}{\emph{Evidence-{Based} {Software} {Engineering}
  and {Systematic} {Reviews}} (\bibinfo{edition}{1st} ed.)}.
\newblock \bibinfo{publisher}{Chapman and Hall/CRC}.
\newblock
\urldef\tempurl%
\url{https://doi.org/10.1201/b19467}
\showDOI{\tempurl}


\bibitem[Kononenko et~al\mbox{.}(2018)]%
        {kononenko_studying_2018}
\bibfield{author}{\bibinfo{person}{Oleksii Kononenko}, \bibinfo{person}{Tresa
  Rose}, \bibinfo{person}{Olga Baysal}, \bibinfo{person}{Michael Godfrey},
  \bibinfo{person}{Dennis Theisen}, {and} \bibinfo{person}{Bart de Water}.}
  \bibinfo{year}{2018}\natexlab{}.
\newblock \showarticletitle{Studying pull request merges: a case study of
  {Shopify}'s {Active} {Merchant}}. In \bibinfo{booktitle}{\emph{Proceedings of
  the 40th {International} {Conference} on {Software} {Engineering}: {Software}
  {Engineering} in {Practice} ({ICSE}-{SEIP} 2018)}}.
  \bibinfo{pages}{124--133}.
\newblock
\urldef\tempurl%
\url{https://doi.org/10.1145/3183519.3183542}
\showDOI{\tempurl}


\bibitem[Kuznetsova et~al\mbox{.}(2017)]%
        {kuznetsova_lmertest_2017}
\bibfield{author}{\bibinfo{person}{Alexandra Kuznetsova},
  \bibinfo{person}{Per~B. Brockhoff}, {and} \bibinfo{person}{Rune H.~B.
  Christensen}.} \bibinfo{year}{2017}\natexlab{}.
\newblock \showarticletitle{{lmerTest} package: tests in linear mixed effects
  models}.
\newblock \bibinfo{journal}{\emph{Journal of Statistical Software}}
  \bibinfo{volume}{82}, \bibinfo{number}{13} (\bibinfo{year}{2017}),
  \bibinfo{pages}{1--26}.
\newblock
\urldef\tempurl%
\url{https://doi.org/10.18637/jss.v082.i13}
\showDOI{\tempurl}


\bibitem[Lenarduzzi et~al\mbox{.}(2021)]%
        {lenarduzzi_does_2021}
\bibfield{author}{\bibinfo{person}{Valentina Lenarduzzi}, \bibinfo{person}{Vili
  Nikkola}, \bibinfo{person}{Nyyti Saarimäki}, {and} \bibinfo{person}{Davide
  Taibi}.} \bibinfo{year}{2021}\natexlab{}.
\newblock \showarticletitle{Does code quality affect pull request acceptance?
  {An} empirical study}.
\newblock \bibinfo{journal}{\emph{Journal of Systems and Software}}
  \bibinfo{volume}{171} (\bibinfo{year}{2021}), \bibinfo{pages}{1--14}.
\newblock
\urldef\tempurl%
\url{https://doi.org/10.1016/j.jss.2020.110806}
\showDOI{\tempurl}


\bibitem[Li et~al\mbox{.}(2022)]%
        {li_are_2022}
\bibfield{author}{\bibinfo{person}{Zhixing Li}, \bibinfo{person}{Yue Yu},
  \bibinfo{person}{Tao Wang}, \bibinfo{person}{Gang Yin},
  \bibinfo{person}{ShanShan Li}, {and} \bibinfo{person}{Huaimin Wang}.}
  \bibinfo{year}{2022}\natexlab{}.
\newblock \showarticletitle{Are you still working on this? {An} empirical study
  on pull request abandonment}.
\newblock \bibinfo{journal}{\emph{IEEE Transactions on Software Engineering}}
  \bibinfo{volume}{48}, \bibinfo{number}{6} (\bibinfo{year}{2022}),
  \bibinfo{pages}{2173--2188}.
\newblock
\urldef\tempurl%
\url{https://doi.org/10.1109/TSE.2021.3053403}
\showDOI{\tempurl}


\bibitem[Liu et~al\mbox{.}(2020)]%
        {liu_understanding_2020}
\bibfield{author}{\bibinfo{person}{Dongyu Liu}, \bibinfo{person}{Micah~J.
  Smith}, {and} \bibinfo{person}{Kalyan Veeramachaneni}.}
  \bibinfo{year}{2020}\natexlab{}.
\newblock \showarticletitle{Understanding user-bot interactions for small-scale
  automation in open-source development}. In \bibinfo{booktitle}{\emph{Extended
  {Abstracts} of the {CHI} {Conference} on {Human} {Factors} in {Computing}
  {Systems} ({CHI} {EA} 2020)}}. \bibinfo{pages}{1--8}.
\newblock
\urldef\tempurl%
\url{https://doi.org/10.1145/3334480.3382998}
\showDOI{\tempurl}


\bibitem[Lüdecke et~al\mbox{.}(2021)]%
        {ludecke_performance_2021}
\bibfield{author}{\bibinfo{person}{Daniel Lüdecke}, \bibinfo{person}{Mattan~S.
  Ben-Shachar}, \bibinfo{person}{Indrajeet Patil}, \bibinfo{person}{Philip
  Waggoner}, {and} \bibinfo{person}{Dominique Makowski}.}
  \bibinfo{year}{2021}\natexlab{}.
\newblock \showarticletitle{performance: an {R} package for assessment,
  comparison and testing of statistical models}.
\newblock \bibinfo{journal}{\emph{Journal of Open Source Software}}
  \bibinfo{volume}{6}, \bibinfo{number}{60} (\bibinfo{year}{2021}),
  \bibinfo{pages}{1--8}.
\newblock
\urldef\tempurl%
\url{https://doi.org/10.21105/joss.03139}
\showDOI{\tempurl}


\bibitem[Mann and Whitney(1947)]%
        {mann_test_1947}
\bibfield{author}{\bibinfo{person}{Henry~B. Mann} {and}
  \bibinfo{person}{Donald~R. Whitney}.} \bibinfo{year}{1947}\natexlab{}.
\newblock \showarticletitle{On a test of whether one of two random variables is
  stochastically larger than the other}.
\newblock \bibinfo{journal}{\emph{Annals of Mathematical Statistics}}
  \bibinfo{volume}{18}, \bibinfo{number}{1} (\bibinfo{year}{1947}),
  \bibinfo{pages}{50--60}.
\newblock
\urldef\tempurl%
\url{https://doi.org/10.1214/aoms/1177730491}
\showDOI{\tempurl}


\bibitem[Nadri et~al\mbox{.}(2021)]%
        {nadri_insights_2021}
\bibfield{author}{\bibinfo{person}{Reza Nadri}, \bibinfo{person}{Gema
  Rodríguez-Pérez}, {and} \bibinfo{person}{Meiyappan Nagappan}.}
  \bibinfo{year}{2021}\natexlab{}.
\newblock \showarticletitle{Insights into nonmerged pull requests in {GitHub}:
  is there evidence of bias based on perceptible race?}
\newblock \bibinfo{journal}{\emph{IEEE Software}} \bibinfo{volume}{38},
  \bibinfo{number}{2} (\bibinfo{year}{2021}), \bibinfo{pages}{51--57}.
\newblock
\urldef\tempurl%
\url{https://doi.org/10.1109/MS.2020.3036758}
\showDOI{\tempurl}


\bibitem[Nadri et~al\mbox{.}(2022)]%
        {nadri_relationship_2022}
\bibfield{author}{\bibinfo{person}{Reza Nadri}, \bibinfo{person}{Gema
  Rodríguez-Pérez}, {and} \bibinfo{person}{Meiyappan Nagappan}.}
  \bibinfo{year}{2022}\natexlab{}.
\newblock \showarticletitle{On the relationship between the developer's
  perceptible race and ethnicity and the evaluation of contributions in {OSS}}.
\newblock \bibinfo{journal}{\emph{IEEE Transactions on Software Engineering}}
  \bibinfo{volume}{48}, \bibinfo{number}{8} (\bibinfo{year}{2022}),
  \bibinfo{pages}{2955--2968}.
\newblock
\urldef\tempurl%
\url{https://doi.org/10.1109/TSE.2021.3073773}
\showDOI{\tempurl}


\bibitem[Nakagawa et~al\mbox{.}(2017)]%
        {nakagawa_coefficient_2017}
\bibfield{author}{\bibinfo{person}{Shinichi Nakagawa}, \bibinfo{person}{Paul
  C.~D. Johnson}, {and} \bibinfo{person}{Holger Schielzeth}.}
  \bibinfo{year}{2017}\natexlab{}.
\newblock \showarticletitle{The coefficient of determination
  \textit{{R}}$^{\textrm{2}}$ and intra-class correlation coefficient from
  generalized linear mixed-effects models revisited and expanded}.
\newblock \bibinfo{journal}{\emph{Journal of The Royal Society Interface}}
  \bibinfo{volume}{14}, \bibinfo{number}{134} (\bibinfo{year}{2017}),
  \bibinfo{pages}{1--11}.
\newblock
\urldef\tempurl%
\url{https://doi.org/10.1098/rsif.2017.0213}
\showDOI{\tempurl}


\bibitem[Patil(2021)]%
        {patil_visualizations_2021}
\bibfield{author}{\bibinfo{person}{Indrajeet Patil}.}
  \bibinfo{year}{2021}\natexlab{}.
\newblock \showarticletitle{Visualizations with statistical details: the
  'ggstatsplot' approach}.
\newblock \bibinfo{journal}{\emph{Journal of Open Source Software}}
  \bibinfo{volume}{6}, \bibinfo{number}{61} (\bibinfo{year}{2021}),
  \bibinfo{pages}{1--5}.
\newblock
\urldef\tempurl%
\url{https://doi.org/10.21105/joss.03167}
\showDOI{\tempurl}


\bibitem[Pinto et~al\mbox{.}(2018)]%
        {pinto_who_2018}
\bibfield{author}{\bibinfo{person}{Gustavo Pinto}, \bibinfo{person}{Luiz~Felipe
  Dias}, {and} \bibinfo{person}{Igor Steinmacher}.}
  \bibinfo{year}{2018}\natexlab{}.
\newblock \showarticletitle{Who gets a patch accepted first? {Comparing} the
  contributions of employees and volunteers}. In
  \bibinfo{booktitle}{\emph{Proceedings of the 11th {International} {Workshop}
  on {Cooperative} and {Human} {Aspects} of {Software} {Engineering} ({CHASE}
  2018)}}. \bibinfo{pages}{110--113}.
\newblock
\urldef\tempurl%
\url{https://doi.org/10.1145/3195836.3195858}
\showDOI{\tempurl}


\bibitem[{R Core Team}(2023)]%
        {r_core_team_r_2023}
\bibfield{author}{\bibinfo{person}{{R Core Team}}.}
  \bibinfo{year}{2023}\natexlab{}.
\newblock \bibinfo{title}{R: a language and environment for statistical
  computing}.
\newblock
\newblock
\urldef\tempurl%
\url{https://www.R-project.org}
\showURL{%
\tempurl}


\bibitem[Rahman et~al\mbox{.}(2022)]%
        {rahman_towards_2022}
\bibfield{author}{\bibinfo{person}{Akond Rahman},
  \bibinfo{person}{Farzana~Ahamed Bhuiyan}, \bibinfo{person}{Mohammad~Mehedi
  Hassan}, \bibinfo{person}{Hossain Shahriar}, {and} \bibinfo{person}{Fan Wu}.}
  \bibinfo{year}{2022}\natexlab{}.
\newblock \showarticletitle{Towards automation for {MLOps}: an exploratory
  study of bot usage in deep learning libraries}. In
  \bibinfo{booktitle}{\emph{Proceedings of the {IEEE} 46th {Annual}
  {Computers}, {Software}, and {Applications} {Conference} ({COMPSAC} 2022)}}.
  \bibinfo{pages}{1093--1097}.
\newblock
\urldef\tempurl%
\url{https://doi.org/10.1109/COMPSAC54236.2022.00171}
\showDOI{\tempurl}


\bibitem[Rastogi(2016)]%
        {rastogi_biases_2016}
\bibfield{author}{\bibinfo{person}{Ayushi Rastogi}.}
  \bibinfo{year}{2016}\natexlab{}.
\newblock \showarticletitle{Do biases related to geographical location
  influence work-related decisions in {GitHub}?}. In
  \bibinfo{booktitle}{\emph{Proceedings of the 38th {International}
  {Conference} on {Software} {Engineering} {Companion} ({ICSE}-{C} 2016)}}.
  \bibinfo{pages}{665--667}.
\newblock
\urldef\tempurl%
\url{https://doi.org/10.1145/2889160.2891035}
\showDOI{\tempurl}


\bibitem[Rastogi et~al\mbox{.}(2018)]%
        {rastogi_relationship_2018}
\bibfield{author}{\bibinfo{person}{Ayushi Rastogi}, \bibinfo{person}{Nachiappan
  Nagappan}, \bibinfo{person}{Georgios Gousios}, {and} \bibinfo{person}{André
  van~der Hoek}.} \bibinfo{year}{2018}\natexlab{}.
\newblock \showarticletitle{Relationship between geographical location and
  evaluation of developer contributions in {GitHub}}. In
  \bibinfo{booktitle}{\emph{Proceedings of the 12th {ACM}/{IEEE}
  {International} {Symposium} on {Empirical} {Software} {Engineering} and
  {Measurement} ({ESEM} 2018)}}. \bibinfo{pages}{1--8}.
\newblock
\urldef\tempurl%
\url{https://doi.org/10.1145/3239235.3240504}
\showDOI{\tempurl}


\bibitem[Soares et~al\mbox{.}(2015)]%
        {soares_acceptance_2015}
\bibfield{author}{\bibinfo{person}{Daricélio~Moreira Soares},
  \bibinfo{person}{Manoel~Limeira de Lima~Júnior}, \bibinfo{person}{Leonardo
  Murta}, {and} \bibinfo{person}{Alexandre Plastino}.}
  \bibinfo{year}{2015}\natexlab{}.
\newblock \showarticletitle{Acceptance factors of pull requests in open-source
  projects}. In \bibinfo{booktitle}{\emph{Proceedings of the 30th {Annual}
  {ACM} {Symposium} on {Applied} {Computing} ({SAC} 2015)}}.
  \bibinfo{pages}{1541--1546}.
\newblock
\urldef\tempurl%
\url{https://doi.org/10.1145/2695664.2695856}
\showDOI{\tempurl}


\bibitem[Steinmacher et~al\mbox{.}(2014)]%
        {steinmacher_preliminary_2014}
\bibfield{author}{\bibinfo{person}{Igor Steinmacher},
  \bibinfo{person}{Ana~Paula Chaves}, \bibinfo{person}{Tayana~Uchoa Conte},
  {and} \bibinfo{person}{Marco~Aurélio Gerosa}.}
  \bibinfo{year}{2014}\natexlab{}.
\newblock \showarticletitle{Preliminary empirical identification of barriers
  faced by newcomers to open source software projects}. In
  \bibinfo{booktitle}{\emph{Proceedings of the {Brazilian} {Symposium} on
  {Software} {Engineering} ({SBES} 2014)}}. \bibinfo{pages}{51--60}.
\newblock
\urldef\tempurl%
\url{https://doi.org/10.1109/SBES.2014.9}
\showDOI{\tempurl}


\bibitem[Steinmacher et~al\mbox{.}(2015)]%
        {steinmacher_social_2015}
\bibfield{author}{\bibinfo{person}{Igor Steinmacher}, \bibinfo{person}{Tayana
  Conte}, \bibinfo{person}{Marco~Aurélio Gerosa}, {and} \bibinfo{person}{David
  Redmiles}.} \bibinfo{year}{2015}\natexlab{}.
\newblock \showarticletitle{Social barriers faced by newcomers placing their
  first contribution in open source software projects}. In
  \bibinfo{booktitle}{\emph{Proceedings of the 18th {ACM} {Conference} on
  {Computer} {Supported} {Cooperative} {Work} \& {Social} {Computing} ({CSCW}
  2015)}}. \bibinfo{pages}{1379--1392}.
\newblock
\urldef\tempurl%
\url{https://doi.org/10.1145/2675133.2675215}
\showDOI{\tempurl}


\bibitem[Steinmacher et~al\mbox{.}(2013)]%
        {steinmacher_why_2013}
\bibfield{author}{\bibinfo{person}{Igor Steinmacher}, \bibinfo{person}{Igor
  Wiese}, \bibinfo{person}{Ana~Paula Chaves}, {and}
  \bibinfo{person}{Marco~Aurélio Gerosa}.} \bibinfo{year}{2013}\natexlab{}.
\newblock \showarticletitle{Why do newcomers abandon open source software
  projects?}. In \bibinfo{booktitle}{\emph{Proceedings of the 6th
  {International} {Workshop} on {Cooperative} and {Human} {Aspects} of
  {Software} {Engineering} ({CHASE} 2013)}}. \bibinfo{pages}{25--32}.
\newblock
\urldef\tempurl%
\url{https://doi.org/10.1109/CHASE.2013.6614728}
\showDOI{\tempurl}


\bibitem[Storey and Zagalsky(2016)]%
        {storey_disrupting_2016}
\bibfield{author}{\bibinfo{person}{Margaret-Anne Storey} {and}
  \bibinfo{person}{Alexey Zagalsky}.} \bibinfo{year}{2016}\natexlab{}.
\newblock \showarticletitle{Disrupting developer productivity one bot at a
  time}. In \bibinfo{booktitle}{\emph{Proceedings of the 24th {ACM} {SIGSOFT}
  {International} {Symposium} on {Foundations} of {Software} {Engineering}
  ({FSE} 2016)}}. \bibinfo{pages}{928--931}.
\newblock
\urldef\tempurl%
\url{https://doi.org/10.1145/2950290.2983989}
\showDOI{\tempurl}


\bibitem[Sørensen(2014)]%
        {sorensen_pull_2014}
\bibfield{author}{\bibinfo{person}{Poul~Kjeldager Sørensen}.}
  \bibinfo{year}{2014}\natexlab{}.
\newblock \bibinfo{title}{Pull {Request} \#2143 -
  {DefinitelyTyped}/{DefinitelyTyped}}.
\newblock
\newblock
\urldef\tempurl%
\url{https://github.com/DefinitelyTyped/DefinitelyTyped/pull/2143}
\showURL{%
\tempurl}


\bibitem[Terrell et~al\mbox{.}(2017)]%
        {terrell_gender_2017}
\bibfield{author}{\bibinfo{person}{Josh Terrell}, \bibinfo{person}{Andrew
  Kofink}, \bibinfo{person}{Justin Middleton}, \bibinfo{person}{Clarissa
  Rainear}, \bibinfo{person}{Emerson Murphy-Hill}, \bibinfo{person}{Chris
  Parnin}, {and} \bibinfo{person}{Jon Stallings}.}
  \bibinfo{year}{2017}\natexlab{}.
\newblock \showarticletitle{Gender differences and bias in open source: pull
  request acceptance of women versus men}.
\newblock \bibinfo{journal}{\emph{PeerJ Computer Science}}
  \bibinfo{volume}{2017}, \bibinfo{number}{3} (\bibinfo{year}{2017}),
  \bibinfo{pages}{1--30}.
\newblock
\urldef\tempurl%
\url{https://doi.org/10.7717/peerj-cs.111}
\showDOI{\tempurl}


\bibitem[Tsay et~al\mbox{.}(2014)]%
        {tsay_influence_2014}
\bibfield{author}{\bibinfo{person}{Jason Tsay}, \bibinfo{person}{Laura
  Dabbish}, {and} \bibinfo{person}{James Herbsleb}.}
  \bibinfo{year}{2014}\natexlab{}.
\newblock \showarticletitle{Influence of social and technical factors for
  evaluating contribution in {GitHub}}. In
  \bibinfo{booktitle}{\emph{Proceedings of the 36th {International}
  {Conference} on {Software} {Engineering} ({ICSE} 2014)}}.
  \bibinfo{pages}{356--366}.
\newblock
\urldef\tempurl%
\url{https://doi.org/10.1145/2568225.2568315}
\showDOI{\tempurl}
\newblock
\shownote{Issue: 1}.


\bibitem[Wagner et~al\mbox{.}(2002)]%
        {wagner_segmented_2002}
\bibfield{author}{\bibinfo{person}{Anita~K. Wagner},
  \bibinfo{person}{Stephen~B. Soumerai}, \bibinfo{person}{Fang Zhang}, {and}
  \bibinfo{person}{Dennis Ross-Degnan}.} \bibinfo{year}{2002}\natexlab{}.
\newblock \showarticletitle{Segmented regression analysis of interrupted time
  series studies in medication use research}.
\newblock \bibinfo{journal}{\emph{Journal of Clinical Pharmacy and
  Therapeutics}} \bibinfo{volume}{27}, \bibinfo{number}{4}
  (\bibinfo{year}{2002}), \bibinfo{pages}{299--309}.
\newblock
\urldef\tempurl%
\url{https://doi.org/10.1046/j.1365-2710.2002.00430.x}
\showDOI{\tempurl}


\bibitem[Wang et~al\mbox{.}(2019)]%
        {wang_why_2019}
\bibfield{author}{\bibinfo{person}{Qingye Wang}, \bibinfo{person}{Xin Xia},
  \bibinfo{person}{David Lo}, {and} \bibinfo{person}{Shanping Li}.}
  \bibinfo{year}{2019}\natexlab{}.
\newblock \showarticletitle{Why is my code change abandoned?}
\newblock \bibinfo{journal}{\emph{Information and Software Technology}}
  \bibinfo{volume}{110} (\bibinfo{year}{2019}), \bibinfo{pages}{108--120}.
\newblock
\urldef\tempurl%
\url{https://doi.org/10.1016/j.infsof.2019.02.007}
\showDOI{\tempurl}


\bibitem[Wessel et~al\mbox{.}(2022a)]%
        {wessel_bots_2022}
\bibfield{author}{\bibinfo{person}{Mairieli Wessel}, \bibinfo{person}{Ahmad
  Abdellatif}, \bibinfo{person}{Igor Wiese}, \bibinfo{person}{Tayana Conte},
  \bibinfo{person}{Emad Shihab}, \bibinfo{person}{Marco~Aurélio Gerosa}, {and}
  \bibinfo{person}{Igor Steinmacher}.} \bibinfo{year}{2022}\natexlab{a}.
\newblock \showarticletitle{Bots for pull requests: the good, the bad, and the
  promising}. In \bibinfo{booktitle}{\emph{Proceedings of the 44th
  {International} {Conference} on {Software} {Engineering} ({ICSE} 2022)}}.
  \bibinfo{pages}{274--286}.
\newblock
\urldef\tempurl%
\url{https://doi.org/10.1145/3510003.3512765}
\showDOI{\tempurl}


\bibitem[Wessel et~al\mbox{.}(2018)]%
        {wessel_power_2018}
\bibfield{author}{\bibinfo{person}{Mairieli Wessel},
  \bibinfo{person}{Bruno~Mendes de Souza}, \bibinfo{person}{Igor Steinmacher},
  \bibinfo{person}{Igor~S. Wiese}, \bibinfo{person}{Ivanilton Polato},
  \bibinfo{person}{Ana~Paula Chaves}, {and} \bibinfo{person}{Marco~Aurélio
  Gerosa}.} \bibinfo{year}{2018}\natexlab{}.
\newblock \showarticletitle{The power of bots: characterizing and understanding
  bots in {OSS} projects}.
\newblock \bibinfo{journal}{\emph{Proceedings of the ACM on Human-Computer
  Interaction}} \bibinfo{volume}{2}, \bibinfo{number}{CSCW}
  (\bibinfo{year}{2018}), \bibinfo{pages}{1--19}.
\newblock
\urldef\tempurl%
\url{https://doi.org/10.1145/3274451}
\showDOI{\tempurl}


\bibitem[Wessel et~al\mbox{.}(2022b)]%
        {wessel_quality_2022}
\bibfield{author}{\bibinfo{person}{Mairieli Wessel}, \bibinfo{person}{Alexander
  Serebrenik}, \bibinfo{person}{Igor Wiese}, \bibinfo{person}{Igor
  Steinmacher}, {and} \bibinfo{person}{Marco~Aurélio Gerosa}.}
  \bibinfo{year}{2022}\natexlab{b}.
\newblock \showarticletitle{Quality gatekeepers: investigating the effects of
  code review bots on pull request activities}.
\newblock \bibinfo{journal}{\emph{Empirical Software Engineering}}
  \bibinfo{volume}{27}, \bibinfo{number}{5} (\bibinfo{year}{2022}),
  \bibinfo{pages}{1--36}.
\newblock
\urldef\tempurl%
\url{https://doi.org/10.1007/s10664-022-10130-9}
\showDOI{\tempurl}


\bibitem[Wessel and Steinmacher(2020)]%
        {wessel_inconvenient_2020}
\bibfield{author}{\bibinfo{person}{Mairieli Wessel} {and} \bibinfo{person}{Igor
  Steinmacher}.} \bibinfo{year}{2020}\natexlab{}.
\newblock \showarticletitle{The inconvenient side of software bots on pull
  requests}. In \bibinfo{booktitle}{\emph{Proceedings of the {IEEE}/{ACM} 42nd
  {International} {Conference} on {Software} {Engineering} {Workshops} ({ICSEW}
  2020)}}. \bibinfo{pages}{51--55}.
\newblock
\urldef\tempurl%
\url{https://doi.org/10.1145/3387940.3391504}
\showDOI{\tempurl}


\bibitem[Wessel et~al\mbox{.}(2019)]%
        {wessel_should_2019}
\bibfield{author}{\bibinfo{person}{Mairieli Wessel}, \bibinfo{person}{Igor
  Steinmacher}, \bibinfo{person}{Igor Wiese}, {and} \bibinfo{person}{Marco~A.
  Gerosa}.} \bibinfo{year}{2019}\natexlab{}.
\newblock \showarticletitle{Should {I} stale or should {I} close? {An} analysis
  of a bot that closes abandoned issues and pull requests}. In
  \bibinfo{booktitle}{\emph{Proceedings of the {IEEE}/{ACM} 1st {International}
  {Workshop} on {Bots} in {Software} {Engineering} ({BotSE} 2019)}}.
  \bibinfo{pages}{38--42}.
\newblock
\urldef\tempurl%
\url{https://doi.org/10.1109/BotSE.2019.00018}
\showDOI{\tempurl}


\bibitem[Wessel et~al\mbox{.}(2022c)]%
        {wessel_github_2022}
\bibfield{author}{\bibinfo{person}{Mairieli Wessel}, \bibinfo{person}{Joseph
  Vargovich}, \bibinfo{person}{Marco~Aurélio Gerosa}, {and}
  \bibinfo{person}{Christoph Treude}.} \bibinfo{year}{2022}\natexlab{c}.
\newblock \bibinfo{title}{{GitHub} {Actions}: the impact on the pull request
  process}.
\newblock
\newblock
\urldef\tempurl%
\url{https://doi.org/10.48550/arXiv.2206.14118}
\showDOI{\tempurl}


\bibitem[Wessel et~al\mbox{.}(2021)]%
        {wessel_dont_2021}
\bibfield{author}{\bibinfo{person}{Mairieli Wessel}, \bibinfo{person}{Igor
  Wiese}, \bibinfo{person}{Igor Steinmacher}, {and}
  \bibinfo{person}{Marco~Aurélio Gerosa}.} \bibinfo{year}{2021}\natexlab{}.
\newblock \showarticletitle{Don't disturb me: challenges of interacting with
  software bots on open source software projects}.
\newblock \bibinfo{journal}{\emph{Proceedings of the ACM on Human-Computer
  Interaction}} \bibinfo{volume}{5}, \bibinfo{number}{CSCW2}
  (\bibinfo{year}{2021}), \bibinfo{pages}{1--21}.
\newblock
\urldef\tempurl%
\url{https://doi.org/10.1145/3476042}
\showDOI{\tempurl}


\bibitem[West et~al\mbox{.}(2014)]%
        {west_linear_2014}
\bibfield{author}{\bibinfo{person}{Brady~T. West}, \bibinfo{person}{Kathleen~B.
  Welch}, {and} \bibinfo{person}{Andrzej~T. Gałecki}.}
  \bibinfo{year}{2014}\natexlab{}.
\newblock \bibinfo{booktitle}{\emph{Linear {Mixed} {Models}: {A} {Practical}
  {Guide} {Using} {Statistical} {Software}} (\bibinfo{edition}{2nd} ed.)}.
\newblock \bibinfo{publisher}{Chapman and Hall/CRC}.
\newblock
\urldef\tempurl%
\url{https://doi.org/10.1201/b17198}
\showDOI{\tempurl}


\bibitem[Winding(2021)]%
        {winding_github_2021}
\bibfield{author}{\bibinfo{person}{Ben Winding}.}
  \bibinfo{year}{2021}\natexlab{}.
\newblock \bibinfo{title}{{GitHub} {Stale} bots: a false economy}.
\newblock
\newblock
\urldef\tempurl%
\url{https://blog.benwinding.com/github-stale-bots/index.html}
\showURL{%
\tempurl}


\bibitem[Yu et~al\mbox{.}(2016)]%
        {yu_determinants_2016}
\bibfield{author}{\bibinfo{person}{Yue Yu}, \bibinfo{person}{Gang Yin},
  \bibinfo{person}{Tao Wang}, \bibinfo{person}{Cheng Yang}, {and}
  \bibinfo{person}{Huaimin Wang}.} \bibinfo{year}{2016}\natexlab{}.
\newblock \showarticletitle{Determinants of pull-based development in the
  context of continuous integration}.
\newblock \bibinfo{journal}{\emph{Science China Information Sciences}}
  \bibinfo{volume}{59}, \bibinfo{number}{8} (\bibinfo{year}{2016}),
  \bibinfo{pages}{1--14}.
\newblock
\urldef\tempurl%
\url{https://doi.org/10.1007/s11432-016-5595-8}
\showDOI{\tempurl}


\bibitem[Zhang et~al\mbox{.}(2020)]%
        {zhang_shoulders_2020}
\bibfield{author}{\bibinfo{person}{Xunhui Zhang}, \bibinfo{person}{Ayushi
  Rastogi}, {and} \bibinfo{person}{Yue Yu}.} \bibinfo{year}{2020}\natexlab{}.
\newblock \showarticletitle{On the shoulders of giants: a new dataset for
  pull-based development research}. In \bibinfo{booktitle}{\emph{Proceedings of
  the 17th {International} {Conference} on {Mining} {Software} {Repositories}
  ({MSR} 2020)}}. \bibinfo{pages}{543--547}.
\newblock
\urldef\tempurl%
\url{https://doi.org/10.1145/3379597.3387489}
\showDOI{\tempurl}


\bibitem[Zhang et~al\mbox{.}(2023)]%
        {zhang_pull_2023}
\bibfield{author}{\bibinfo{person}{Xunhui Zhang}, \bibinfo{person}{Yue Yu},
  \bibinfo{person}{Georgios Gousios}, {and} \bibinfo{person}{Ayushi Rastogi}.}
  \bibinfo{year}{2023}\natexlab{}.
\newblock \showarticletitle{Pull request decisions explained: an empirical
  overview}.
\newblock \bibinfo{journal}{\emph{IEEE Transactions on Software Engineering}}
  \bibinfo{volume}{49}, \bibinfo{number}{2} (\bibinfo{year}{2023}),
  \bibinfo{pages}{849--871}.
\newblock
\urldef\tempurl%
\url{https://doi.org/10.1109/TSE.2022.3165056}
\showDOI{\tempurl}


\bibitem[Zhang et~al\mbox{.}(2022)]%
        {zhang_pull_2022}
\bibfield{author}{\bibinfo{person}{Xunhui Zhang}, \bibinfo{person}{Yue Yu},
  \bibinfo{person}{Tao Wang}, \bibinfo{person}{Ayushi Rastogi}, {and}
  \bibinfo{person}{Huaimin Wang}.} \bibinfo{year}{2022}\natexlab{}.
\newblock \showarticletitle{Pull request latency explained: an empirical
  overview}.
\newblock \bibinfo{journal}{\emph{Empirical Software Engineering}}
  \bibinfo{volume}{27}, \bibinfo{number}{6} (\bibinfo{year}{2022}),
  \bibinfo{pages}{1--38}.
\newblock
\urldef\tempurl%
\url{https://doi.org/10.1007/s10664-022-10143-4}
\showDOI{\tempurl}


\bibitem[Zhao et~al\mbox{.}(2017)]%
        {zhao_impact_2017}
\bibfield{author}{\bibinfo{person}{Yangyang Zhao}, \bibinfo{person}{Alexander
  Serebrenik}, \bibinfo{person}{Yuming Zhou}, \bibinfo{person}{Vladimir
  Filkov}, {and} \bibinfo{person}{Bogdan Vasilescu}.}
  \bibinfo{year}{2017}\natexlab{}.
\newblock \showarticletitle{The impact of continuous integration on other
  software development practices: a large-scale empirical study}. In
  \bibinfo{booktitle}{\emph{Proceedings of the 32nd {IEEE}/{ACM}
  {International} {Conference} on {Automated} {Software} {Engineering} ({ASE}
  2017)}}. \bibinfo{pages}{60--71}.
\newblock
\urldef\tempurl%
\url{https://doi.org/10.1109/ASE.2017.8115619}
\showDOI{\tempurl}


\bibitem[Zhou and Mockus(2012)]%
        {zhou_what_2012}
\bibfield{author}{\bibinfo{person}{Minghui Zhou} {and} \bibinfo{person}{Audris
  Mockus}.} \bibinfo{year}{2012}\natexlab{}.
\newblock \showarticletitle{What make long term contributors: willingness and
  opportunity in {OSS} community}. In \bibinfo{booktitle}{\emph{Proceedings of
  the 34th {International} {Conference} on {Software} {Engineering} ({ICSE}
  2012)}}. \bibinfo{pages}{518--528}.
\newblock
\urldef\tempurl%
\url{https://doi.org/10.1109/ICSE.2012.6227164}
\showDOI{\tempurl}


\bibitem[Zhu et~al\mbox{.}(2016)]%
        {zhu_effectiveness_2016}
\bibfield{author}{\bibinfo{person}{Jiaxin Zhu}, \bibinfo{person}{Minghui Zhou},
  {and} \bibinfo{person}{Audris Mockus}.} \bibinfo{year}{2016}\natexlab{}.
\newblock \showarticletitle{Effectiveness of code contribution: from
  patch-based to pull-request-based tools}. In
  \bibinfo{booktitle}{\emph{Proceedings of the 24th {ACM} {SIGSOFT}
  {International} {Symposium} on {Foundations} of {Software} {Engineering}
  ({FSE} 2016)}}. \bibinfo{pages}{871--882}.
\newblock
\urldef\tempurl%
\url{https://doi.org/10.1145/2950290.2950364}
\showDOI{\tempurl}


\bibitem[Ziborov(2020)]%
        {ziborov_commit_2020}
\bibfield{author}{\bibinfo{person}{Alexander Ziborov}.}
  \bibinfo{year}{2020}\natexlab{}.
\newblock \bibinfo{title}{Commit \#439b8735fa93709ec32602bb32944bf9214ce785 -
  {DevExpress}/{DevExtreme}}.
\newblock
\newblock
\urldef\tempurl%
\url{https://github.com/DevExpress/DevExtreme/commit/439b8735fa93709ec32602bb32944bf9214ce785}
\showURL{%
\tempurl}


\bibitem[Zou et~al\mbox{.}(2019)]%
        {zou_how_2019}
\bibfield{author}{\bibinfo{person}{Weiqin Zou}, \bibinfo{person}{Jifeng Xuan},
  \bibinfo{person}{Xiaoyuan Xie}, \bibinfo{person}{Zhenyu Chen}, {and}
  \bibinfo{person}{Baowen Xu}.} \bibinfo{year}{2019}\natexlab{}.
\newblock \showarticletitle{How does code style inconsistency affect pull
  request integration? {An} exploratory study on 117 {GitHub} projects}.
\newblock \bibinfo{journal}{\emph{Empirical Software Engineering}}
  \bibinfo{volume}{24}, \bibinfo{number}{6} (\bibinfo{year}{2019}),
  \bibinfo{pages}{3871--3903}.
\newblock
\urldef\tempurl%
\url{https://doi.org/10.1007/s10664-019-09720-x}
\showDOI{\tempurl}


\end{thebibliography}

\clearpage
\section*{Appendix}
\appendix
\section{Models for Estimating the Impact of Stale Bot}
\label{appendix:models}

\begin{table}[H]
    \caption{Models for estimating the impact of adopting Stale bot on the number of merged PRs (\textit{merged\_pulls}) and closed PRs (\textit{closed\_pulls}) in the studied projects.}
    \label{tab:resolved_pulls}
    \resizebox{0.63\textwidth}{!}{%
        \begin{threeparttable}
            \begin{tabular}{@{}lllll@{}}
                \toprule
                                                & \multicolumn{2}{c}{\textbf{log(merged\_pulls)}} & \multicolumn{2}{c}{\textbf{log(closed\_pulls)}} \\
                \cmidrule(l){2-3}
                \cmidrule(l){4-5}
                                                & \textbf{Coefficient} & \textbf{Sum Sq.}         & \textbf{Coefficient} & \textbf{Sum Sq.}         \\
                \midrule
                \textbf{time}                   & 0.041***             & 4.852***                 & 0.038***             & 4.159***                 \\
                \textbf{adoption}               & -0.129*              & 0.496*                   & 0.187*               & 1.040*                   \\
                \textbf{time\_since\_adoption}  & -0.021*              & 0.653*                   & -0.044**             & 2.721**                  \\
                log(age\_at\_adoption)          & -0.627**             & 1.397**                  & -0.678**             & 2.483**                  \\
                log(pulls\_at\_adoption)        & 0.883***             & 1.792***                 & 0.279                & 0.271                    \\
                log(contributors\_at\_adoption) & 0.152                & 0.334                    & 0.379**              & 3.158**                  \\
                log(maintainers\_at\_adoption)  & -0.210               & 0.124                    & 0.226                & 0.216                    \\
                intercept                       & -0.688               &                          & 0.299                &                          \\
                \midrule
                Marginal $R^2$                  & \multicolumn{2}{l}{0.57}                        & \multicolumn{2}{l}{0.44}                        \\
                Conditional $R^2$               & \multicolumn{2}{l}{0.82}                        & \multicolumn{2}{l}{0.70}                        \\
                \bottomrule
            \end{tabular}
            \begin{tablenotes}
                \item *** $p < 0.001$, ** $p < 0.01$, * $p < 0.05$.
            \end{tablenotes}
        \end{threeparttable}
    }
\end{table}

\begin{table}[H]
    \caption{Models for estimating the impact of adopting Stale bot on the first response latency of merged PRs (\textit{first\_latency\_m}) and closed PRs (\textit{first\_latency\_c}) in the studied projects.}
    \label{tab:first_latency}
    \resizebox{0.63\textwidth}{!}{%
        \begin{threeparttable}
            \begin{tabular}{@{}lllll@{}}
                \toprule
                                                & \multicolumn{2}{c}{\textbf{log(first\_latency\_m)}} & \multicolumn{2}{c}{\textbf{log(first\_latency\_c)}} \\
                \cmidrule(l){2-3}
                \cmidrule(l){4-5}
                                                & \textbf{Coefficient} & \textbf{Sum Sq.}             & \textbf{Coefficient} & \textbf{Sum Sq.}             \\
                \midrule
                \textbf{time}                   & 0.029*               & 2.398*                       & 0.085***             & 20.577***                    \\
                \textbf{adoption}               & 0.064                & 0.123                        & 0.071                & 0.151                        \\
                \textbf{time\_since\_adoption}  & -0.045*              & 2.920*                       & -0.051               & 3.708                        \\
                log(age\_at\_adoption)          & 0.983                & 0.786                        & 1.162                & 2.381                        \\
                log(pulls\_at\_adoption)        & 0.662                & 0.230                        & 1.342                & 2.050                        \\
                log(contributors\_at\_adoption) & -0.228               & 0.173                        & -0.748               & 4.027                        \\
                log(maintainers\_at\_adoption)  & -0.950               & 0.577                        & -1.257               & 2.186                        \\
                intercept                       & -2.599               &                              & -4.939               &                              \\
                \midrule
                Marginal $R^2$                  & \multicolumn{2}{l}{0.08}                            & \multicolumn{2}{l}{0.14}                            \\
                Conditional $R^2$               & \multicolumn{2}{l}{0.87}                            & \multicolumn{2}{l}{0.77}                            \\
                \bottomrule
            \end{tabular}
            \begin{tablenotes}
                \item *** $p < 0.001$, ** $p < 0.01$, * $p < 0.05$.
            \end{tablenotes}
        \end{threeparttable}
    }
\end{table}

\begin{table}[H]
    \caption{Models for estimating the impact of adopting Stale bot on the mean response latency of merged PRs (\textit{mean\_latency\_m}) and closed PRs (\textit{mean\_latency\_c}) in the studied projects.}
    \label{tab:mean_latency}
    \resizebox{0.63\textwidth}{!}{%
        \begin{threeparttable}
            \begin{tabular}{@{}lllll@{}}
                \toprule
                                                & \multicolumn{2}{c}{\textbf{log(mean\_latency\_m)}} & \multicolumn{2}{c}{\textbf{log(mean\_latency\_c)}} \\
                \cmidrule(l){2-3}
                \cmidrule(l){4-5}
                                                & \textbf{Coefficient} & \textbf{Sum Sq.}            & \textbf{Coefficient} & \textbf{Sum Sq.}            \\
                \midrule
                \textbf{time}                   & 0.011                & 0.318                       & 0.031                & 2.728                       \\
                \textbf{adoption}               & -0.023               & 0.015                       & 0.128                & 0.486                       \\
                \textbf{time\_since\_adoption}  & -0.006               & 0.052                       & -0.025               & 0.901                       \\
                log(age\_at\_adoption)          & 0.677*               & 1.240*                      & 0.770                & 2.698                       \\
                log(pulls\_at\_adoption)        & 0.133                & 0.031                       & 0.467                & 0.640                       \\
                log(contributors\_at\_adoption) & 0.093                & 0.095                       & -0.131               & 0.319                       \\
                log(maintainers\_at\_adoption)  & -0.274               & 0.160                       & -0.504               & 0.907                       \\
                intercept                       & -0.158               &                             & -0.142               &                             \\
                \midrule
                Marginal $R^2$                  & \multicolumn{2}{l}{0.22}                           & \multicolumn{2}{l}{0.15}                           \\
                Conditional $R^2$               & \multicolumn{2}{l}{0.72}                           & \multicolumn{2}{l}{0.58}                           \\
                \bottomrule
            \end{tabular}
            \begin{tablenotes}
                \item *** $p < 0.001$, ** $p < 0.01$, * $p < 0.05$.
            \end{tablenotes}
        \end{threeparttable}
    }
\end{table}

\begin{table}[H]
    \caption{Models for estimating the impact of adopting Stale bot on the resolution time of merged PRs (\textit{resolution\_time\_m}) and closed PRs (\textit{resolution\_time\_c}) in the studied projects.}
    \label{tab:resolution_time}
    \resizebox{0.63\textwidth}{!}{%
        \begin{threeparttable}
            \begin{tabular}{@{}lllll@{}}
                \toprule
                                                & \multicolumn{2}{c}{\textbf{log(resolution\_time\_m)}} & \multicolumn{2}{c}{\textbf{log(resolution\_time\_c)}} \\
                \cmidrule(l){2-3}
                \cmidrule(l){4-5}
                                                & \textbf{Coefficient} & \textbf{Sum Sq.}               & \textbf{Coefficient} & \textbf{Sum Sq.}               \\
                \midrule
                \textbf{time}                   & -0.001               & 0.004                          & 0.029*               & 2.366*                         \\
                \textbf{adoption}               & 0.052                & 0.080                          & 0.267                & 2.124                          \\
                \textbf{time\_since\_adoption}  & -0.001               & 0.001                          & -0.043*              & 2.602*                         \\
                log(age\_at\_adoption)          & 0.509                & 0.460                          & 0.547                & 1.007                          \\
                log(pulls\_at\_adoption)        & -0.345               & 0.136                          & -0.029               & 0.002                          \\
                log(contributors\_at\_adoption) & 0.228                & 0.376                          & 0.156                & 0.336                          \\
                log(maintainers\_at\_adoption)  & 0.333                & 0.154                          & 0.102                & 0.027                          \\
                intercept                       & 3.482                &                                & 3.021                &                                \\
                \midrule
                Marginal $R^2$                  & \multicolumn{2}{l}{0.25}                              & \multicolumn{2}{l}{0.18}                              \\
                Conditional $R^2$               & \multicolumn{2}{l}{0.80}                              & \multicolumn{2}{l}{0.66}                              \\
                \bottomrule
            \end{tabular}
            \begin{tablenotes}
                \item *** $p < 0.001$, ** $p < 0.01$, * $p < 0.05$.
            \end{tablenotes}
        \end{threeparttable}
    }
\end{table}

\begin{table}[H]
    \caption{Models for estimating the impact of adopting Stale bot on the number of comments in merged PRs (\textit{comments\_m}) and closed PRs (\textit{comments\_c}) in the studied projects.}
    \label{tab:comments}
    \resizebox{0.63\textwidth}{!}{%
        \begin{threeparttable}
            \begin{tabular}{@{}lllll@{}}
                \toprule
                                                & \multicolumn{2}{c}{\textbf{log(comments\_m)}} & \multicolumn{2}{c}{\textbf{log(comments\_c)}} \\
                \cmidrule(l){2-3}
                \cmidrule(l){4-5}
                                                & \textbf{Coefficient} & \textbf{Sum Sq.}       & \textbf{Coefficient} & \textbf{Sum Sq.}       \\
                \midrule
                \textbf{time}                   & -0.009               & 0.216                  & -0.008               & 0.195                  \\
                \textbf{adoption}               & 0.023                & 0.016                  & -0.001               & 0.000                  \\
                \textbf{time\_since\_adoption}  & 0.001                & 0.002                  & -0.003               & 0.015                  \\
                log(age\_at\_adoption)          & -0.165               & 0.020                  & -0.075               & 0.009                  \\
                log(pulls\_at\_adoption)        & -0.603               & 0.171                  & -0.438               & 0.209                  \\
                log(contributors\_at\_adoption) & 0.221                & 0.145                  & 0.300*               & 0.617*                 \\
                log(maintainers\_at\_adoption)  & 0.826*               & 0.390*                 & 0.375                & 0.186                  \\
                intercept                       & 3.252                &                        & 2.589                &                        \\
                \midrule
                Marginal $R^2$                  & \multicolumn{2}{l}{0.26}                      & \multicolumn{2}{l}{0.19}                      \\
                Conditional $R^2$               & \multicolumn{2}{l}{0.90}                      & \multicolumn{2}{l}{0.79}                      \\
                \bottomrule
            \end{tabular}
            \begin{tablenotes}
                \item *** $p < 0.001$, ** $p < 0.01$, * $p < 0.05$.
            \end{tablenotes}
        \end{threeparttable}
    }
\end{table}

\begin{table}[H]
    \caption{Models for estimating the impact of adopting Stale bot on the number of commits in merged PRs (\textit{commits\_m}) and closed PRs (\textit{commits\_c}) in the studied projects.}
    \label{tab:commits}
    \resizebox{0.63\textwidth}{!}{%
        \begin{threeparttable}
            \begin{tabular}{@{}lllll@{}}
                \toprule
                                                & \multicolumn{2}{c}{\textbf{log(commits\_m)}} & \multicolumn{2}{c}{\textbf{log(commits\_c)}} \\
                \cmidrule(l){2-3}
                \cmidrule(l){4-5}
                                                & \textbf{Coefficient} & \textbf{Sum Sq.}      & \textbf{Coefficient} & \textbf{Sum Sq.}      \\
                \midrule
                \textbf{time}                   & 0.010                & 0.271                 & 0.030*               & 2.501*                \\
                \textbf{adoption}               & -0.021               & 0.013                 & -0.215               & 1.369                 \\
                \textbf{time\_since\_adoption}  & -0.015*              & 0.335*                & -0.026               & 0.942                 \\
                log(age\_at\_adoption)          & 0.189                & 0.094                 & 0.545**              & 7.487**               \\
                log(pulls\_at\_adoption)        & -0.252               & 0.108                 & -0.120               & 0.234                 \\
                log(contributors\_at\_adoption) & -0.019               & 0.004                 & 0.085                & 0.741                 \\
                log(maintainers\_at\_adoption)  & 0.064                & 0.009                 & -0.188               & 0.700                 \\
                intercept                       & 2.580                &                       & 0.783                &                       \\
                \midrule
                Marginal $R^2$                  & \multicolumn{2}{l}{0.09}                     & \multicolumn{2}{l}{0.16}                     \\
                Conditional $R^2$               & \multicolumn{2}{l}{0.67}                     & \multicolumn{2}{l}{0.27}                     \\
                \bottomrule
            \end{tabular}
            \begin{tablenotes}
                \item *** $p < 0.001$, ** $p < 0.01$, * $p < 0.05$.
            \end{tablenotes}
        \end{threeparttable}
    }
\end{table}

\begin{table}[H]
    \caption{Models for estimating the impact of adopting Stale bot on the number of active contributors (\textit{contributors}) in the studied projects.}
    \label{tab:contributors}
    \resizebox{0.44\textwidth}{!}{%
        \begin{threeparttable}
            \begin{tabular}{@{}lll@{}}
                \toprule
                                                & \multicolumn{2}{c}{\textbf{log(contributors)}} \\
                \cmidrule(l){2-3}
                                                & \textbf{Coefficient} & \textbf{Sum Sq.}        \\
                \midrule
                \textbf{time}                   & 0.026***             & 1.867***                \\
                \textbf{adoption}               & 0.003                & 0.000                   \\
                \textbf{time\_since\_adoption}  & -0.022***            & 0.708***                \\
                log(age\_at\_adoption)          & -0.535**             & 0.493**                 \\
                log(pulls\_at\_adoption)        & 0.250                & 0.070                   \\
                log(contributors\_at\_adoption) & 0.697***             & 3.411***                \\
                log(maintainers\_at\_adoption)  & 0.114                & 0.018                   \\
                intercept                       & -1.033               &                         \\
                \midrule
                Marginal $R^2$                  & \multicolumn{2}{l}{0.84}                       \\
                Conditional $R^2$               & \multicolumn{2}{l}{0.96}                       \\
                \bottomrule
            \end{tabular}
            \begin{tablenotes}
                \item *** $p < 0.001$, ** $p < 0.01$, * $p < 0.05$.
            \end{tablenotes}
        \end{threeparttable}
    }
\end{table}

\clearpage
\section{Variation of the Performance Indicators}
\label{appendix:variation}

\begin{figure}[H]
    \begin{subfigure}{0.325\textwidth}
        \includegraphics[width=\textwidth]{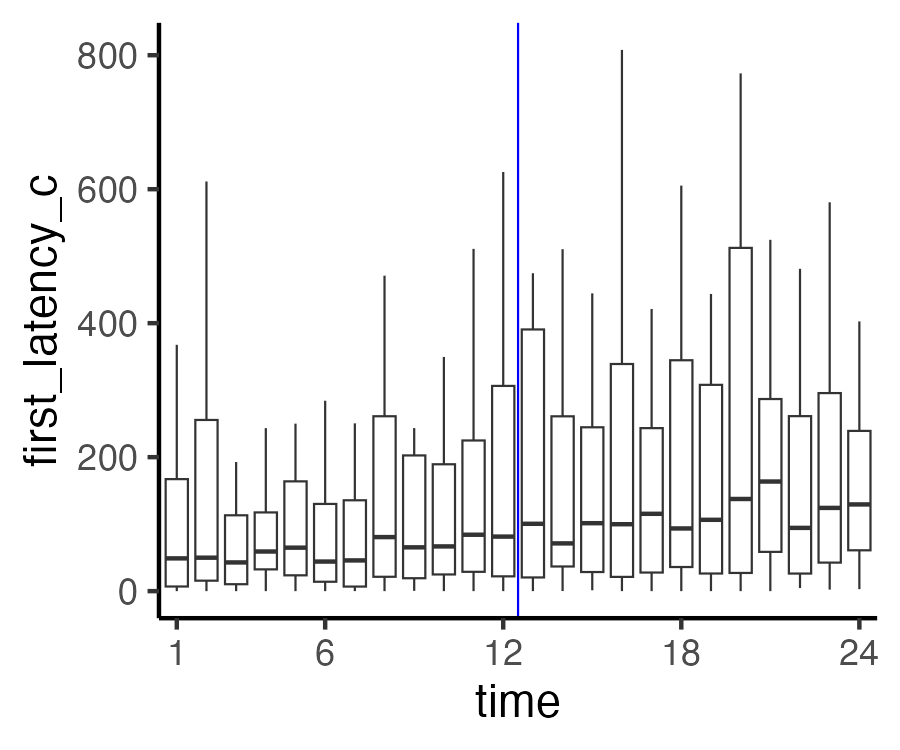}
        \caption{}
    \end{subfigure}
    \begin{subfigure}{0.325\textwidth}
        \includegraphics[width=\textwidth]{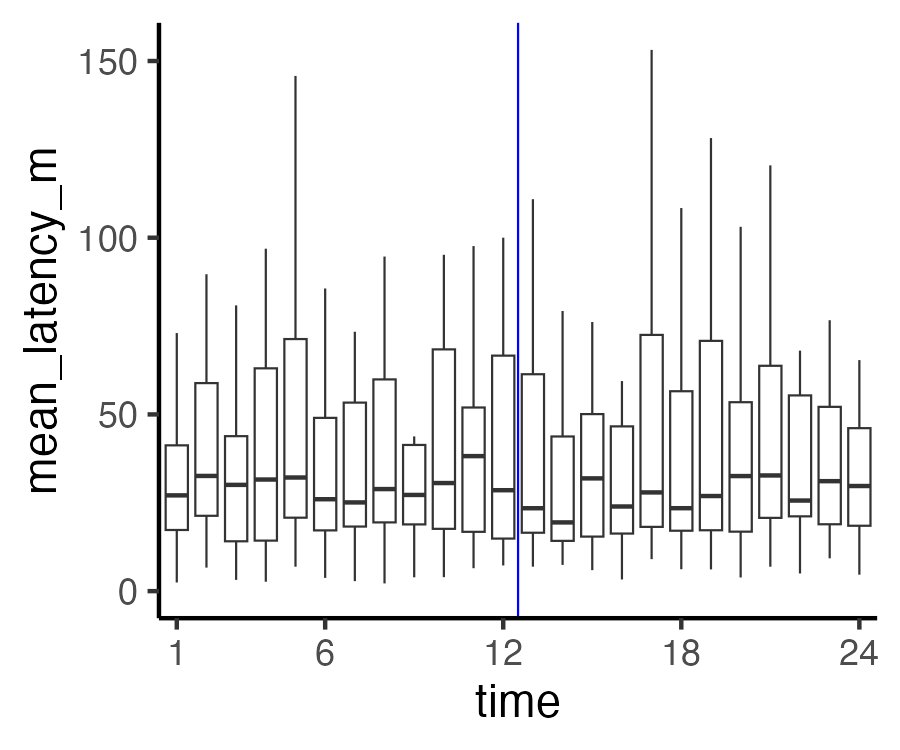}
        \caption{}
    \end{subfigure}
    \begin{subfigure}{0.325\textwidth}
        \includegraphics[width=\textwidth]{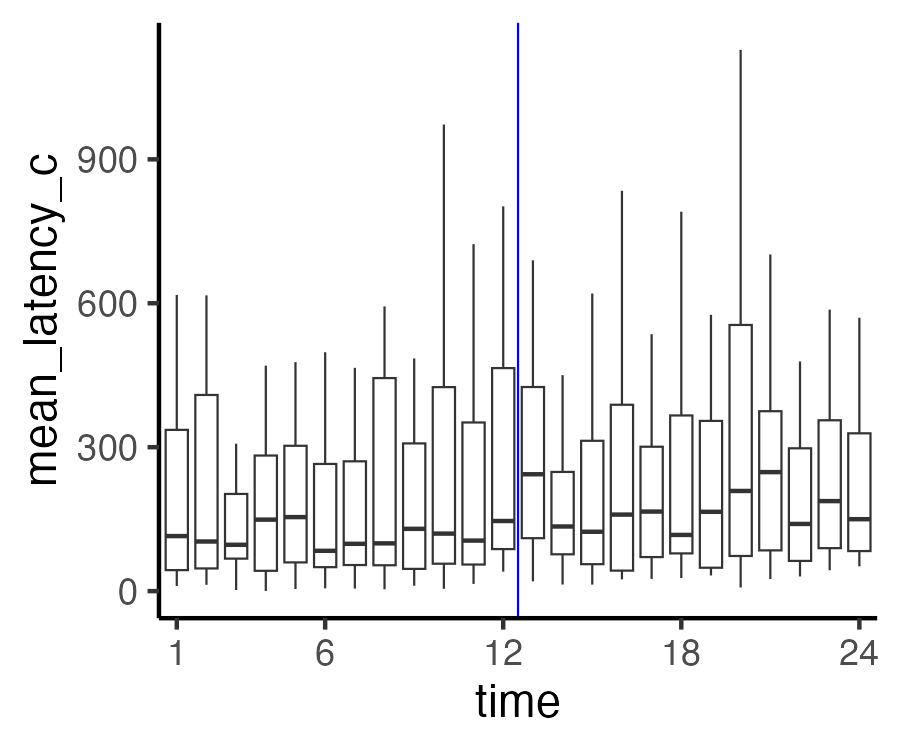}
        \caption{}
    \end{subfigure}
    \begin{subfigure}{0.325\textwidth}
        \includegraphics[width=\textwidth]{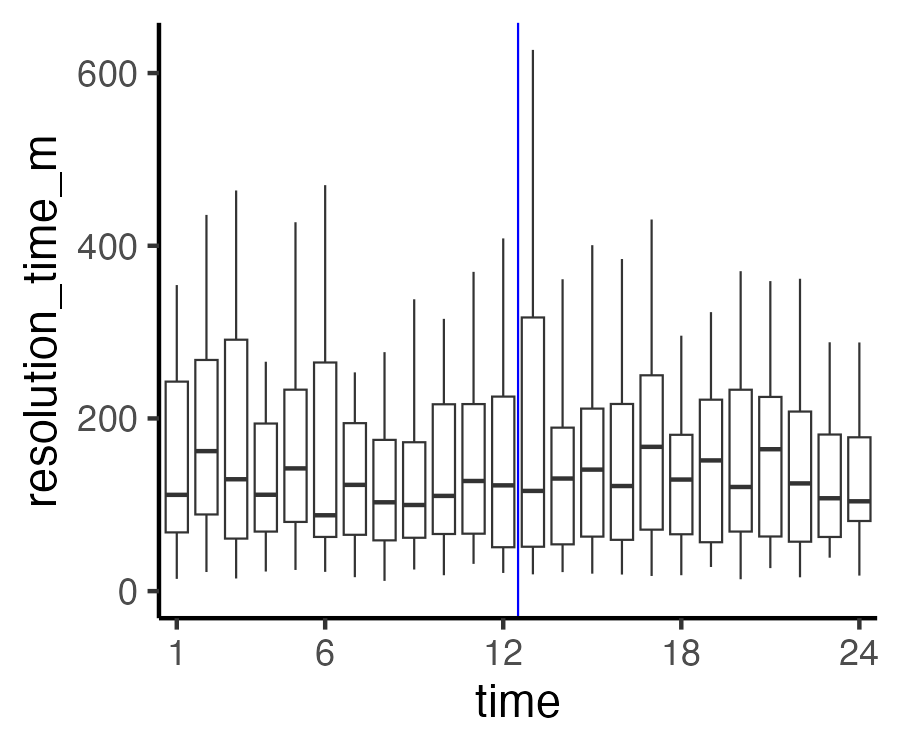}
        \caption{}
    \end{subfigure}
    \begin{subfigure}{0.325\textwidth}
        \includegraphics[width=\textwidth]{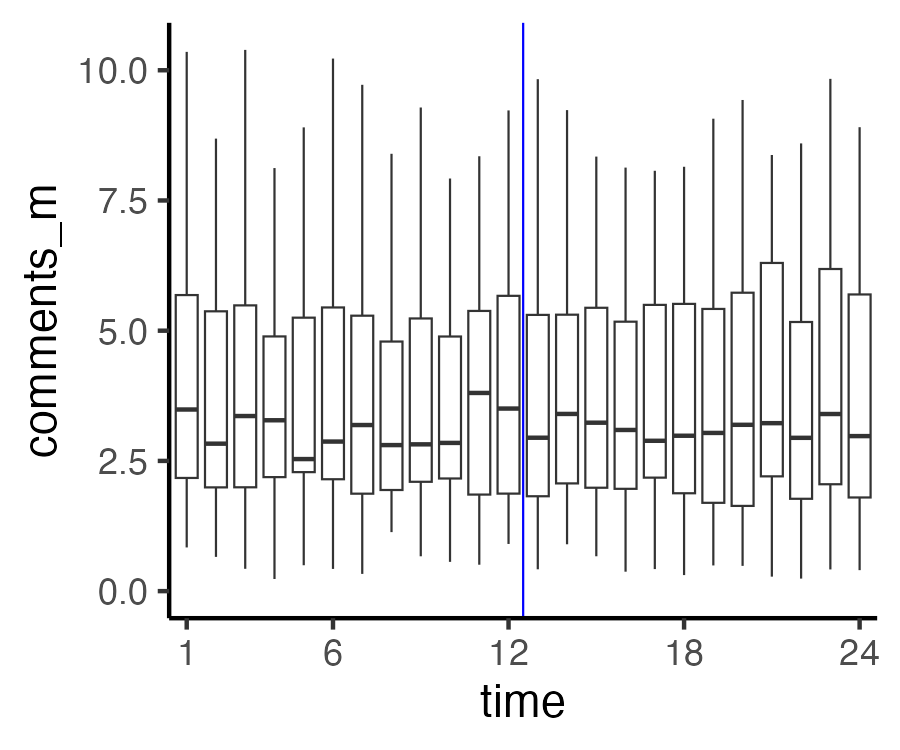}
        \caption{}
    \end{subfigure}
    \begin{subfigure}{0.325\textwidth}
        \includegraphics[width=\textwidth]{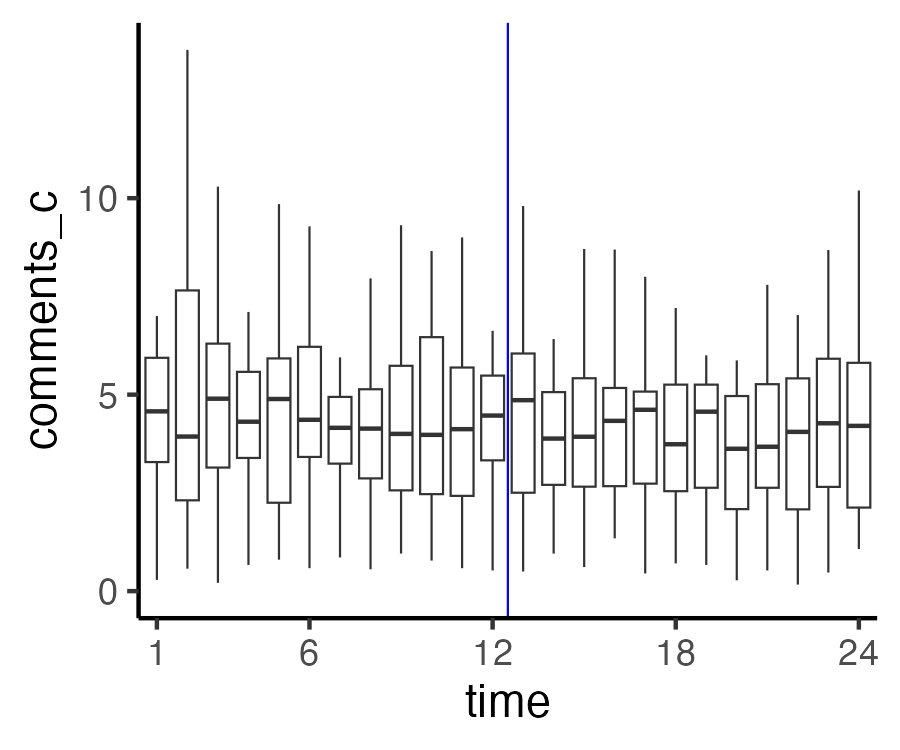}
        \caption{}
    \end{subfigure}
    \begin{subfigure}{0.325\textwidth}
        \includegraphics[width=\textwidth]{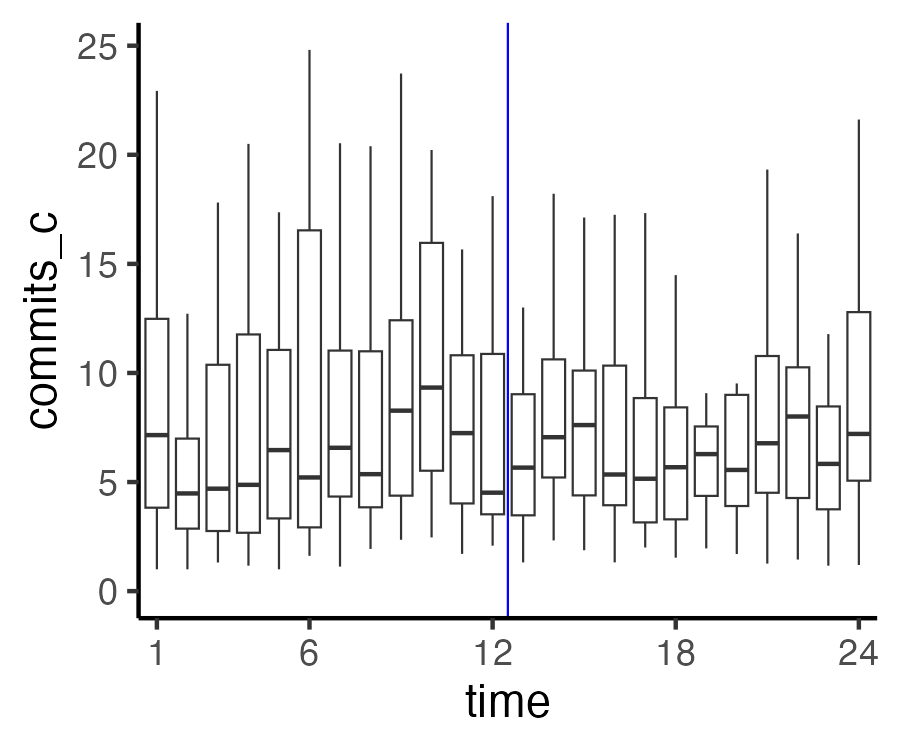}
        \caption{}
    \end{subfigure}
    \caption{Variation in (a) the first response latency of closed PRs, (b) the mean response latency of merged PRs, (c) the mean response latency of closed PRs, (d) the resolution time of merged PRs, (e) the number of comments in merged PRs, (f) the number of comments in closed PRs, and (g) the number of commits in closed PRs each month during our observation period. The \textcolor{blue}{blue lines} show the adoption time.}
\end{figure}

\clearpage
\section{Characteristics of PRs Intervened by Stale Bot}
\label{appendix:stats}

\begin{figure}[H]
    \includegraphics[width=\textwidth]{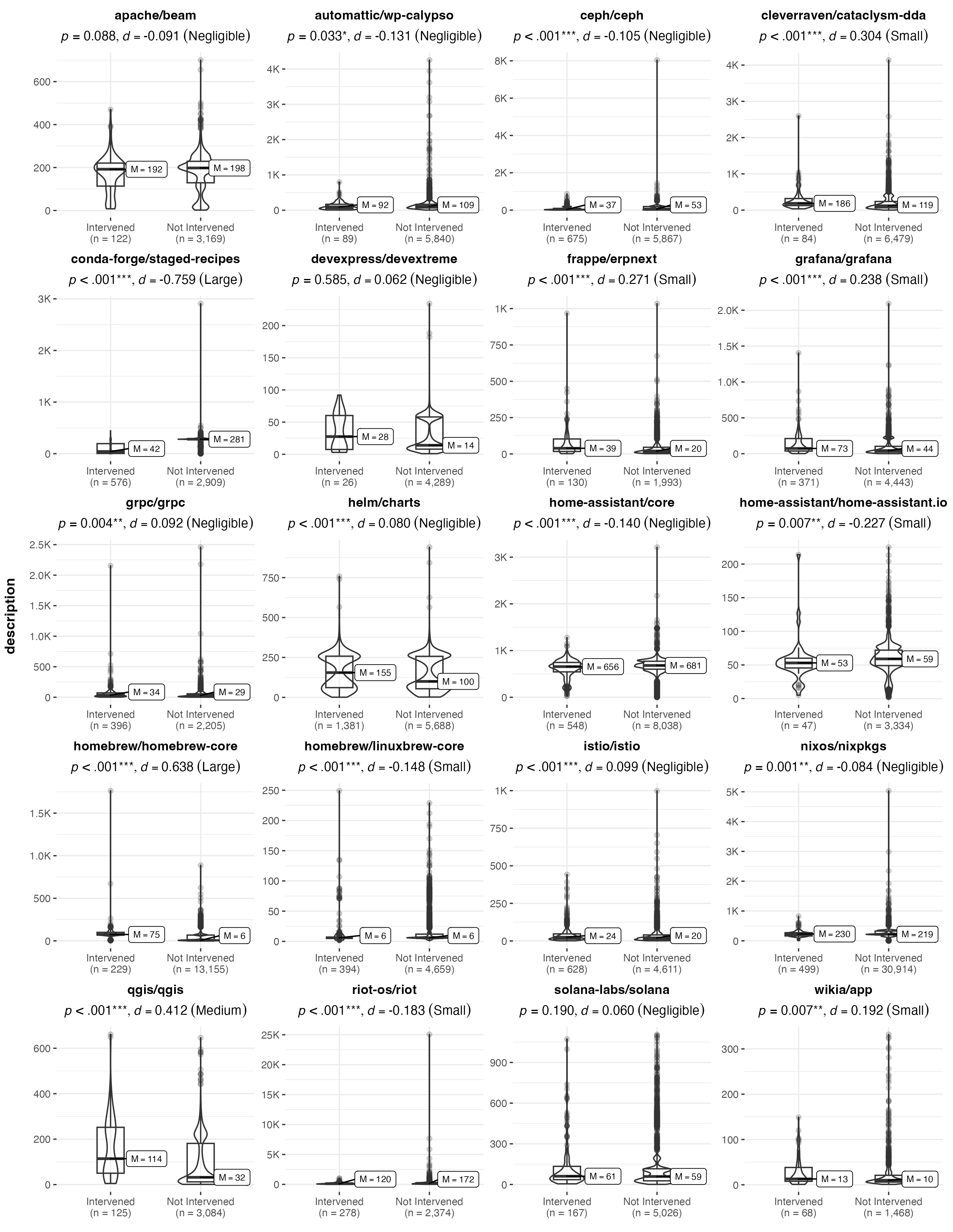}
    \caption{Comparison of intervened and not intervened PRs regarding their description length across the studied projects. *** $p < 0.001$, ** $p < 0.01$, * $p < 0.05$}
\end{figure}

\begin{figure}[H]
    \includegraphics[width=\textwidth]{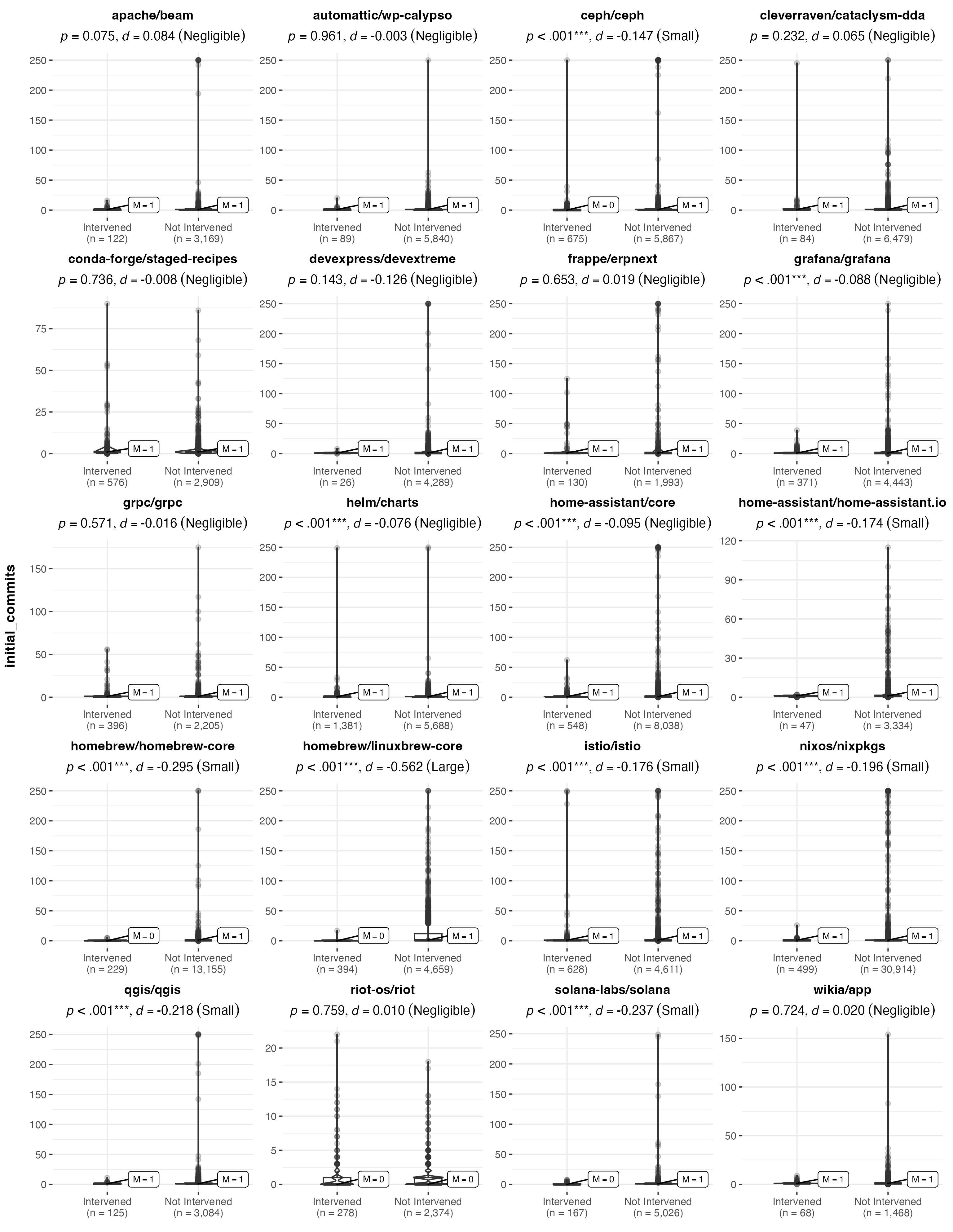}
    \caption{Comparison of intervened and not intervened PRs regarding their number of initial commits across the studied projects. *** $p < 0.001$, ** $p < 0.01$, * $p < 0.05$}
\end{figure}

\begin{figure}[H]
    \includegraphics[width=\textwidth]{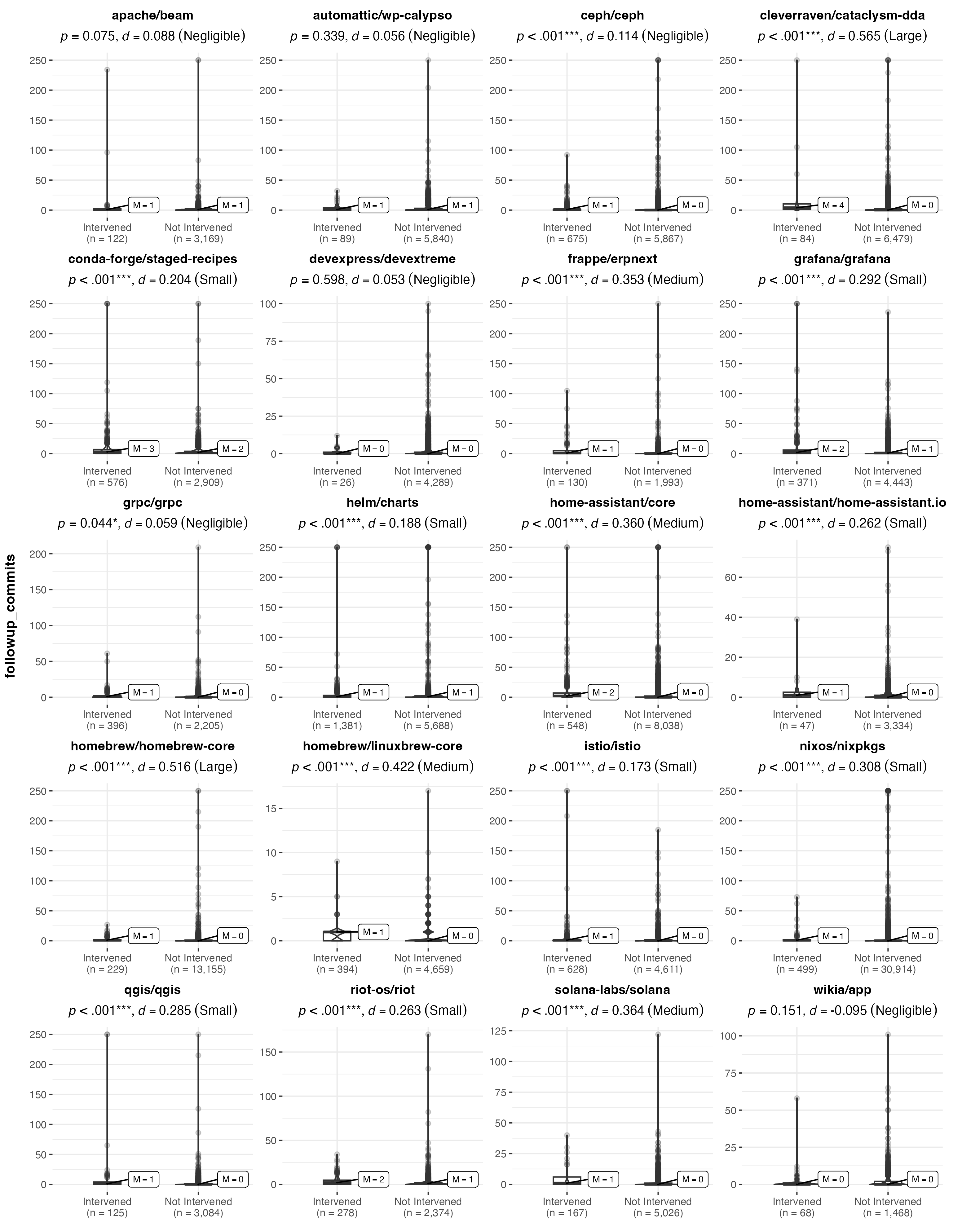}
    \caption{Comparison of intervened and not intervened PRs regarding their number of follow-up commits across the studied projects. *** $p < 0.001$, ** $p < 0.01$, * $p < 0.05$}
\end{figure}

\begin{figure}[H]
    \includegraphics[width=\textwidth]{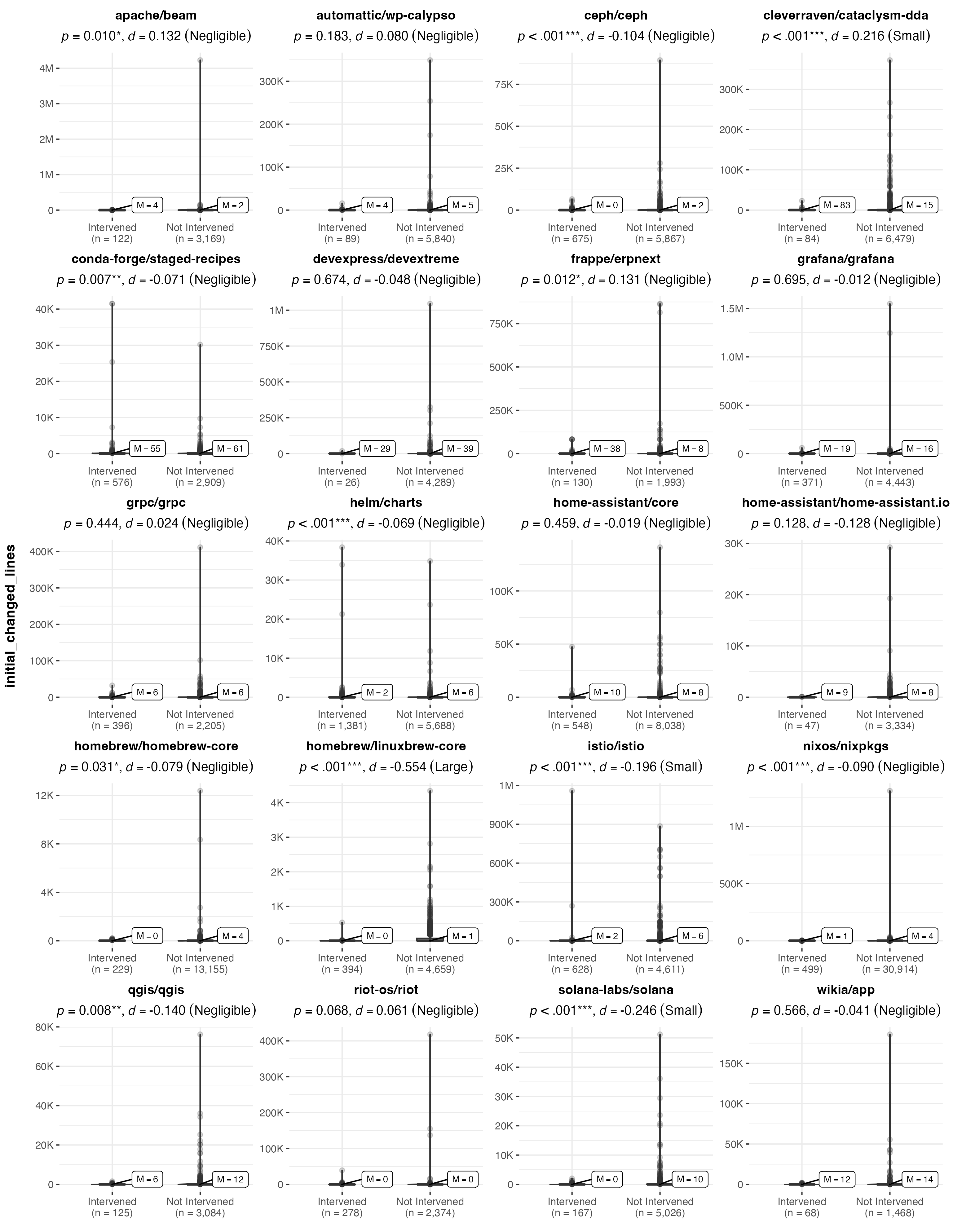}
    \caption{Comparison of intervened and not intervened PRs regarding their number of initial changed lines across the studied projects. *** $p < 0.001$, ** $p < 0.01$, * $p < 0.05$}
\end{figure}

\begin{figure}[H]
    \includegraphics[width=\textwidth]{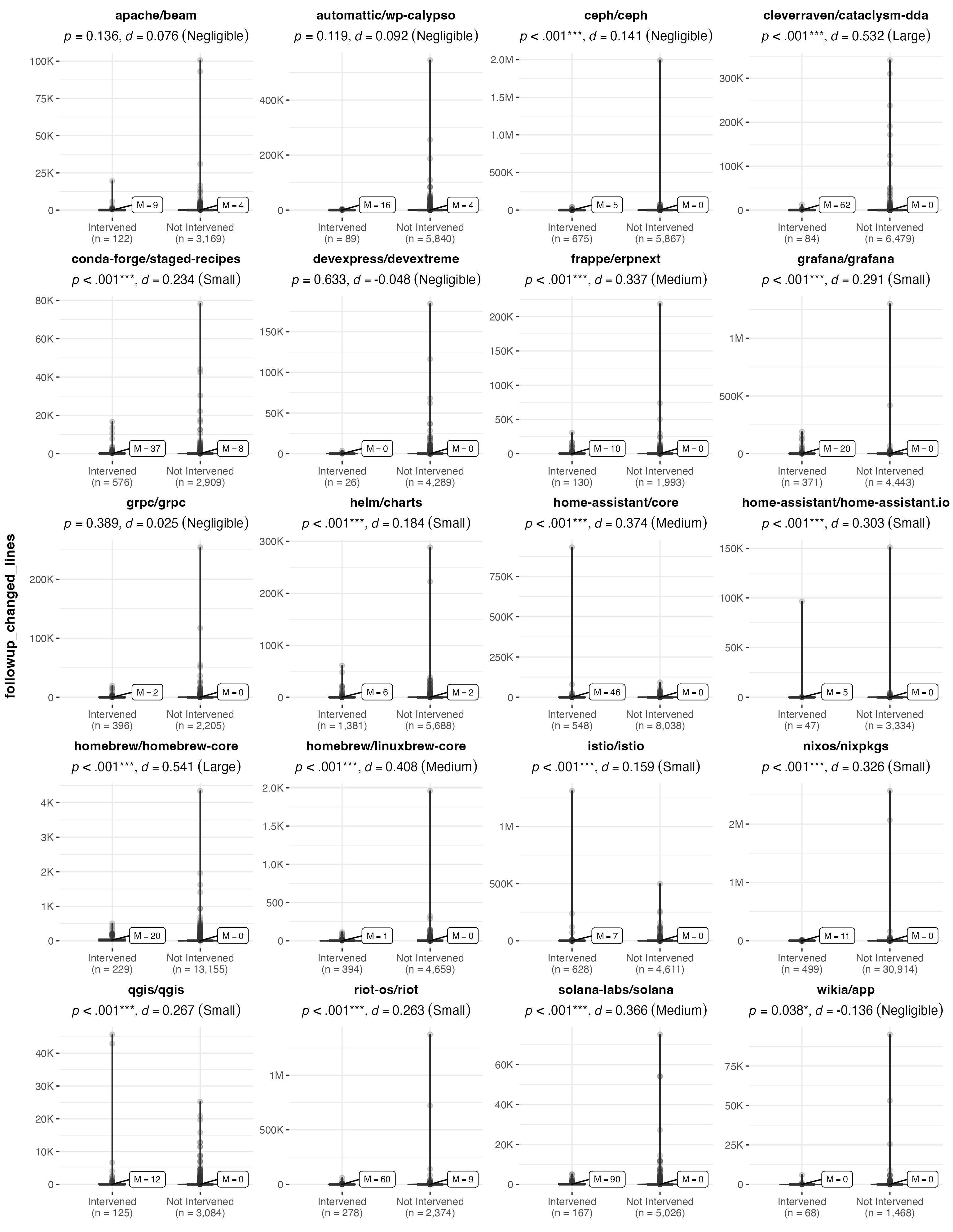}
    \caption{Comparison of intervened and not intervened PRs regarding their number of follow-up changed lines across the studied projects. *** $p < 0.001$, ** $p < 0.01$, * $p < 0.05$}
\end{figure}

\begin{figure}[H]
    \includegraphics[width=\textwidth]{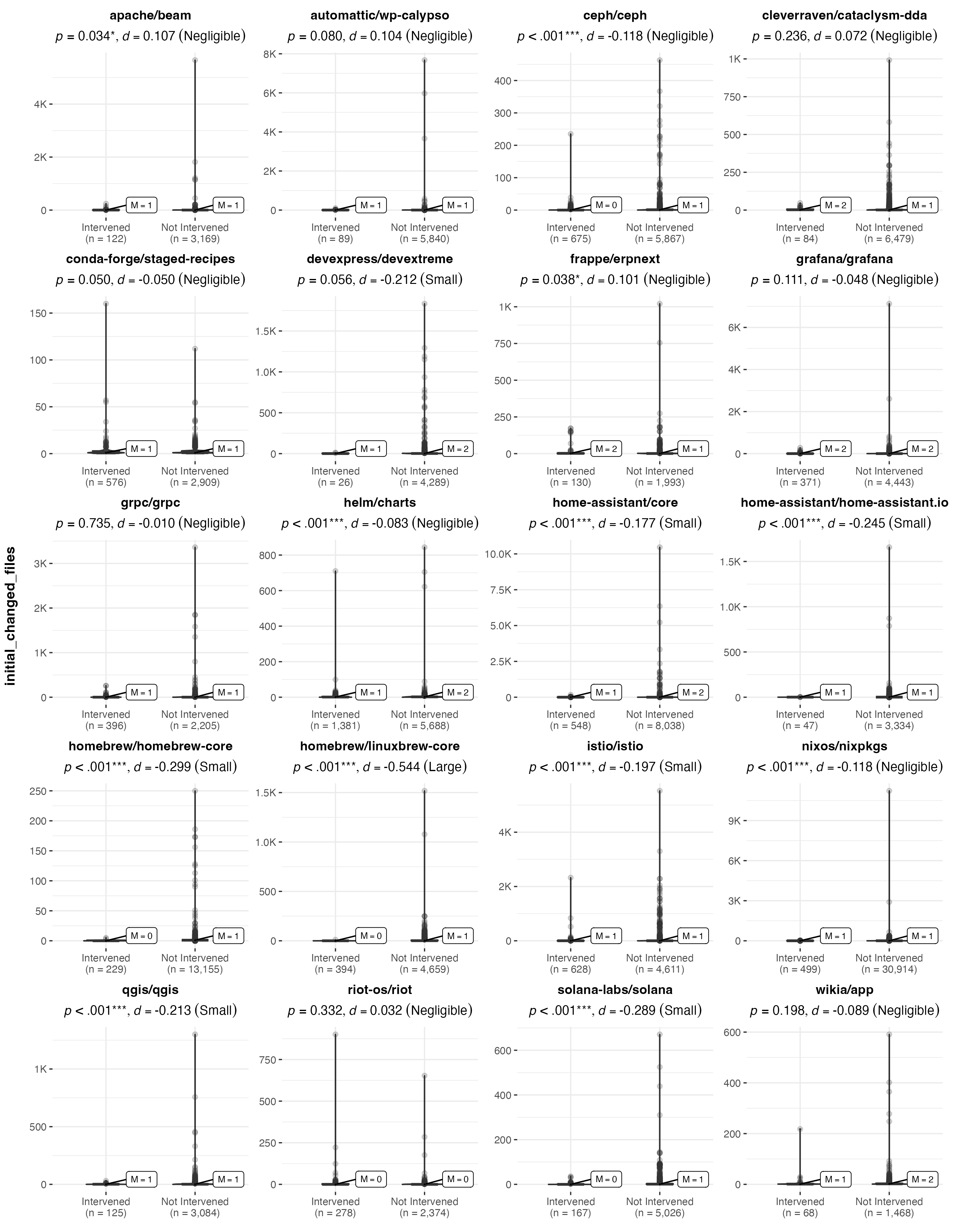}
    \caption{Comparison of intervened and not intervened PRs regarding their number of initial changed files across the studied projects. *** $p < 0.001$, ** $p < 0.01$, * $p < 0.05$}
\end{figure}

\begin{figure}[H]
    \includegraphics[width=\textwidth]{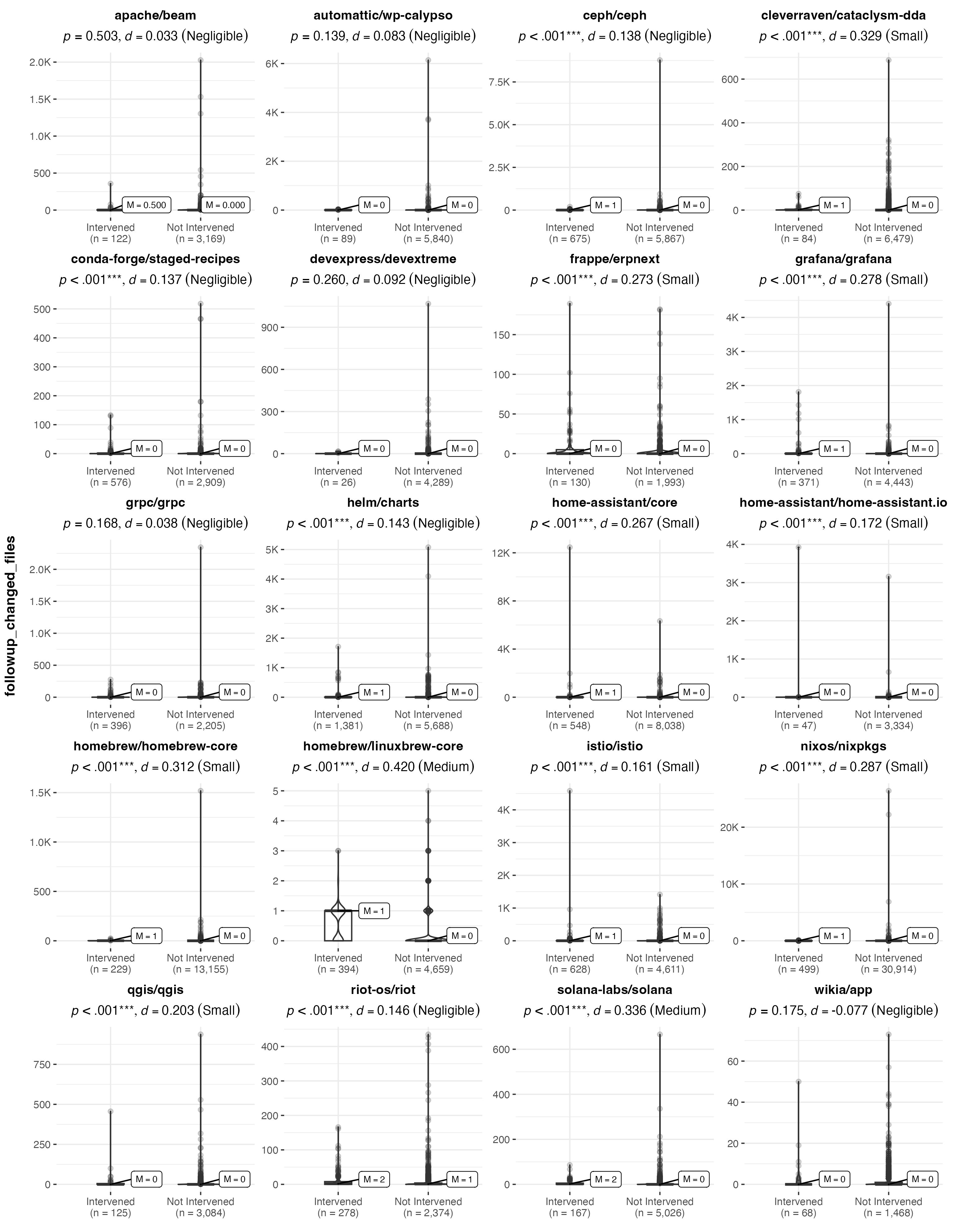}
    \caption{Comparison of intervened and not intervened PRs regarding their number of follow-up changed files across the studied projects. *** $p < 0.001$, ** $p < 0.01$, * $p < 0.05$}
\end{figure}

\begin{figure}[H]
    \includegraphics[width=\textwidth]{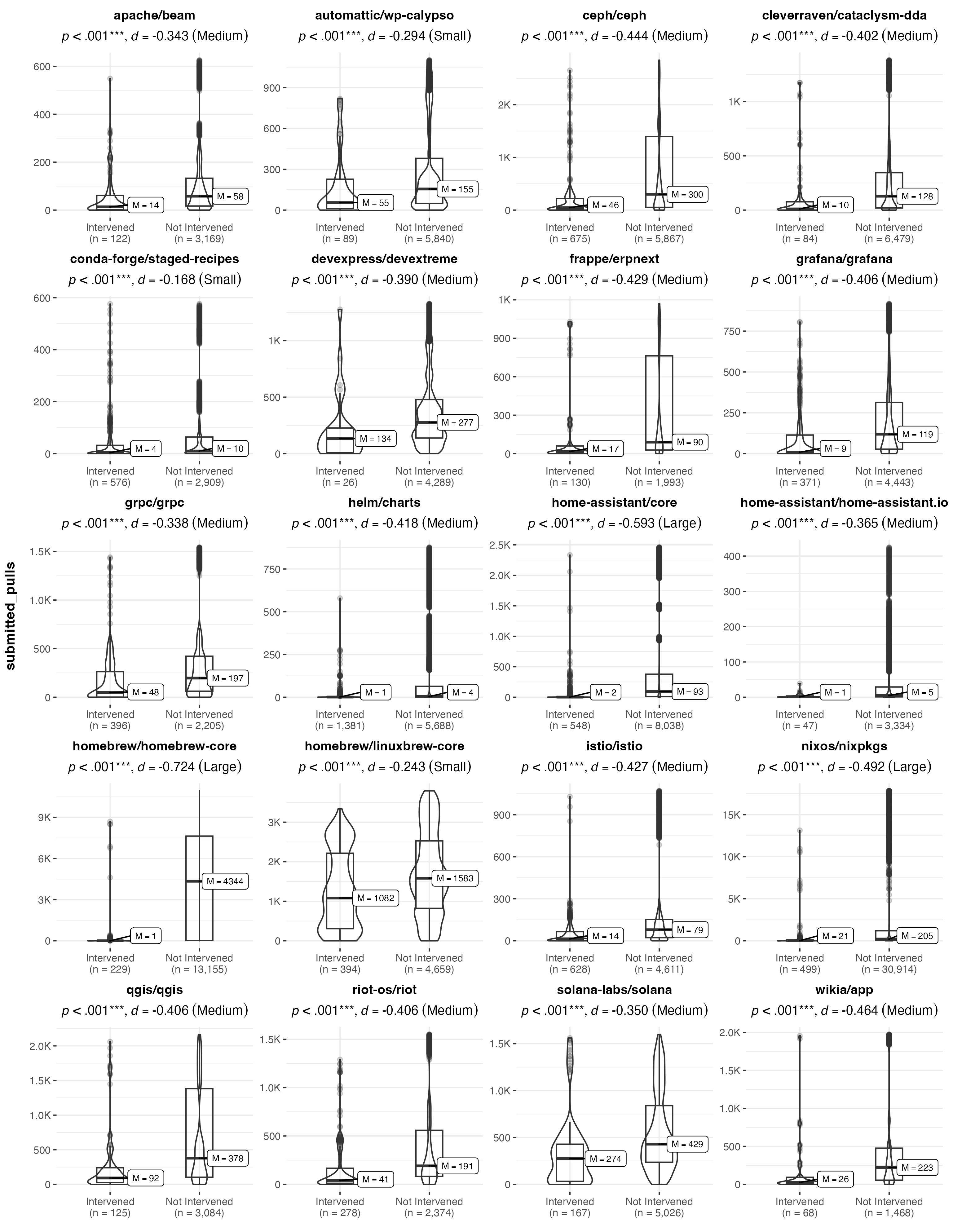}
    \caption{Comparison of intervened and not intervened PRs regarding the number of prior PRs by their contributors across the studied projects. *** $p < 0.001$, ** $p < 0.01$, * $p < 0.05$}
\end{figure}

\begin{figure}[H]
    \includegraphics[width=\textwidth]{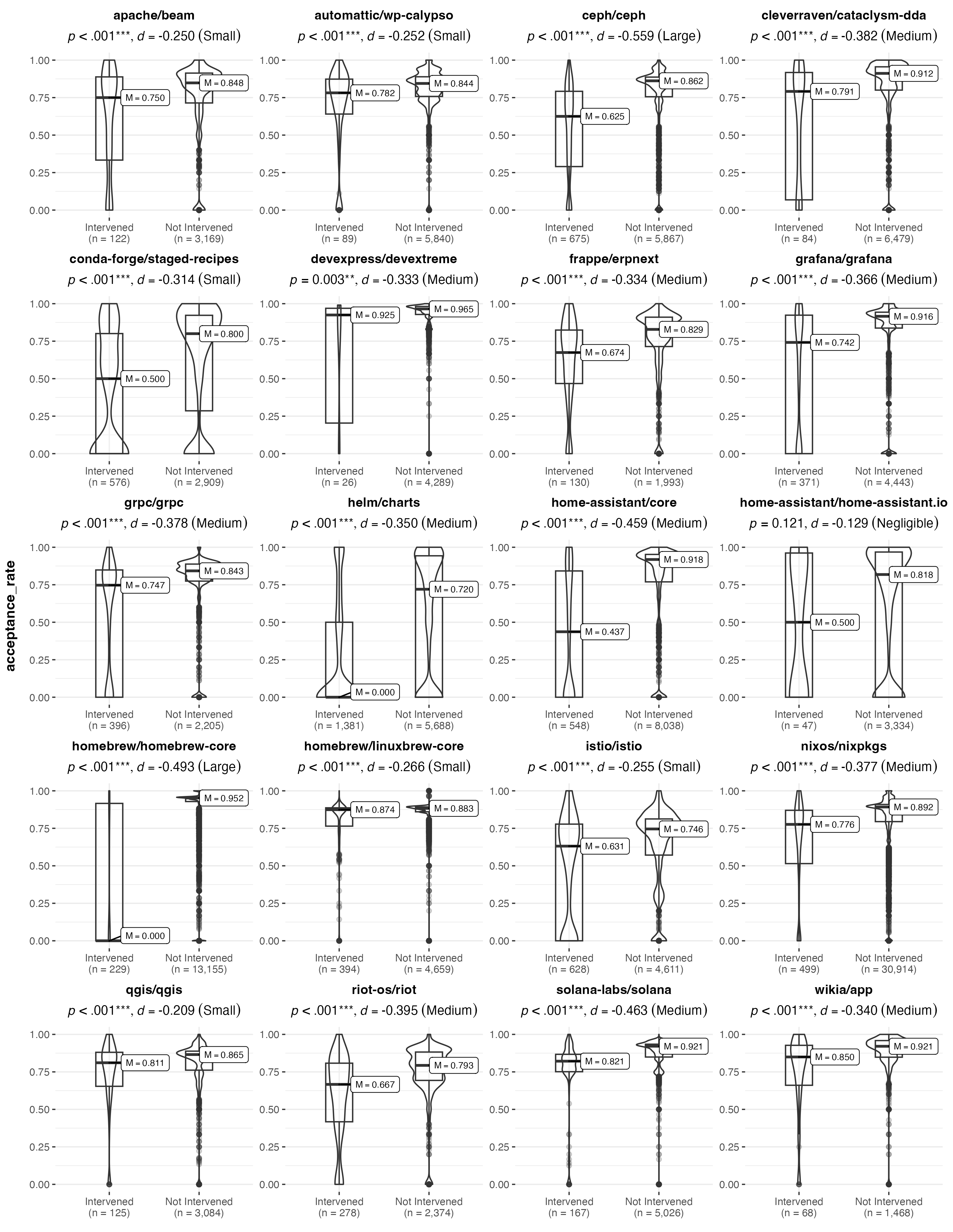}
    \caption{Comparison of intervened and not intervened PRs regarding the acceptance rate of their contributors across the studied projects. *** $p < 0.001$, ** $p < 0.01$, * $p < 0.05$}
\end{figure}

\begin{figure}[H]
    \includegraphics[width=\textwidth]{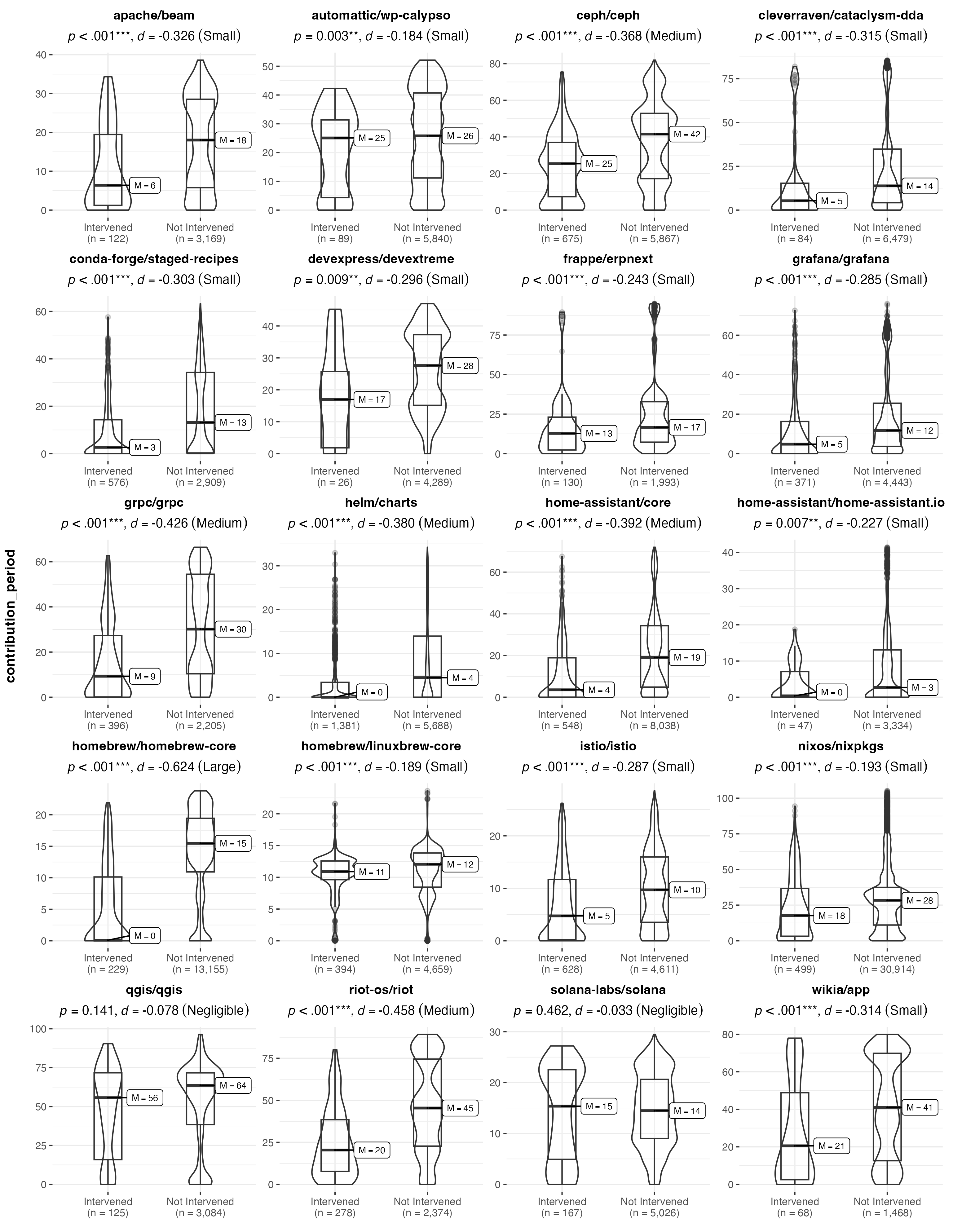}
    \caption{Comparison of intervened and not intervened PRs regarding the contribution period of their contributors across the studied projects. *** $p < 0.001$, ** $p < 0.01$, * $p < 0.05$}
\end{figure}

\begin{figure}[H]
    \includegraphics[width=\textwidth]{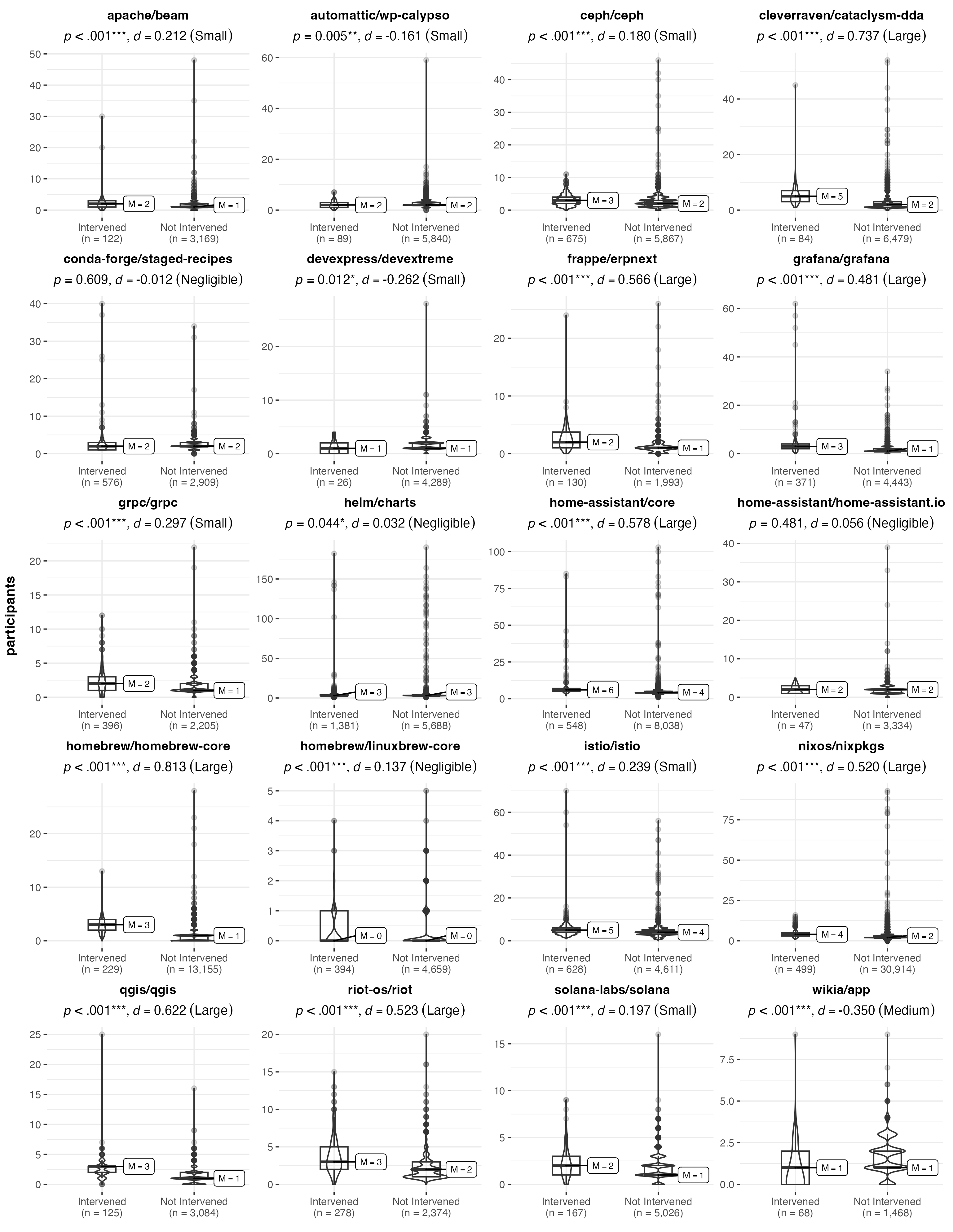}
    \caption{Comparison of intervened and not intervened PRs regarding the number of participants in their review process across the studied projects. *** $p < 0.001$, ** $p < 0.01$, * $p < 0.05$}
\end{figure}

\begin{figure}[H]
    \includegraphics[width=\textwidth]{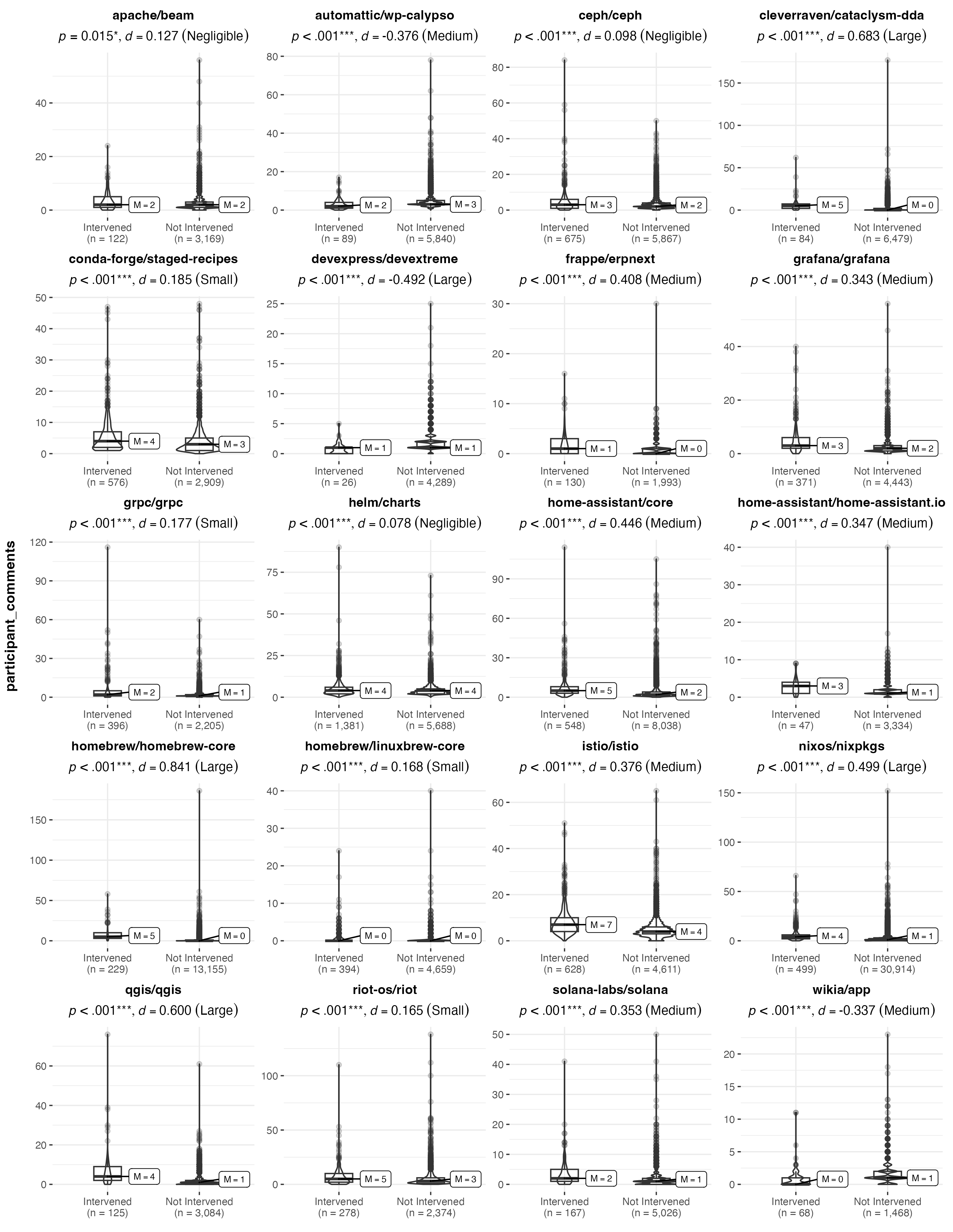}
    \caption{Comparison of intervened and not intervened PRs regarding the number of participant comments in their review process across the studied projects. *** $p < 0.001$, ** $p < 0.01$, * $p < 0.05$}
\end{figure}

\begin{figure}[H]
    \includegraphics[width=\textwidth]{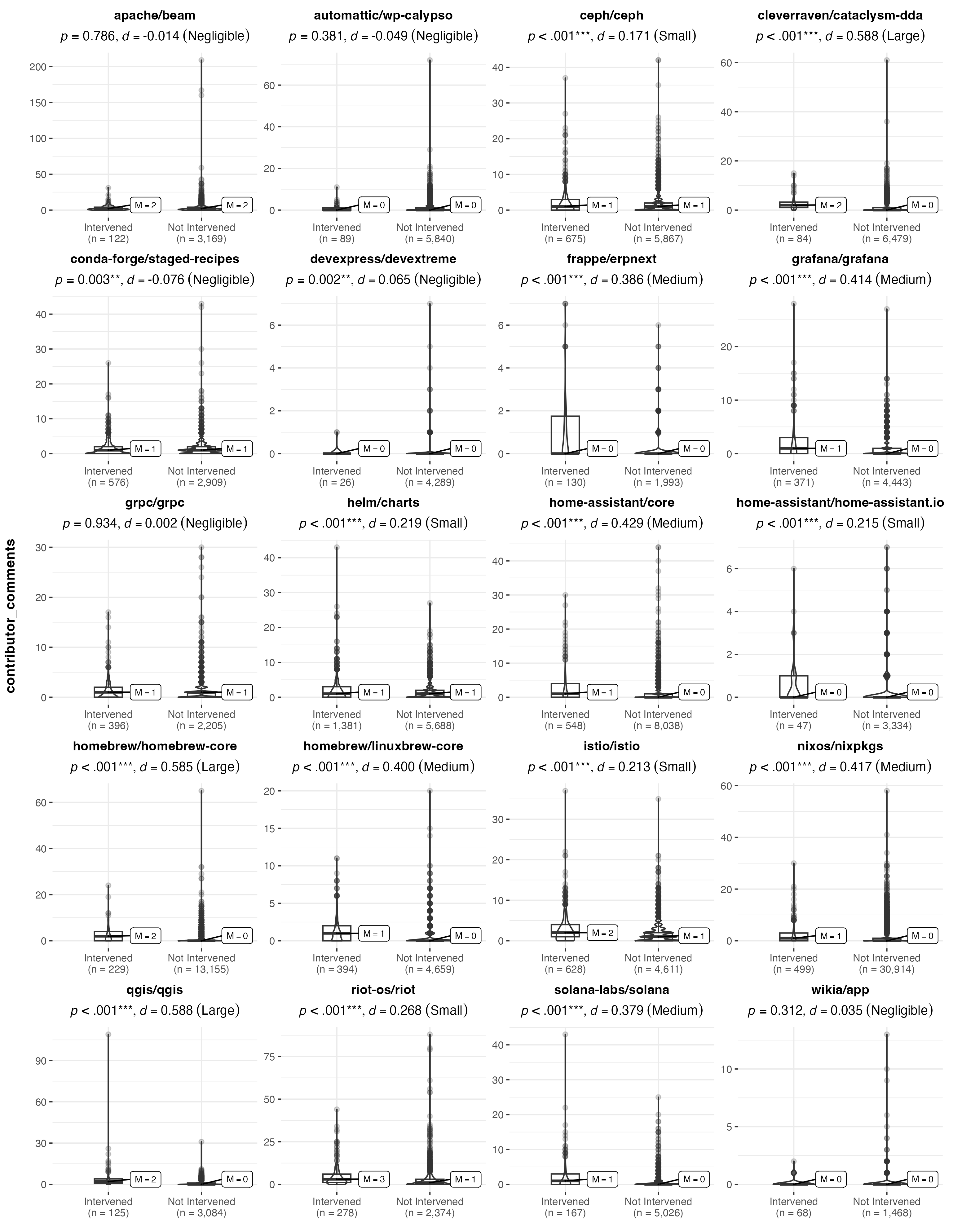}
    \caption{Comparison of intervened and not intervened PRs regarding the number of contributor comments in their review process across the studied projects. *** $p < 0.001$, ** $p < 0.01$, * $p < 0.05$}
\end{figure}

\begin{figure}[H]
    \includegraphics[width=\textwidth]{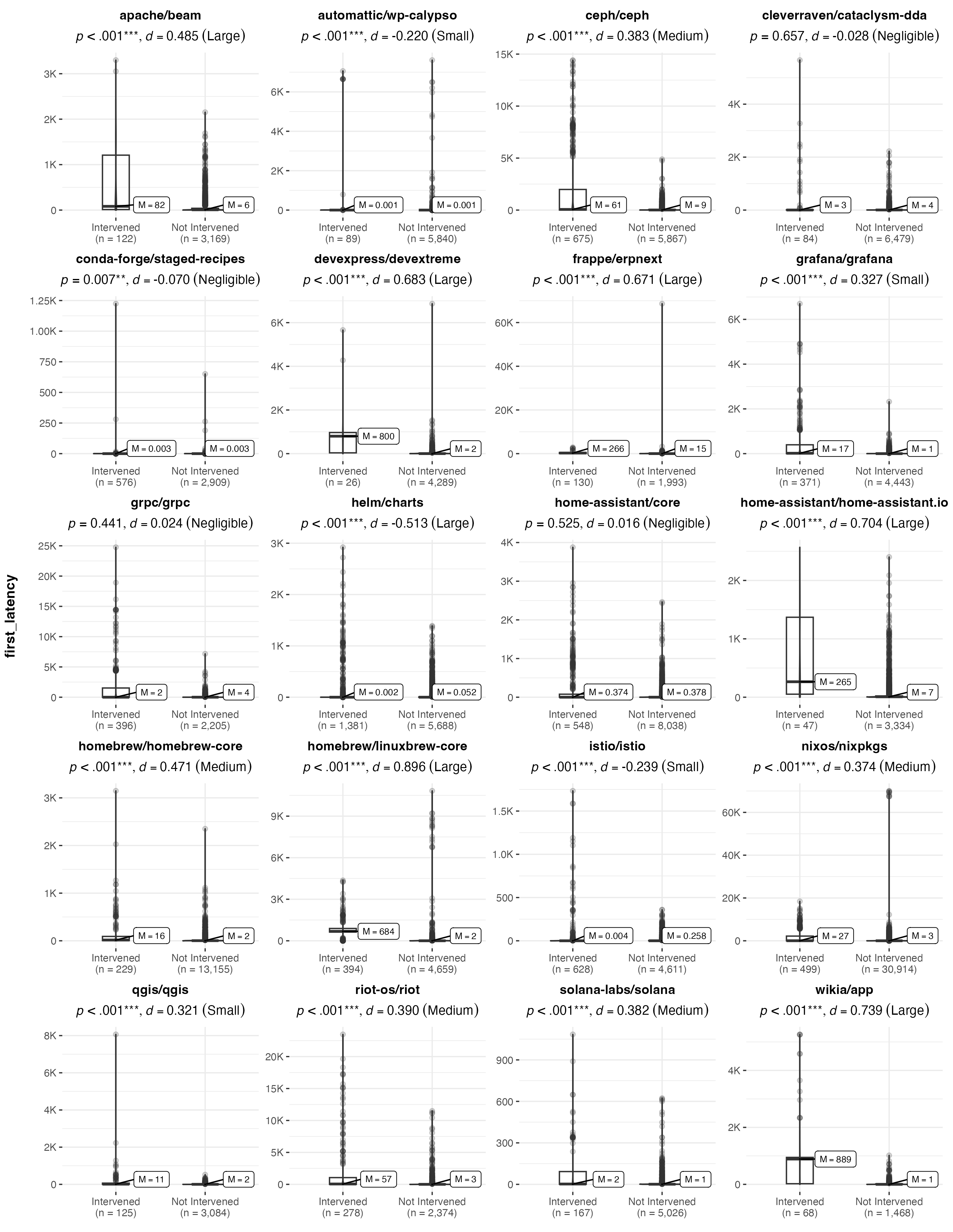}
    \caption{Comparison of intervened and not intervened PRs regarding their first response latency across the studied projects. *** $p < 0.001$, ** $p < 0.01$, * $p < 0.05$}
\end{figure}

\begin{figure}[H]
    \includegraphics[width=\textwidth]{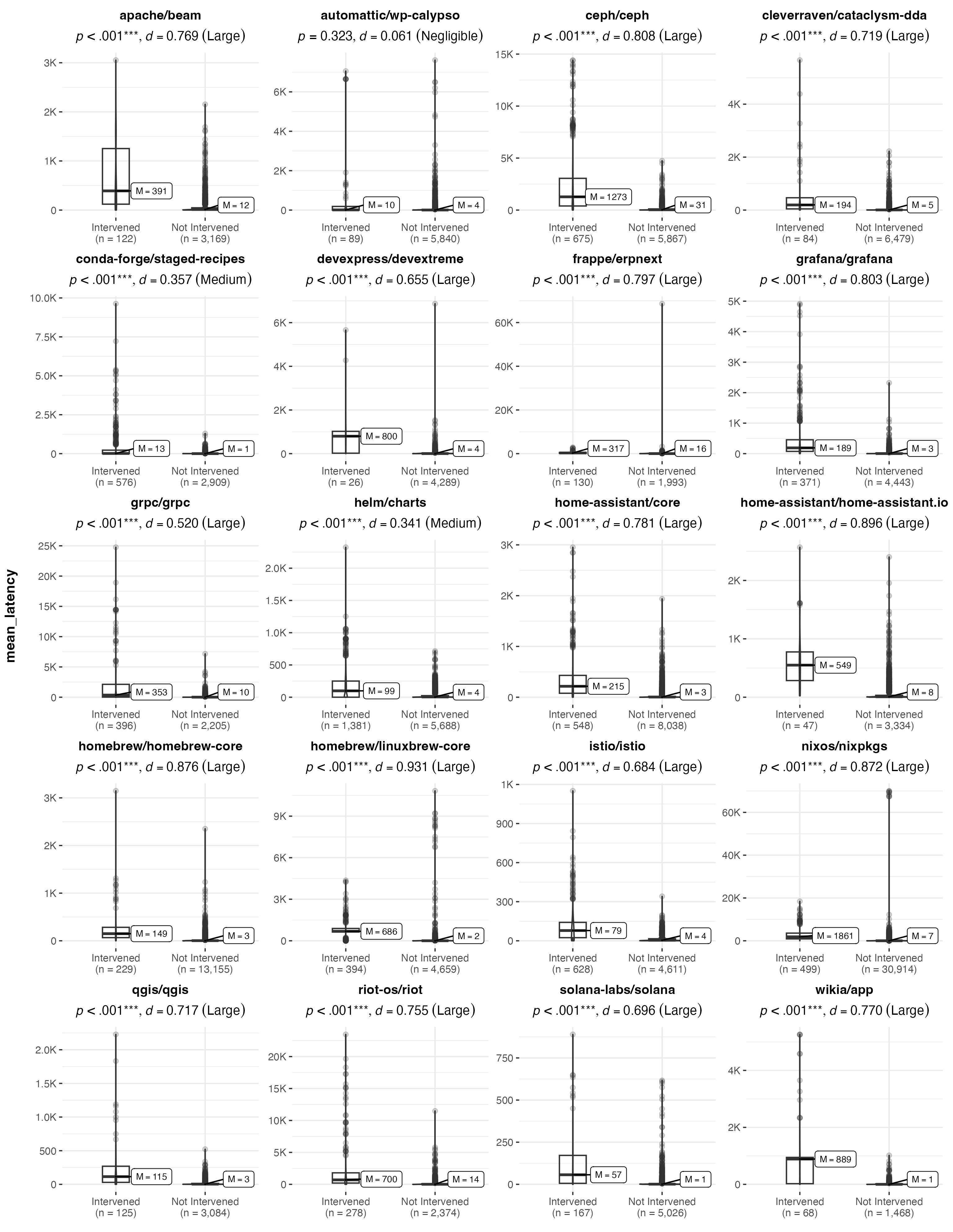}
    \caption{Comparison of intervened and not intervened PRs regarding their mean response latency across the studied projects. *** $p < 0.001$, ** $p < 0.01$, * $p < 0.05$}
\end{figure}

\begin{figure}[H]
    \includegraphics[width=\textwidth]{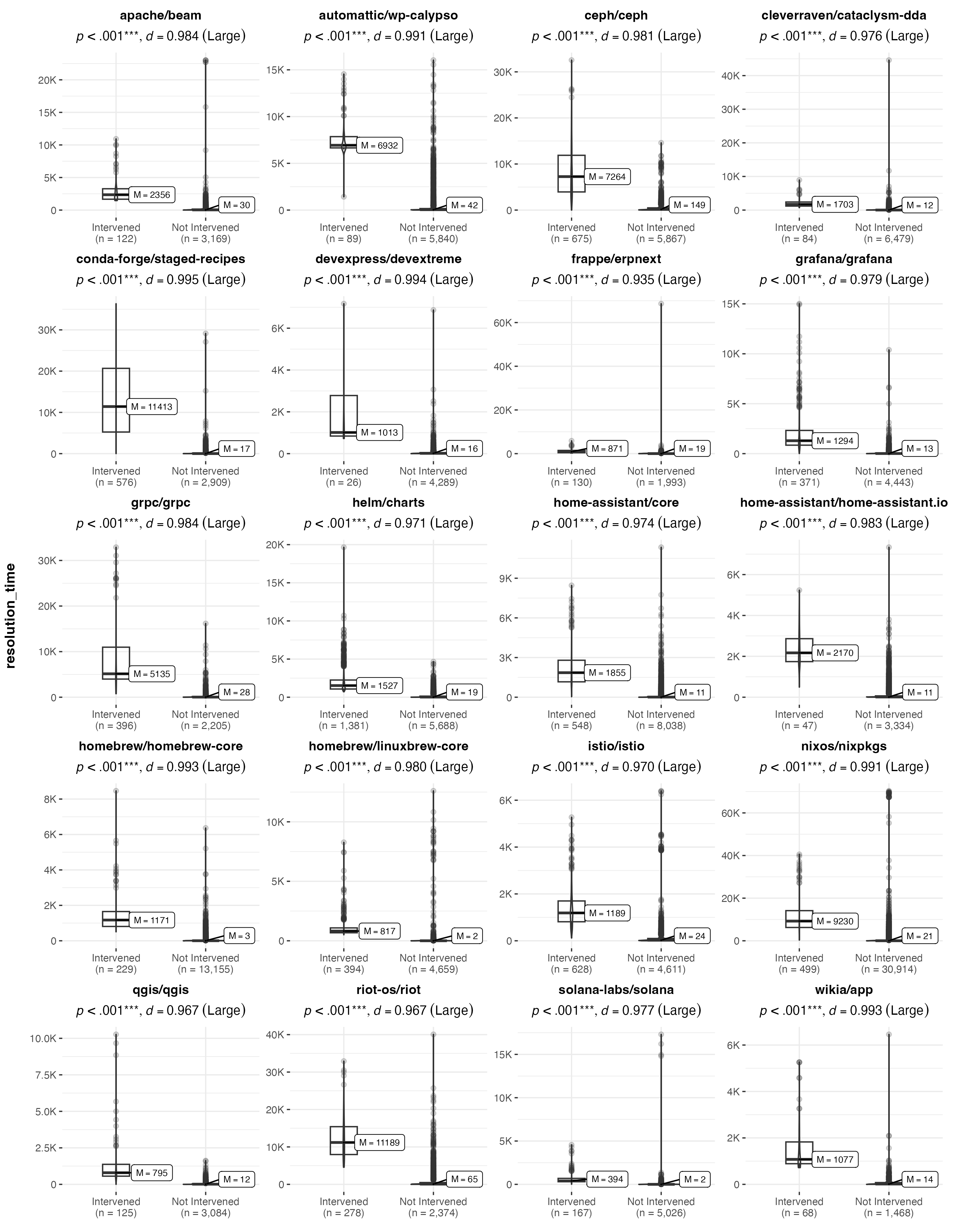}
    \caption{Comparison of intervened and not intervened PRs regarding their resolution time across the studied projects. *** $p < 0.001$, ** $p < 0.01$, * $p < 0.05$}
\end{figure}

\end{document}